\documentclass[11pt,a4paper,twoside,dvips]{report}

\usepackage{fancyheadings}
\usepackage[english]{babel}
\usepackage{amsmath,amssymb}
\usepackage[footnotesize,bf]{caption}
\usepackage{array,dcolumn}
\usepackage{enumerate}
\usepackage{graphicx}
\usepackage{feynmp}

\DeclareMathAlphabet   {\mathsc}{OT1}{cmr}{m}{sc}

\newcommand{\nochapter}[1]{
  \chapter*{#1}
  \markboth{#1}{#1}
  \addcontentsline{toc}{chapter}{#1}}

\newcommand{\+}        {\hphantom{+}}
\newcommand{\clearemptydoublepage}{\clearpage{\thispagestyle{empty}\cleardoublepage}}
\newcommand{\enref}    {\eqref}

\newcommand{\rhs}      {\mbox{r.h.s.}}
\newcommand{\hc}       {\mathrm{h.c.}}

\newcommand{\LT}       {\left}
\newcommand{\RT}       {\right}
\newcommand{\GZ}       {{\gamma Z}}
\newcommand{\pCL}      {\%~\mathrm{C.L.}}
\newcommand{\MeV}      {~\mathrm{MeV}}
\newcommand{\GeV}      {~\mathrm{GeV}}
\newcommand{\TeV}      {~\mathrm{TeV}}
\newcommand{\albar}    {\bar\alpha}
\newcommand{\altw}     {\delta\alpha_{t,W}}

\newcommand{\MS}       {\mathsc{\overline{MS}}}
\newcommand{\Ms}       {\mathsc{\overline{ms}}}
\newcommand{\SM}       {\mathsc{sm}}
\newcommand{\ND}       {\mathsc{nd}}
\newcommand{\NP}       {\mathsc{np}}
\newcommand{\GUT}      {\mathsc{gut}}
\newcommand{\SUSY}     {\mathsc{susy}}
\newcommand{\altwx}    {\delta\alpha_{t,W,X}}

\newcommand{\tdB}      {{\Tilde{B}}}
\newcommand{\tdW}      {{\Tilde{W}}}
\newcommand{\tdZ}      {{\Tilde{Z}}}
\newcommand{\tdH}      {{\Tilde{H}}}
\newcommand{\tdC}      {{\Tilde{\chi}}}

\newenvironment{TempGraph}{}{}

\begin{document}
\setlength{\unitlength}{1mm}

\setlength{\normalbaselineskip} {1.0905\baselineskip}
\normalbaselines

\allowdisplaybreaks

\pagenumbering{roman}

\author{Michele Maltoni}
\title{Precision measurements of IVB parameters and bounds on new physics}

\thispagestyle{empty}

\vspace{-3cm}
\begin{center}
    \underline{\LARGE\sc Universit\`a degli studi di Ferrara} \\[3mm]
    {\Large\sc Facolt\`a di Scienze Matematiche, Fisiche e Naturali} \\[3mm]
    {\Large\sc Dipartimento di Fisica}
    
    \vfill

    \raisebox{0mm}[0mm][0mm]{\Huge\bf Precision measurements} \\[8mm]
    \raisebox{0mm}[0mm][0mm]{\Huge\bf of IVB parameters} \\[8mm]
    \raisebox{0mm}[0mm][0mm]{\Huge\bf and bounds on New Physics}
\end{center}

\vfill

{\it Relatore \hfill Candidato} \\
{\rm Chiar.mo Prof. \bf Mikhail~I.~Vysotsky \hfill
  \rm Dott. \bf Michele~Maltoni}

\vspace{1.5cm}

{\it Relatore} \\
{\rm Chiar.mo Prof. \bf Giovanni~Fiorentini}

\vfill

\begin{center}
    {\Large\sc Dottorato di Ricerca in Fisica - XII ciclo}
\end{center}

\clearemptydoublepage

{ 
  \renewcommand{\MakeUppercase}[1]{#1}
  
  \tableofcontents
  \clearemptydoublepage
  
  \listoftables
  \clearemptydoublepage
  
  \listoffigures
  \clearemptydoublepage
}

\pagenumbering{arabic}

\nochapter{Introduction}

In the last 30 years, since the formulation of the \emph{Standard Model}
(SM) as a unified theory for electromagnetic and weak interactions, particle
physics has enjoyed a period of great success. All the predictions of the
SM, except for the existence of the Higgs boson which still has to be
observed, have so far been confirmed with very good accuracy, and recent
fits including LEP~I, LEP~II and SLC data are characterized by the excellent
value $\chi^2 / \text{d.o.f} = 14.4/14$, which cannot be better.

However, despite of its success, the SM is currently not believed to be the
ultimate theory of Nature. The absence of a relation between the electroweak
and strong coupling constants, the need of a fine tuning to protect scalar
masses, the large number of free parameters are all problems which since
many years are leading theoretical physicists towards the formulation of
``more fundamental'' theories, of course having the SM as a low-energy
limit. Among these theories, the concept of \emph{supersymmetry} plays a
central role, and despite of the fact that no evidence of superpartners has
so far been found it is widely believed that supersymmetry will ultimately
occur at some energy scale.

Many of the present experimental confirmations of the SM, as well as most of
the present bounds on its extensions, come from accelerator physics.
\emph{Direct search} experiments have reached a very high level of accuracy,
and recent results from LEP~II and Tevatron allow to put stringent bounds on
New Physics. However, there are special situations where direct search
analysis, for some peculiar reason, fails, and in these cases experimental
bounds become sensibly weaker. When this happens, the study of
\emph{radiative corrections} to electroweak observables emerges as the only
way to investigate the existence of New Physics in these domains, and in the
last years the analysis of \emph{precision measurements} has proved to be a
strong tool for constraining wide regions of parameter space.

The main purpose of this thesis is to discuss the impact of precision
measurements of \emph{intermediate vector boson} (IVB) parameters on the
present knowledge of particle physics, both in the framework of the Standard
Model and of its most straightforward extensions. The research presented in
the following chapters is organized in a natural way, starting from the SM
and then moving towards New Physics.

In Chapter~\ref{sec:DEF}, we give a general overview of the Standard Model
and of its supersymmetric extension. We do not present here any original
result, nor we pretend to be exhaustive, but simply provide the technical
background for the next chapters and discuss the main motivations for
looking beyond the Standard Model. The structure of this chapter closely
resembles that of the whole thesis, following on a smaller scale the same
logical line.

The rest of this thesis is mainly oriented into three complementary
subjects:

\begin{enumerate}[(1)]
  \item\label{misc:INT.a} the study of the relations between the
    phenomenological electroweak mixing angle $\theta$, the $\MS$ parameter
    $\hat\theta$ and the effective angle $\theta_l$ describing decay of $Z$
    boson into charged leptons;
    
  \item\label{misc:INT.b} the analysis of contributions of ``charginos
    almost degenerate with the lightest neutralino'' to oblique radiative
    corrections, and bounds on mass of these particles from precision data;
    
  \item\label{misc:INT.c} the possibility to have extra fermion generations
    without spoiling precision data description, both in the framework of
    the SM and of its minimal supersymmetric extension (MSSM).
\end{enumerate}

Concerning point~\enref{misc:INT.a}, in Chapter~\ref{sec:ANG} we analyze the
effects of radiative corrections in the electroweak sector of the SM, with
particular attention to various definitions of the electroweak mixing angle
$\theta$ (or equivalently of its sine squared, $s^2$). It happens that $s^2$
and $\hat s^2$ are equal with $0.1\%$ accuracy, though they are split by
radiative corrections and a natural estimate for their difference is $1\%$. We
study the origin of this degeneracy and show that it occurs only if the top
mass is close to $170\GeV$, so no deep physical reason can be attributed to
it. However, another puzzle of the Standard Model, the degeneracy of $s_l^2$
and $s^2$, is not independent of the previous one since a good physical reason
exists for $s_l^2$ and $\hat s^2$ degeneracy. We also present explicit
formulas relating these three angles.

The analysis of the relations between different definitions of the
electroweak mixing angle is then extended in Chapter~\ref{sec:RUN} from the
SM to New Physics, dealing with some aspects of the running of $\MS$
quantities from low (electroweak) to high (unification) energy scale. In
particular, we emphasize how two apparently different approaches to the
running of coupling constants, the one in which high-energy new physics is
NOT decoupled from low-energy scale quantities, and the one in which these
contributions are decoupled but thresholds are introduced, are essentially
equivalent. As an application of the techniques described here, we conclude
discussing what is the value of $\hat{\alpha}_s$ which can be predicted from
the demand of SUSY Grand Unification.

Concerning point~\enref{misc:INT.b}, in Chapter~\ref{sec:GAU} we study the
case of a chargino almost degenerate with the lightest neutralino in the
framework of the MSSM. In the general case, gaugino contributions to oblique
radiative corrections cannot be written analytically, but for the special
case considered here this is possible and the corresponding expressions are
reported. Then we analyze effects on precision measurements, giving lower
bounds for nearly degenerate chargino/neutralino. It is important to note
that for the considered case experimental bounds from direct search are
rather weak, just half of the $Z$ boson mass, and the study of oblique
corrections allows a concrete improvements of these bounds.

Finally, in Chapter~\ref{sec:GEN} we deal with point~\enref{misc:INT.c},
investigating the effects of new fermion generations on precision
measurements. We show that even one extra generation with all particles
heavier than $Z$ boson is strongly disfavored by experimental data. However,
for the specific case of a heavy neutrino (around $50\GeV$ in mass), the
situation changes and the quality of the fit is not worse than the SM.
Moreover, as a direct application of the results of the previous chapter,
effects of almost degenerate chargino/neutralino on a new generation are
discussed, showing that also in this case experimental bounds are relaxed.

The results discussed in this thesis are presented in the published
papers~\cite{Maltoni98b,Maltoni99a,Maltoni99b} and in
Ref.~\cite{Maltoni99c}, currently submitted for publication on PLB. Also,
results quoted in Chapters~\ref{sec:GAU} and~\ref{sec:GEN} were presented at
the conference PASCOS99, and will appear in the
proceedings~\cite{Maltoni99d,Maltoni99e}. Other results, not contained here,
are in Refs.~\cite{Burigana97,Burigana99,Maltoni98a}.

\clearemptydoublepage

\begin{fmffile}{Feynman/definitions}

\chapter{General properties of the Standard Model and SUSY extensions}
\label{sec:DEF}

The purpose of this chapter is to provide a general overview on some
concepts which will be widely used in the following chapters. Its structure
closely resembles that of the present thesis: we start from the general
properties of the tree-level Lagrangean of the electroweak interactions in
Sec.~\ref{sec:DEF.10}, then we extend our analysis to include effects of
radiative corrections in Sec.~\ref{sec:DEF.20}, and finally we outline the
main problems of the Standard Model and introduce supersymmetry in
Sec.~\ref{sec:DEF.30}. We do not intend to be exhaustive, and we will only
consider topics which are relevant in the framework of the present
dissertation.

\section{The Standard Model} \label{sec:DEF.10}

The present theory of the electroweak interactions, known as ``Standard
Model'', is the Glashow-Salam-Weinberg theory~\cite{Glashow61} of leptons,
extended to quarks~\cite{Cabibbo63} and made anomaly free through the
introduction of the concept of color. Intermediate vector bosons get mass
through the \emph{Higgs mechanism}, and Fermi model arises as a low energy
effective theory; \emph{spontaneous symmetry breaking} is also used to
generate leptons and quark masses. All the predictions of the SM are
actually in perfect agreement with all the experimental data.

\subsection{The gauge and higgs sector}

The core of the Standard Model is a renormalizable Yang-Mills
theory~\cite{Yang54} based on the non-Abelian gauge group $SU(2)_L \times
U(1)_Y$~\cite{Hollik93,Guadagnini97}. The generators of the corresponding
Lie algebra are the three isospin operators $I_1$, $I_2$, $I_3$ and the
hypercharge $Y$. Each of these operators is associated with a vector field,
so the model includes the isotriplet $W_\mu^a$ and the isosinglet $B_\mu$.
The corresponding field strengths are:
\begin{align}
    W_{\mu\nu}^a & = \partial_\mu W_\nu^a - \partial_\nu W_\mu^a
      - g \epsilon^{abc} W_\mu^b W_\nu^c, \\
    B_{\mu\nu} & = \partial_\mu B_\nu - \partial_\nu B_\mu.
\end{align}
The Lagrangean for the pure gauge fields can be written in the usual way:
\begin{equation}
    \mathcal{L}_G = -\frac{1}{4} W_{\mu\nu}^a W^{\mu\nu}_a
    - \frac{1}{4} B_{\mu\nu} B^{\mu\nu}.
\end{equation}
In addition to these vector particles, the model also includes an $SU(2)_L$
doublet $\Phi$ of complex scalar fields with hypercharge $Y = 1$. This field
is known as \emph{Higgs} field and the corresponding Lagrangean is:
\begin{gather}
    \label{eq:DEF.120l} \mathcal{L}_H = \LT( D_\mu \Phi \RT)^\dagger
      \LT( D_\mu \Phi \RT) - V(\Phi^\dagger \Phi), \\
    \label{eq:DEF.120v} V(\Phi^\dagger \Phi) = \frac{\lambda}{4}
      \LT( \Phi^\dagger \Phi - \frac{\eta^2}{2} \RT)^2,
\end{gather}
with the covariant derivative:
\begin{equation} \label{eq:DEF.130}
    D_\mu \Phi = \LT( \partial_\mu + i g W_\mu^a I_a
    + \frac{1}{2} i g' B_\mu \RT) \Phi.
\end{equation}
The whole Lagrangean $\mathcal{L}_G + \mathcal{L}_H$ is invariant under
$SU(2)_L \times U(1)_Y$ gauge transformations:
\begin{align}
    \label{eq:DEF.140b} B_\mu & \to B'_\mu = B_\mu - \frac{1}{g'} 
      \partial_\mu \alpha, \\
    \label{eq:DEF.140w} W_\mu \equiv W_\mu^a I_a & \to W'_\mu =
      e^{i \beta_a I_a} \LT( W_\mu - \frac{i}{g} \partial_\mu \RT)
      e^{-i \beta_a I_a}, \\
    \label{eq:DEF.140h} \Phi & \to \Phi' =
      e^{i \alpha Y/2 + i \beta_a I_a} \Phi.
\end{align}
The presence of a gauge symmetry is responsible for the renormalizability of
the model. However, the scalar potential $V$, which describes the Higgs
self-interaction, is constructed in such a way that its minima occur for a
non-vanishing value of $\Phi$:
\begin{equation} \label{eq:DEF.150}
    V(\Phi^\dagger \Phi) = 0 \quad\Leftrightarrow\quad
    \Phi^\dagger \Phi = \frac{\eta^2}{2} \quad\Rightarrow\quad
    \LT< \Phi \RT> \ne 0.
\end{equation}
It is convenient to rewrite $\Phi$ in the following way:
\begin{equation} \label{eq:DEF.160}
    \Phi = \frac{1}{\sqrt{2}} e^{-i \theta_a I_a}
    \begin{pmatrix}
	0 \\ \eta + H
    \end{pmatrix},
\end{equation}
where $\theta_a$ and $H$ are real scalar fields. Comparing
Eq.~\eqref{eq:DEF.160} with~\eqref{eq:DEF.150}, it is straightforward to see
that any field configuration $\bar{\Phi}$ satisfying the
condition~\eqref{eq:DEF.150} is characterized by a specific choice of the
three fields $\bar{\theta}_a$ and has $\bar{H} \equiv 0$. Now, looking at
Eq.~\eqref{eq:DEF.140h} it is clear that such a ground state $\bar{\Phi}$ is
\emph{not} invariant under a generic gauge transformation. Therefore, the
gauge symmetry obeyed by the \emph{Lagrangean} is not respected by the
\emph{vacuum} of the theory; this mechanism is called \emph{spontaneous
symmetry breaking}, and within the framework of the SM it is responsible for
all the gauge bosons (as well as leptons and quarks) to acquire a non-zero
mass. To understand how it happens, let us substitute Eq.~\eqref{eq:DEF.160}
into~\eqref{eq:DEF.120l}, so to rewrite $\mathcal{L}_H$ in terms of the new
fields $\theta_a$ and $H$. It is easy to see that all the $\theta_a$ can be
reabsorbed into $W_\mu^a$ by means of a gauge
transformation~(\ref{eq:DEF.140b}-\ref{eq:DEF.140h}) having $\beta_a =
\theta_a$ and $\alpha = 0$:
\begin{align}
    \label{eq:DEF.170b} B'_\mu & = B_\mu, \\
    \label{eq:DEF.170w} W'_\mu & = e^{i \theta_a I_a}
      \LT( W_\mu - \frac{i}{g} \partial_\mu \RT) e^{-i \theta_a I_a}, \\
    \label{eq:DEF.170h} \Phi' & = \frac{1}{\sqrt{2}}
    \begin{pmatrix}
	0 \\ \eta + H
    \end{pmatrix}.
\end{align}
Since both $\mathcal{L}_G$ and $\mathcal{L}_H$ are invariant under
transformations~(\ref{eq:DEF.140b}-\ref{eq:DEF.140h}), their new expressions
in terms of $B'_\mu$, ${W'_\mu}^a$ and $\Phi'$ are identical to the old ones
in terms of $B_\mu$, $W_\mu^a$ and $\Phi$, so the only visible effect
of~(\ref{eq:DEF.170b}-\ref{eq:DEF.170h}) is to make the fields $\theta_a$ to
disappear from the Lagrangean. However, to achieve this result we have to
pay the price of \emph{fixing} three of the four gauge parameters in a
proper way, and as a consequence the symmetry of the model decreases. The
only gauge transformation which is still allowed is the one preserving the
form~\eqref{eq:DEF.170h} of $\Phi'$, i.e.\ the $U(1)_\mathrm{em}$ subgroup
of transformations~(\ref{eq:DEF.140b}-\ref{eq:DEF.140h}) having $\beta_1 =
\beta_2 = 0$ and $\beta_3 = -\alpha$. In this way, the original $SU(2)_L
\times U(1)_Y$ symmetry has been broken to $U(1)_\mathrm{em}$.

For simplicity, let us drop the prime
in Eqs.~(\ref{eq:DEF.170b}-\ref{eq:DEF.170h}) and denote the new fields
$B'_\mu$, ${W'_\mu}^a$ and $\Phi'$ by $B_\mu$, $W_\mu^a$ and $\Phi$. The scalar
potential $V$ given in Eq.~\eqref{eq:DEF.120v} can be written as:
\begin{equation} \label{eq:DEF.180}
    V(H) = \frac{\lambda^2 \eta^2}{4} H^2 + \frac{\lambda^2 \eta}{4} H^3 
    + \frac{\lambda^2}{16} H^4,
\end{equation}
while from Eq.~\eqref{eq:DEF.130} and~\eqref{eq:DEF.170h} we have:
\begin{equation} \begin{split} \label{eq:DEF.190}
    \LT( D_\mu \Phi \RT)^\dagger \LT( D_\mu \Phi \RT) =
    \frac{1}{2} \partial_\mu H \partial^\mu H
    & + \frac{g^2}{8} \LT( W_\mu^1 W^\mu_1 + W_\mu^2 W^\mu_2 \RT)
    \LT( \eta + H \RT)^2 \\
    & + \frac{1}{8} \LT( g W_\mu^3 - g' B_\mu \RT)^2 
    \LT( \eta + H \RT)^2. \\
\end{split} \end{equation}
Looking at this expression we see that the real field $H$ (called the
\emph{physical} Higgs field) describes a particle with mass $m_H = \eta
\sqrt{\lambda/2}$. Also, Eq.~\eqref{eq:DEF.190} provides the mass terms
for the $W_\mu^a$ and $B_\mu$ fields, which therefore are no longer massless
fields. The mass Lagrangean is:
\begin{equation} \label{eq:DEF.200}
    \mathcal{L}_m = \frac{g^2 \eta^2}{8}
    \LT( W_\mu^1 W^\mu_1 + W_\mu^2 W^\mu_2 \RT) + \frac{\eta^2}{8}
    \begin{pmatrix}
	W_\mu & B_\mu
    \end{pmatrix}
    \begin{pmatrix}
	~g  g & -g' g \\
	-g' g & ~g' g'
    \end{pmatrix}
    \begin{pmatrix}
	W^\mu \\ B^\mu
    \end{pmatrix}.
\end{equation}
To diagonalize the mass matrix in the second term of Eq.~\eqref{eq:DEF.200},
let's introduce an angle $\theta$ (known as the \emph{electroweak angle})
and let's denote by $c \equiv \cos\theta$ and $s \equiv \sin\theta$ its sine
and cosine. It is convenient to define the following fields:
\begin{gather}
    \label{eq:DEF.210w} W_\mu^\pm = 
      \frac{W_\mu^1 \mp i W_\mu^2}{\sqrt{2}}, \\[2mm]
    \label{eq:DEF.210z} \begin{pmatrix}
	Z_\mu \\ A_\mu
    \end{pmatrix} = 
    \begin{pmatrix}
	c & -s \\
	s & ~c
    \end{pmatrix}
    \begin{pmatrix}
	W_\mu^3 \\ B_\mu
    \end{pmatrix}.
\end{gather}
Substituting Eq.~\eqref{eq:DEF.210z} into~\eqref{eq:DEF.200} we see that
$\LT( A_\mu, Z_\mu \RT)$ are mass eigenstates \emph{if} $\theta$ is related
to $g$ and $g'$ by the expression:
\begin{equation} \label{eq:DEF.220}
    c = \frac{g}{\sqrt{g^2 + {g'}^2}}, \qquad
    s = \frac{g'}{\sqrt{g^2 + {g'}^2}},
\end{equation}
It is easy to see that the field $A_\mu$ is massless, and that the residual
$U(1)_\mathrm{em}$ gauge symmetry corresponds to transformations $A_\mu \to
A_\mu - \partial_\mu \chi$; therefore, we can identify $A_\mu$ with the
electromagnetic field. The masses of the $W^\pm$ and $Z$ bosons are:
\begin{equation} \label{eq:DEF.230}
    \LT. 
    \begin{aligned}
	m_W & = \frac{1}{2} g \eta \\
	m_Z & = \frac{1}{2} f \eta
    \end{aligned} \RT\} \quad \Rightarrow \quad m_W = m_Z c,
\end{equation}
where $f \equiv \sqrt{g^2 + {g'}^2}$. 

Interactions among $W^\pm$, $Z$, $A$, $H$ are described by the cubic and
quartic terms in the Lagrangeans $\mathcal{L}_G$ and $\mathcal{L}_H$. In
particular, one finds that $Z$ and $H$ do not couple to $A$ (i.e.\ they are
\emph{neutral}), while $W^\pm$ is \emph{charged} and its coupling constant
to the photon field is $gs$.

\subsection{The fermion sector}

Fermions observed in Nature can be divided into two classes, depending on
whether or not they participate to \emph{strong} (color) interactions:
\emph{leptons}, which are color singlets and only have electromagnetic and
weak interactions, and \emph{quarks}, which are color triplets. Both of them
can be further divided into three \emph{families}, or generations, all
carrying the same $SU(3)_c \times SU(2)_L \times U(1)_Y$ quantum numbers and
differing from one another only in masses. The general structure of the
fermionic sector of the SM is shown in Table~\ref{tab:DEF.10}.

\begin{table}[!t] \centering
    \setlength{\extrarowheight}{1mm}
    \begin{tabular}{ccc@{\hspace{15mm}}ccc@{\hspace{15mm}}c}
	\hline\hline
	I & II & III & $SU(3)_c$ & $SU(2)_L$ & $U(1)_Y$ & $Q$ \\
	\hline
	$\LT(\begin{gathered} \nu_e    \\ e_L    \end{gathered}\RT)$ &
	$\LT(\begin{gathered} \nu_\mu  \\ \mu_L  \end{gathered}\RT)$ &
	$\LT(\begin{gathered} \nu_\tau \\ \tau_L \end{gathered}\RT)$ &
	$\mathbf{1}$ & $\mathbf{2}$ & $-1$ & $\LT(\begin{gathered} \+0 \\ -1 \end{gathered}\RT)$ \\
	$e_R$ & $\mu_R$ & $\tau_R$ & $\mathbf{1}$ & $\mathbf{1}$ & $-2$ & $-1$ \\
	\hline
	$\LT(\begin{gathered} u_L \\ d_L \end{gathered}\RT)$ &
	$\LT(\begin{gathered} c_L \\ s_L \end{gathered}\RT)$ &
	$\LT(\begin{gathered} t_L \\ b_L \end{gathered}\RT)$ &
	$\mathbf{3}$ & $\mathbf{2}$ & $\+1/3$ & $\LT(\begin{gathered} \+2/3 \\ -1/3 \end{gathered}\RT)$ \\
	$u_R$ & $c_R$ & $t_R$ & $\mathbf{3}$ & $\mathbf{1}$ & $\+4/3$ & $\+2/3$ \\
	$d_R$ & $s_R$ & $b_R$ & $\mathbf{3}$ & $\mathbf{1}$ & $-2/3$  & $-1/3$ \\
	\hline\hline
    \end{tabular}
    \caption[Particle contents of the fermionic sector of the Standard
    Model]{
      The particle contents of the fermionic sector of the Standard Model.
      If the mixing among quarks is neglected, the three generations are
      identical to one another (apart from particle masses) and independent.
      \emph{Leptons} are $SU(3)_c$ singlets, and \emph{quarks} are $SU(3)_c$
      triplets. Note that no right-handed neutral lepton is present in the
      minimal model.}
    \label{tab:DEF.10}
\end{table}

In the present discussion, we will neglect mixing among quarks, since in
this thesis we will never make use of this concept. With this simplifying
assumption, all the three families are independent from one another, and the
Lagrangean describing interactions of leptons and quarks with $W^\pm$, $Z$,
$A$, $H$ bosons is naturally split into three identical parts. Therefore, to
describe fermions in a simple way we will introduce a generation index
$f=1,2,3$, and denote by $l_L^f$, $q_L^f$ each lepton and quark left
$SU(2)_L$ doublet and by $E_R^f$, $U_R^f$, $D_R^f$ each down-lepton,
up-quark and down-quark right $SU(2)_L$ singlet, respectively. With this
notation, the fermionic part of the SM Lagrangean is:
\begin{align}
    \label{eq:DEF.310l} \mathcal{L}_l & = i \sum_f \LT[ 
      {l_L^f}^\dagger \bar{\sigma}^\mu D_\mu l_L^f 
      + {E_R^f}^\dagger \sigma^\mu D_\mu E_R^f \RT]
      - \sum_f \lambda_E^f \LT[ {l_L^f}^\dagger \Phi E_R^f + \hc \RT], \\
    \begin{split} \label{eq:DEF.310q} 
	\mathcal{L}_q & = i \sum_f \LT[ {q_L^f}^\dagger \bar{\sigma}^\mu
	  D_\mu q_L^f  + {U_R^f}^\dagger \sigma^\mu D_\mu U_R^f
	  + {D_R^f}^\dagger \sigma^\mu D_\mu D_R^f \RT] \\
	& - \sum_f \lambda_U^f \LT[ {q_L^f}^\dagger \tilde{\Phi} U_R^f + \hc \RT]
	  - \sum_f \lambda_D^f \LT[ {q_L^f}^\dagger \Phi D_R^f + \hc \RT],
    \end{split}
\end{align}
where $\sigma^i$ are the Pauli matrices:
\begin{gather}
    \sigma^0 =
    \begin{pmatrix} 
	1 & 0 \\ 
	0 & 1 
    \end{pmatrix}, \qquad \sigma^1 =
    \begin{pmatrix} 
	0 & 1 \\ 
	1 & 0 
    \end{pmatrix}, \qquad \sigma^2 =
    \begin{pmatrix} 
	0 &  -i \\ 
	i & \+0 
    \end{pmatrix}, \qquad \sigma^3 =
    \begin{pmatrix} 
	1 & \+0 \\ 
	0 &  -1 
    \end{pmatrix}, \\[2mm]
    \bar{\sigma}^0 =  \sigma^0, \qquad
    \bar{\sigma}^i = -\sigma^i,
\end{gather}
and the covariant derivatives are:
\begin{align}
    \label{eq:DEF.330d} D_\mu X_L^f & = \LT( \partial_\mu
      + i g W_\mu^a I_a + i g' \frac{Y_X}{2} B_\mu \RT) X_L^f 
      & X & = l, q; \\
    \label{eq:DEF.330s} D_\mu X_R^f & = \LT( \partial_\mu
      + i g' \frac{Y_X}{2} B_\mu \RT) X_L^f & X & = E, U, D.
\end{align}
The field $\tilde{\Phi}$ is used to give mass to the up-quarks, and is
defined as:
\begin{equation}
    \tilde{\Phi} = i \sigma^2 \Phi^*.
\end{equation}
After spontaneous symmetry breaking, $\Phi$ acquire a non-zero vacuum
expectation value and it is convenient to rewrite it as in
Eq.~\eqref{eq:DEF.160}; again, the unphysical fields $\theta_a$ can be
gauged away and we have:
\begin{equation} \label{eq:DEF.350}
    \Phi = \frac{1}{\sqrt{2}}
    \begin{pmatrix}
	0 \\ \eta + H
    \end{pmatrix}, \qquad
    \tilde{\Phi} = \frac{1}{\sqrt{2}}
    \begin{pmatrix}
	\eta + H \\ 0
    \end{pmatrix}.
\end{equation}
Replacing Eq.~\eqref{eq:DEF.350} into $\mathcal{L}_l$ and $\mathcal{L}_q$ we
obtain both fermion masses and fermion-higgs interaction terms:
\begin{equation} \label{eq:DEF.360}
    \mathcal{L}_H^{f\bar{f}} = - \frac{\eta + H}{\sqrt{2}} \sum_f \LT[
    \lambda_E^f \LT( {E_L^f}^\dagger E_R^f + \hc \RT) +
    \lambda_U^f \LT( {U_L^f}^\dagger U_R^f + \hc \RT) +
    \lambda_D^f \LT( {D_L^f}^\dagger D_R^f + \hc \RT) \RT].
\end{equation}
Looking at this expression, we see first of all that all the lepton and
quark masses are proportional to the vacuum expectation value $\eta$ of the
higgs field, just as for the gauge bosons $W$ and $Z$:
\begin{equation} \label{eq:DEF.370}
    m_X^f = \frac{1}{\sqrt{2}} \lambda_X^f \eta 
    \quad \Rightarrow \quad
    \lambda_X^f = \sqrt{2} \frac{m_X^f}{\eta}, \qquad X = E,U,D;
\end{equation}
also, comparing Eq.~\eqref{eq:DEF.370} with~\eqref{eq:DEF.360} it is clear
that the coupling of the physical higgs and with fermions is proportional to
the fermion mass.

To derive an expression for the interaction between fermions and $A$,
$W^\pm$, $Z$ gauge bosons, we must insert
Eqs.~(\ref{eq:DEF.210w}-\ref{eq:DEF.220})
and~(\ref{eq:DEF.330d},~\ref{eq:DEF.330s}) into~\eqref{eq:DEF.310l}
and~\eqref{eq:DEF.310q}. After some calculation, in 4-component notation the
interaction Lagrangean can be written as:
\begin{equation} \label{eq:DEF.380}
    \mathcal{L}_G^{f\bar{f}} = - gs A^\mu J_\mu^\mathrm{em}
    - g \LT[ {W^-}^\mu J_\mu^+ + \hc \RT] - f Z^\mu J_\mu^0,
\end{equation}
where $J_\mu^\mathrm{em}$ is the \emph{electromagnetic} current and
$J_\mu^+$, $J_\mu^0$ are the \emph{charged} and \emph{neutral} weak
currents:
\begin{align}
    \label{eq:DEF.390a} J_\mu^\mathrm{em} & = \sum_f \LT[ 
      - \bar{E}^f \gamma_\mu E^f
      + \frac{2}{3} \bar{U}^f \gamma_\mu U^f 
      - \frac{1}{3} \bar{D}^f \gamma_\mu D^f \RT], \\
    \label{eq:DEF.390w} J_\mu^+ & = \frac{1}{2\sqrt{2}} \sum_f
      \LT[ \bar{E}^f \gamma_\mu \LT( 1 - \gamma_5 \RT) N^f
      + \bar{D}^f \gamma_\mu \LT( 1 - \gamma_5 \RT) U^f \RT], \\
    \label{eq:DEF.390z} J_\mu^0 & = \frac{1}{2} \sum_f \sum_X
      \bar{X}^f \gamma_\mu \LT( g_V^X - g_A^X \gamma_5 \RT) X
      \quad \text{for} \quad X=N,E,U,D,
\end{align}
and the \emph{vector} and \emph{axial} neutral current coupling constant
$g_V$ and $g_A$ are:
\begin{align}
    g_V & = I_3 - 2 Q s^2, \\
    g_A & = I_3.
\end{align}

Looking at Eqs.~\eqref{eq:DEF.380} and~\eqref{eq:DEF.390a}, we see that the
coupling constant between the electron and $A$ is $-gs$. But $A_\mu$ is the
photon field, and we know from QED that the electric charge of the electron
is $-e$. Therefore:
\begin{equation} \label{eq:DEF.410}
    e = g s = g' c.
\end{equation}

In the low-energy limit, the electroweak sector of the SM must produce as an
effective theory the Fermi model of weak interactions. It is easy to see
that this requirement is satisfied only if the following relation between
the Fermi coupling constant $G_F$ and the vacuum expectation value of the
higgs field $\eta$ holds:
\begin{equation} \label{eq:DEF.420}
    G_F = \frac{1}{\sqrt{2} \eta^2} \quad \Rightarrow \quad
    \eta = \frac{1}{\sqrt{\sqrt{2} G_F}},
\end{equation}
and from the numerical value $G_F \approx 1.16639 \cdot 10^{-5} \GeV^{-2}$
we get $\eta \approx 246\GeV$.

\section{Regularization and renormalization} \label{sec:DEF.20}

\subsection{General discussion}

Any field theory, both classical and quantum, is characterized by a
Lagrangean function from which the equations of motion can be derived. This
tree-level Lagrangean involves a certain number of free parameters, $a_i$,
which are not fixed by the theory. To determine them, the common strategy
consists in choosing a suitable set of experimental results $e_j$, deriving
a theoretical prediction $e_j^\mathrm{th}$ (which is a function of the free
parameters: $e_j^\mathrm{th} = e_j^\mathrm{th}(a_i)$ for them, and then
inverting the relations thus obtaining the expressions of the free
parameters $a_i$ in terms of the measured quantities $e_j$. After that has
been done, the bare quantities $a_i$, being now functions of experimental
results, can be viewed as experimental quantities themselves and acquire a
precise and well-defined physical meaning. If the relation between the $a_i$
and $e_j$ is simple enough, it may be convenient to replace $a_i$ with $e_j$
directly into the Lagrangean, so that now the model is formulated from the
very beginning in terms of the results of some fundamental experiments. This
approach is the default one in classical physics.

In quantum field theory, the situation is different. In higher order
perturbation theory, the relations between formal parameters and measurable
quantities get contributions not only from tree graphs but also from loop
diagrams (the so-called \emph{radiative corrections}), so in general they
are different from tree level relations. The first problem one has to deal
with arises when performing loop integrals: many of them turn out to be
divergent, so the whole theory seems to be mathematically inconsistent. To
overcome this problem, it is necessary to introduce a \emph{regularization
procedure}, i.e.\ a consistent way to properly parameterize the divergencies
and to keep them under control by introducing in the model a new
(unphysical) parameter, known as \emph{cutoff}. The simplest way to do this
is to perform integrations in the euclidian momentum space only in the
finite region $p^2 \le \Lambda^2$, rather than up to $p^2 = +\infty$; as a
consequence, all the integrals are now convergent, but the relations between
physical quantities and bare parameters depend also on the cutoff $\Lambda$.
A more sophisticated technique, which is preferred for gauge theories since
it preserves Lorentz and gauge invariance from the very beginning, is
\emph{dimensional regularization}, which consists in replacing the dimension
4 of the space-time by a lower dimension $D=4 - 2\epsilon$ where the
integrals are convergent:
\begin{equation}
    \int \frac{d^4 k}{(2\pi)^4} \quad\to\quad
    \mu^{4-D} \int \frac{d^D k}{(2\pi)^D},
\end{equation}
where the mass parameter $\mu$ has been introduced in order to keep the
dimensions of the coupling constants in front of the integral independent on
$D$. Now all the integrals are mathematically well-defined objects, so we
can safely derive the expressions for the bare parameters $a_i$ in terms of
the measurable quantities $e_j$, as in the classical case. However, if in
these relations we try to perform the limits $\Lambda \to \infty$ or
$\epsilon \to 0$, so to eliminate the cutoff and recover the original
theory, the divergencies which were removed by the regularization procedure
rises again and we find that the $a_i$ parameters have infinite value.
Therefore, it is not possible to write the Lagrangean directly through
experimental quantities, and it is not clear how the model can be
predictive.

The solution is straightforward. Leaving the Lagrangean written in terms of
bare quantities, it is still possible to derive theoretical predictions for
any experiment $e'_k$ expressing it as a function of the $a_i$, and then
using the previously found relations to eliminate $a_i$ in favor of the
measured quantities $e_j$. In this way, we can use our model to derive
(cutoff dependent) relations among the different sets of experimental
quantities $e_j$ and $e'_k$. Now we eliminate the cutoff by performing the
limit $\Lambda \to \infty$ or $\epsilon \to 0$: if for \emph{any} choice of
the measurable quantities $e_j$ and $e'_k$ our relations remain finite,
i.e.\ the divergent pieces automatically cancel among one another, then our
theory is told to be \emph{renormalizable}. In this case, the predictivity
of the model is preserved, although it is no longer possible to give a
physical meaning to the bare parameters.

In general, writing the Lagrangean in terms of experimental quantities is a
convenient and desirable fact, and to make it possible some interesting
techniques have been developed. The most common take advantage of the
concept of \emph{counterterms}, which are extra (divergent) terms inserted
into the tree Lagrangean to compensate the divergencies coming from loop
integrals. Although this method is very useful for proving renormalizability
theorems and deriving relations at two or more loops, when working at the
one-loop level it is simpler to keep the Lagrangean expressed through the
bare quantities and to use the approach described above.

Since in the present thesis we will rarely deal with expressions beyond
one-loop, we won't discuss further details of alternative methods. However,
let us conclude introducing a few ideas that will be widely used in
Chapters~\ref{sec:ANG} and~\ref{sec:RUN}. In the dimensional regularization
approach, all the divergent terms are proportional to some power of
$1/\epsilon$. If the model under construction has been proved to be
renormalizable, we know by sure that all these terms will cancel from the
final relations among physical observables. So there is no reason to keep
them while performing calculations: we can simply drop all divergent terms
from the very beginning. In particular, we can remove them directly in the
expressions of the bare parameters $a_i$ through the experimental quantities
$e_j$, thus defining a new set of parameters $\hat{a}_i$ which are
completely equivalent to $a_i$ but have a finite and cutoff-independent
value. This way of eliminating divergencies is called \emph{Minimal
Subtraction} (MS) scheme, and is very useful since it allows to write the
Renormalization Group equations (RGE) in a very simple form (see
Chapter~\ref{sec:RUN}). A variant of this scheme called \emph{modified} MS
(or $\MS$) scheme, is obtained by reparameterizing the divergent pieces in
terms of $\Delta = 1/\epsilon - \gamma + \ln 4\pi$ and then dropping
$\Delta$ from all formulas. It is worth noting that MS or $\MS$ quantities,
despite of their finite numerical value, are purely mathematical objects
without any obvious physical meaning.

\subsection{Renormalization of the Standard Model}

Having discussed the general properties of renormalization, let us turn our
attention to the Standard Model. The tree level Lagrangean was introduced in
Sec.~\ref{sec:DEF.10}, and it is now intended that all the free parameters
(masses, coupling constants, \mbox{etc.}) appearing into it have to be
regarded to as \emph{bare quantities}. From now on, we will denote them by a
subscript ``0'' ($m_{W0}$, $m_{Z0}$, $\alpha_0$, \mbox{etc.}), so to
distinguish them from physical observables ($m_W$, $m_Z$, $\alpha$,
\mbox{etc.}); all the tree-level expressions derived in
Sec.~\ref{sec:DEF.10} are valid at any order in perturbation theory,
provided that they are intended as relations among bare quantities.

The electroweak sector of the Standard Model can be fully described in terms
of the three experimental quantities $e$, $m_W$ and $m_Z$, to which we must
add the fermion masses $m_f$ and the (still unknown) higgs mass $m_H$. As
usual, the free/dressed propagators are defined as the inverse of the
differential operator entering the quadratic part of the free/effective
Lagrangean. It is convenient to factor out the Lorentz metrics and to write:
\begin{align}
    D_W^{\mu\nu} & = -i \, g^{\mu\nu} \, D_W, && \Rightarrow
      & \LT( D_W^{\mu\nu} \RT)^{-1} & = i \, g^{\mu\nu} \LT( D_W \RT)^{-1}, \\
    \mathbf{D}_\GZ^{\mu\nu} & = -i \, g^{\mu\nu} \, \mathbf{D}_\GZ,
      && \Rightarrow & \LT( \mathbf{D}_\GZ^{\mu\nu} \RT)^{-1} & =
      i \, g^{\mu\nu} \LT( \mathbf{D}_\GZ \RT)^{-1},
\end{align}
where $\mathbf{D}_\GZ$ is a $(2\times 2)$ matrix, which is non-diagonal due
to the mixing between $\gamma$ and $Z$ induced by radiative corrections:
\begin{equation}
    \mathbf{D}_\GZ =
    \begin{pmatrix}
	D_G   & D_\GZ \\
	D_\GZ & D_Z
    \end{pmatrix}.
\end{equation}
The relation between the dressed propagators and the gauge boson
\emph{self-energies} are:
\begin{gather}
    \LT( D_W \RT)^{-1}(k^2) = k^2 - m_{W0}^2 + \Sigma_W(k^2), \\[2mm]
    \LT( \mathbf{D}_\GZ \RT)^{-1}(k^2) =
    \begin{pmatrix}
	k^2 + \Sigma_\gamma(k^2) & - \Sigma_\GZ(k^2) \\[1mm]
	- \Sigma_\GZ(k^2) & k^2 - m_{Z0}^2 + \Sigma_Z(k^2)
    \end{pmatrix},
\end{gather}
from which we can derive the expression for the propagators (at the one-loop
level):
\begin{align}
    \label{eq:DEF.530w} D_W(k^2)
      & = \frac 1{k^2 - m_{W0}^2 + \Sigma_W(k^2)}, 
      & \Sigma_W(k^2) & = m_W^2 \Pi_W(k^2), \\
    \label{eq:DEF.530a} D_\gamma(k^2)
      & = \frac{1}{k^2 + \Sigma_\gamma(k^2)},
      & \Sigma_\gamma(k^2) & = m_Z^2 \Pi_\gamma(k^2), \\
    \label{eq:DEF.530z} D_Z(k^2)
      & = \frac{1}{k^2 -m_{Z0}^2 + \Sigma_Z(k^2)},
      & \Sigma_Z(k^2) & = m_Z^2 \Pi_Z(k^2), \\
    \label{eq:DEF.530x} D_\GZ(k^2)
      & = \frac{\Sigma_\GZ(k^2)}{k^2 \LT[ k^2 - m_{Z0}^2 \RT]},
      & \Sigma_\GZ(k^2) & = m_Z^2 \Pi_\GZ(k^2).
\end{align}
Note that the $\Sigma_i(k^2)$ self-energies have dimensions of a mass
squared, while the $\Pi_i(k^2)$ quantities are adimensional.

Now, let us consider the photon propagator $D_\gamma(k^2)$ shown in
Eq.~\eqref{eq:DEF.530a}. For any given Feynman diagram, each internal photon
line carrying momentum squared $k^2$ contributes to the total amplitude with
a factor $\alpha_0 D_\gamma(k^2)$, where $\alpha_0$ accounts for the (bare)
electric charges at the two vertices to which the photon propagator
$D_\gamma(k^2)$ is attached. We can reabsorb the effects of vacuum
polarization (described by $\Sigma_\gamma(k^2)$) into the fine structure
constant by defining a ``\emph{running}'' coupling constant $\alpha(k^2)$:
\begin{equation}
    \frac{\alpha(k^2)}{k^2} \equiv
    \frac{\alpha_0}{k^2 + \Sigma_\gamma(k^2)}.
\end{equation}
This relation defines $\alpha(k^2)$ for any values of $k^2$. Since the bare
parameter $\alpha_0$ does not depend on $k^2$, we can write:
\begin{equation}
    \alpha_0 = \alpha(k^2) \LT[ 1 + \frac{\Sigma_\gamma(k^2)}{k^2} \RT] =
    \alpha(q^2) \LT[ 1 + \frac{\Sigma_\gamma(q^2)}{q^2} \RT]
    \qquad \forall k,q.
\end{equation}
Setting $q^2 \to 0$, and requiring that $\alpha(0)$ coincides with the fine
structure constant $\alpha$ so to have the correct low-energy limit, we get
(at one-loop):
\begin{equation}
    \alpha(k^2) = \frac{\alpha}{1
      - \LT[ \Sigma'_\gamma(0) - \LT( m_Z^2 / k^2 \RT) \Pi_\gamma(k^2) \RT]}.
\end{equation}
In this thesis, we will deal mainly with quantities which are defined at the
electroweak scale. Since $\alpha$ is essentially a low-energy parameter, it
is more convenient for our purposes to expand the perturbation series in
powers of $\alpha(m_Z^2)$ rather than $\alpha$. However, $\alpha(m_Z^2)$ is
not a purely electromagnetic parameter (it also gains contributions from
charged gauge bosons), and in principle it is sensitive to radiative effects
due to New Physics. To overcome these problems, let's define:
\begin{equation} \label{eq:DEF.570}
    \albar = \frac{\alpha}{1 - \delta\alpha}, \qquad \delta\alpha =
    \LT[ \Sigma'_\gamma(0) - \Pi_\gamma(m_Z^2) \RT]_{\text{light fermions}},
\end{equation}
where $\delta\alpha$ takes into account only the contributions of light
fermions (here ``light'' means ``lighter than Z-boson''), i.e.\ the 3
charged leptons ($e$, $\mu$, $\tau$) and the 5 quarks ($u$, $d$, $c$, $s$,
$b$): top quark ($t$) and charged bosons ($W$ boson and Higgs ghosts) are
excluded. Defined in this way, $\albar$ is a purely electromagnetic
gauge-invariant parameter, but conversely radiative corrections from top and
gauge bosons to $\albar$-related quantities are \emph{not} automatically
included, and must be added by hand~\cite{NOV94a}. For later convenience, we
also define:
\begin{equation} \label{eq:DEF.580}
    \altw = \LT[ \Sigma'_\gamma(0) - \Pi_\gamma(m_Z^2) \RT]_{t,W}
    \qquad\Rightarrow\qquad
    \delta\alpha + \altw = \Sigma'_\gamma(0) - \Pi_\gamma(m_Z^2).
\end{equation}
Numerically, we have $1/\alpha = 137.035\,9895(61)$ and $1/\albar =
128.878(90)$.

As already stated, the gauge sector of the Standard Model can be fully
described in terms of three experimental quantities. The best measured
electroweak observables are presently the Fermi coupling constant $G_F$, the
$Z$ boson mass $m_Z$, and of course the fine structure constant $\alpha$;
for the above mentioned reasons, it is better to replace the last one with
$\albar$. Using Eqs.~\eqref{eq:DEF.230}, \eqref{eq:DEF.410}
and~\eqref{eq:DEF.420}, we define the electroweak angle in the following
way:
\begin{equation} \label{eq:DEF.590}
    s^2 c^2 = \frac{\pi\albar}{\sqrt{2} G_F m_Z^2},
\end{equation}
and substituting the numerical values we get $s^2 = 0.23116(22)$.

When considering the effects of radiative corrections to the relations among
physical observables, a number of different classes of Feynman diagrams must
be taken into account. Although in principle all of them participate to the
final result, in practice for most of the cases the largest contributions
comes from vector boson self-energies, and vertex and box diagrams may
safely be neglected. Since all the $\Sigma_i$ defined in
Eqs.(~\ref{eq:DEF.530w}-\ref{eq:DEF.530x}) contain divergent terms, only
some special combinations of them can enter relations among measurable
quantities, and it can be proved that the total number of divergency-free
independent combinations is 3. To parameterize them, let us consider the $W$
and $Z$ boson mass ratio $m_W/m_Z$, the $Zee$ axial coupling $g_a \equiv
g_a^E$, and the ratio of vector to axial coupling $g_v/g_a$ in the $Zee$
vertex. From the previous section, we know that the tree level value for
these quantities are $c$, $-1/2$, and $1-4s^2$, respectively, so a
convenient parameterization for the higher order effects is:
\begin{align}
    \label{eq:DEF.600m} \frac{m_W}{m_Z} & = c 
      + \frac{3c\albar}{32\pi s^2 \LT( c^2 -s^2 \RT)} V_m, \\
    \label{eq:DEF.600a} g_A & = - \frac{1}{2}
      - \frac{3\albar}{64\pi c^2 s^2} V_A, \\
    \label{eq:DEF.600r} R \equiv \frac{g_V}{g_A} & = 1 - 4s^2
      + \frac{3\albar}{4\pi \LT( c^2 -s^2 \RT)} V_R.
\end{align}
The expressions for the $V_i$ functions in terms of gauge bosons
self-energies, vertex and box diagrams are given in Appendix~\ref{sec:APP}.
However, let us mention that the largest contributions are due to the big
mass splitting (which breaks $SU(2)_L$ symmetry) between top and bottom
quarks, and are proportional to $m_t^2$; the numerical coefficients in front
of $V_i$ in Eqs.~(\ref{eq:DEF.600m}-\ref{eq:DEF.600r}) were chosen in such a
way that the leading top contribution into $V_i$ is simply $\LT( m_t / m_Z
\RT)^2$~\cite{NORV99}.

\section{Supersymmetry} \label{sec:DEF.30}

\subsection{Motivations for New Physics}

Despite of its success in explaining all the present experimental data, the
Standard Model is not believed to be the ultimate theory of Nature; this is
due mainly to three ``problems'', or at least ``unpleasantnesses'', which
occur within its framework~\cite{Haber85}. The first one is the large number
of arbitrary assumptions and parameters it exhibits: besides the 3 charged
lepton and 6 quark masses, we have the higgs boson mass, the higgs vacuum
expectation value, the 3 gauge coupling constants, and the 4 parameters of
the CKM quark mixing matrix (which we didn't discuss in
Sec.~\ref{sec:DEF.10}). The recently found evidence for neutrino
oscillations adds to this set also neutrino masses and mixing angles.
Moreover, many features of the model~- for example the number of fermion
generations, see Chapter~\ref{sec:GEN}~- have no theoretical explanation,
and must be accepted as a matter of fact. This is not a ``problem'' in a
proper sense, but it is generally believed that a fundamental theory should
make as little unexplained assumptions as possible (the best being no
assumptions at all), and from this point of view the SM is clearly
unsatisfactory.

\begin{figure}[!t] \centering
    \includegraphics[width=0.9\textwidth]{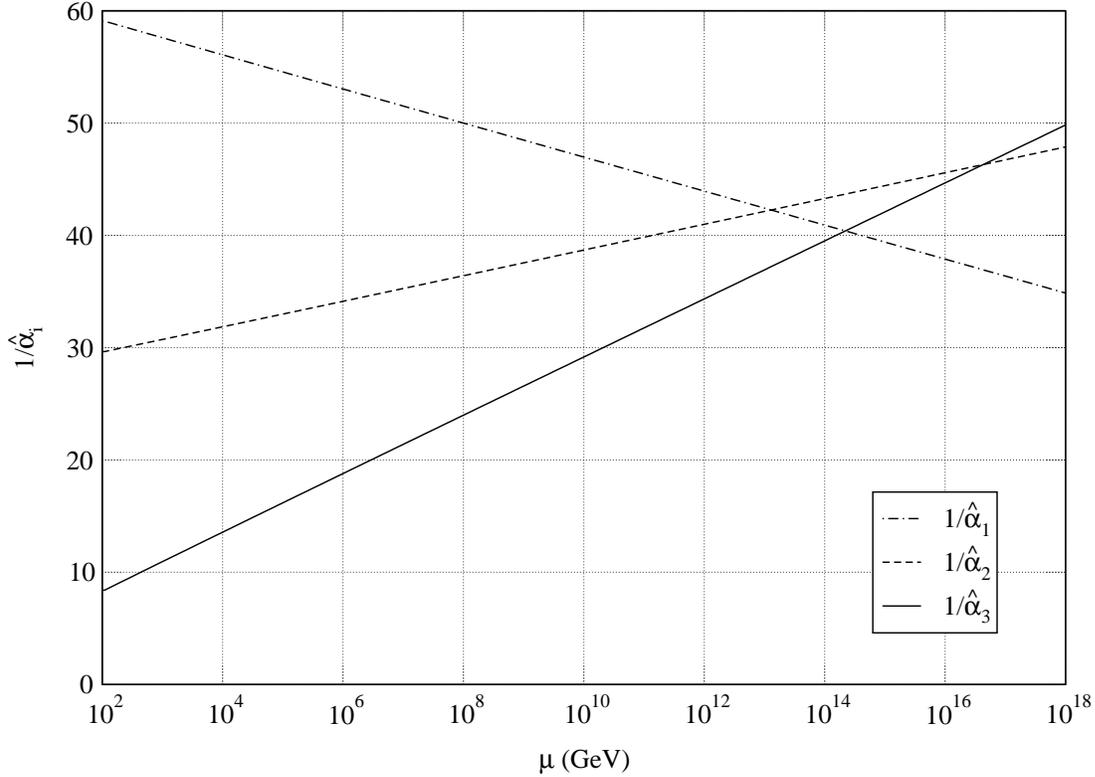}
    \caption[Running of $\hat{\alpha}_i$ in the Standard Model]{
      Running of the $\hat{\alpha}_1$, $\hat{\alpha}_2$ and $\hat{\alpha}_3$
      coupling constants in the framework of the Standard Model, as a
      function of the mass scale parameter $\mu$. The starting point is
      $m_Z=91.1867\GeV$, $\hat\alpha(m_Z)=1/128.1$, $\hat{s}^2=0.23147$,
      $\hat{\alpha}_3(m_Z)=0.119$, $\hat{m}_t(m_Z)=175\GeV$.}
    \label{fig:DEF.10}
\end{figure}

The second problem is that the Standard Model, due to the invariant $U(1)_Y$
subgroup of hypercharge, is not asymptotically free, so ultimately, at some
energy scale, its interactions will become strong~\cite{Haber85}. This
suggests that the SM is the low-energy limit of a more fundamental theory,
eventually unifying all the three gauge coupling constants into one and
accommodating all known particles into a few representations of its gauge
group. This idea has led to the concept of \emph{Grand Unified Theory}, or
GUT, whose prototype is $SU(5)$. In this model, the 12 generators of the
Standard Model $SU(3)_c \times SU(2)_L \times U(1)_Y$ group are embedded
into the 24-dimensional algebra of $SU(5)$, and the 15 Weyl spinors which
form \emph{each} SM family ($N_L$, $E_L$, $E_R$ and $3_c \times (U_L, U_R,
D_L, D_R)$) perfectly fit into a $\mathbf{5}$ and $\mathbf{10}$
representation of $SU(5)$~- in particular, no right-handed neutrino is
required. When the model was first suggested~\cite{Georgi74}, the
poor-quality experimental data were in agreement with the hypothesis of
unification of the three SM coupling constants $\alpha_i$ into a single one
$\alpha_5$ at a mass scale $m_\GUT \sim 10^{14} \div 10^{15}\GeV$.
Unfortunately, more accurate measurements proved that within the SM the
coupling constants never get unified (or equivalently forcing their match at
a single point lead to a wrong prediction for $s^2$ or $\alpha_s$), as we
show in Fig.~\ref{fig:DEF.10}; however, the basic idea of GUT is still alive
and widely accepted, provided that some New Physics exists at a scale
\emph{lower} than $m_\GUT$ and yields unification.

The third problem, known as \emph{the hierarchy problem}, is related to the
protection of the higgs mass from effects of radiative corrections due to
superheavy particles. It is well known that when the higgs mass is of the
order of a few TeV, then the higgs self-coupling gets too strong, and we
should not be observing the apparently successful perturbation theory at low
energies. Any GUT model predicts the existence of heavy gauge bosons and
Higgs scalars, whose mass is of the same order of the scale at which the GUT
symmetry is broken (i.e.\ $\sim m_\GUT$); the reason why such a GUT scale
($\sim 10^{16}\GeV$ for SUSY $SU(5)$) and the electroweak scale ($\sim
10^2\GeV$) are so different should be explained by the model. However, even
if we choose the hierarchy in such a way, the radiative corrections will
destroy it. To see this, let us consider the contribution to the SM higgs
self-energy due to a superheavy boson of mass $M$~\cite{Kazakov96}: it is
given by the following Feynman diagram, and it is proportional to the mass
squared of the heavy particle:
\begin{equation}
    \raisebox{-5.5mm}{
      \fmfframe(2,6)(0,-4){
	\begin{fmfgraph*}(45,20) \fmfkeep{boson}
	    \fmfleft{i}
	    \fmfright{o}
	    \fmf{dashes}{i,v,v,o}
	    \fmfv{label=$\lambda^2$,label.angle=-90}{v}
	    \fmfdot{v}
	\end{fmfgraph*}}}
    \quad\Rightarrow\quad \delta m^2 = \lambda^2 M^2.
\end{equation}
This correction, if not canceled by some mechanism, obviously spoil the
hierarchy, unless a completely unnatural fine tuning of the order of
$10^{-14}$ for the coupling constant $\lambda^2$ is introduced.

The solution to the last problem is the main motivation to introduce
supersymmetry (SUSY). The most elegant way to prevent a physical quantity to
acquire an uncontrolled value due to radiative corrections is to introduce a
symmetry which protects it. In (unbroken) supersymmetric models, each boson
has a fermionic parter of equal mass, whose contribution should be accounted
for when evaluating radiative corrections. For the case of the SM higgs in
the framework of GUT, the joint contributions of superheavy bosons and their
fermionic superpartners are free of quadratic divergencies:
\begin{equation}
    \raisebox{-5.5mm}{
      \fmfframe(1,6)(0,-4){
	\fmfreuse{boson}}} 
    \; + \;
    \raisebox{-7.5mm}{
      \fmfframe(1,-2)(0,-2){
	\begin{fmfgraph*}(45,20)
	    \fmfleft{i}
	    \fmfright{o}
	    \fmf{dashes}{i,v1}
	    \fmf{dashes}{v2,o}
	    \fmf{plain,left,tension=0.5}{v1,v2,v1}
	    \fmfv{label=$\lambda$,label.angle=-120}{v1}
	    \fmfv{label=$\lambda$,label.angle=-60}{v2}
	    \fmfdot{v1,v2}
	\end{fmfgraph*}}} 
    \quad\Rightarrow\quad \delta m^2 = 0.
\end{equation}
This mechanism for cancellation of quadratic divergencies is peculiar to
SUSY models, and it can be proved that it occurs at any order in
perturbation theory.

\subsection{The Minimal Supersymmetric Standard Model}

From a mathematical point of view, the basic idea of supersymmetry consists
in the extension of the Poincar\'e group of space-time transformations
through the introduction of new \emph{anticommuting} generators
$Q_\alpha^i$, $\bar{Q}_{\dot{\alpha}}^i$ (here $\alpha,\dot{\alpha} = 1,2$
and $i = 1,\ldots, N$)~\cite{Bagger83}. This extra generators have the
remarkable property of transforming a boson into a fermion and vice versa,
thus any supermultiplet corresponding to an irreducible representation of
the SUSY algebra contains both bosonic and fermionic degrees of freedom (in
equal number). For the simplest case $N=1$ we have only 4 new generators,
and in addition to the usual commutation relations among Poincar\'e
generators we have~\cite{Kazakov96}:
\begin{align}
    \label{eq:DEF.800qb} \LT\{ Q_\alpha, \bar{Q}_{\dot{\beta}} \RT\} 
      & = 2 \LT( \sigma^\mu \RT)_{\alpha\dot{\beta}} P_\mu, \\
    \label{eq:DEF.800qq} \LT\{ Q_\alpha, Q_\beta \RT\} = 
      \LT\{ \bar{Q}_{\dot{\alpha}}, \bar{Q}_{\dot{\beta}} \RT\} & = 0, \\
    \label{eq:DEF.800p} \LT[ Q_\alpha, P_\mu \RT] = 
      \LT[ \bar{Q}_{\dot{\alpha}}, P_\mu \RT] & = 0, \\
    \label{eq:DEF.800m} \LT[ Q_\alpha, M_{\mu\nu} \RT] & = \frac{1}{2}
      \LT( \sigma_{\mu\nu} \RT)_\alpha^\beta Q_\beta, \\
    \label{eq:DEF.800mb} \LT[ \bar{Q}_{\dot{\alpha}}, M_{\mu\nu} \RT] 
      & = - \frac{1}{2} \bar{Q}_{\dot{\beta}} 
      \LT( \bar{\sigma}_{\mu\nu} \RT)_{\dot{\alpha}}^{\dot{\beta}}, \\
    \alpha, \dot{\alpha}, \beta, \dot{\beta} & = 1,2,
\end{align}
where $P_\mu$ and $M_{\mu\nu}$ are four-momentum and angular momentum
operators, respectively. From these relations we see that $Q_\alpha$ and
$\bar{Q}_{\dot{\alpha}}$ are spinors under rotations and invariant under
spatial translations; in particular, from Eq.~\eqref{eq:DEF.800p} we observe
that even under the SUSY group $P^2$ is a Casimir operator, thus all
particles within the same irreducible supermultiplet share the same mass.

The simplest supersymmetric model including the SM is called Minimal
Supersymmetric Standard Model (MSSM). Since in the $N=1$ case the SUSY
generators commute with the generators of internal symmetries, all
superpartners have the same quantum numbers as the corresponding SM
particles. Within the SM, the only couple of particles with different spin
but the same gauge quantum numbers is neutrino/higgs boson; however,
assuming that the higgs is the superpartner of a neutral lepton would imply
that it also carries a lepton number, and this is phenomenologically
unacceptable. Therefore, in the MSSM \emph{all} the superpartners are
introduced as \emph{new} particles.

\begin{table}[!t] \centering
    \begin{tabular}{lc@{\hspace{15mm}}lc@{\hspace{15mm}}cccc}
	\hline\hline
	\textbf{Bosons} && \textbf{Fermions} &&	$SU(3)_c$ & $SU(2)_L$ & $U(1)_Y$ \\
	\hline
	gluon & $g^k$ & gluino & $\tilde{g}^k$ & $\mathbf{8}$ & $\mathbf{1}$ & $0$ \\
	weak  & $W^a$ & wino   & $\tilde{W}^a$ & $\mathbf{1}$ & $\mathbf{3}$ & $0$ \\
	hypercharge & $B$ & bino & $\tilde{B}$ & $\mathbf{1}$ & $\mathbf{1}$ & $0$ \\
	\hline
	sleptons & $\LT(\begin{gathered} \tilde{N}_L \\ \tilde{E}_L \end{gathered}\RT)$
	& leptons & $\LT(\begin{gathered} N_L \\ E_L \end{gathered}\RT)$
	& $\mathbf{1}$ & $\mathbf{2}$ & $-1$ \\
	~ & $\tilde{E}_R$ & ~ & $E_R$ & $\mathbf{1}$ & $\mathbf{1}$ & $-2$ \\
	\hline
	squarks & $\LT(\begin{gathered} \tilde{U}_L \\ \tilde{D}_L \end{gathered}\RT)$
	& quarks & $\LT(\begin{gathered} U_L \\ D_L \end{gathered}\RT)$
	& $\mathbf{3}$ & $\mathbf{2}$ & $\+1/3$ \\
	~ & $\tilde{U}_R$ & ~ & $U_R$ & $\mathbf{3}$ & $\mathbf{1}$ & $\+4/3$ \\
	~ & $\tilde{D}_R$ & ~ & $D_R$ & $\mathbf{3}$ & $\mathbf{1}$ & $-2/3$ \\
	\hline
	higgs & $H_1$ & higgsino & $\tilde{H}_1$ 
	& $\mathbf{1}$ & $\mathbf{2}$ & $\+1$ \\
	~ & $H_2$ & ~ & $\tilde{H}_2$ & $\mathbf{1}$ & $\mathbf{2}$ & $-1$ \\
	\hline\hline
    \end{tabular}
    \caption[Particle contents of the Minimal Supersymmetric Standard
    Model]{
      The particle contents of the MSSM. Supersymmetric particles are
      denoted adding a tilde to the letter identifying the corresponding SM
      partner. Note that \emph{two} higgs doublets (rather than \emph{one})
      are included in the model.}
    \label{tab:DEF.20}
\end{table}

The particle contents of the MSSM is shown in Table~\ref{tab:DEF.20}. Unlike
the SM, \emph{two} higgs doublets rather than \emph{one} are introduced,
each having of course a fermionic counterpart. This is due to the fact that
the mechanism used in the SM to give mass to the up-quarks, i.e.\ the
construction of the $\tilde{\Phi}$ field by means of $\Phi$ alone, cannot be
applied within the MSSM since it would require the introduction of
explicitly SUSY-breaking terms into the Lagrangean. As a consequence, the
Higgs sector of the MSSM is richer than the SM one: we start with 8 bosonic
degrees of freedom, and after incorporating 3 of them into $W^\pm$ and $Z$
(which gain their mass in the usual way) we are left with 3 neutral scalar
particles (2 CP-even, $h$ and $H$, and 1 CP-odd, $A$) and 1 charged one
($H^\pm$). Without entering the details of calculations, let us mention that
the low-energy ground state of the MSSM is characterized by two different
vacuum expectation values, $\eta_1$ and $\eta_2$, and that the breaking of
the $SU(2)_L$ symmetry produces at tree level the following relations:
\begin{gather}
    \eta^2 = \eta_1^2 + \eta_2^2, \qquad \tan\beta \equiv
    \frac{\eta_2}{\eta_1}; \\
    \LT. \begin{aligned}
	m_W & = \frac{1}{2} g \eta \\
	m_Z & = \frac{1}{2} f \eta
    \end{aligned} \RT\} \quad \Rightarrow \quad \eta \approx 246\GeV,
\end{gather}
where we see that the parameter $\eta$ plays the same role as in the SM.
Also, the masses of the higgs bosons are not independent from one another:
\begin{align}
    \label{eq:DEF.820c} m_{H^\pm} & = m_A^2 + m_W^2, \\
    \label{eq:DEF.820n} m_{h,H}^2 & = \frac{1}{2} \LT[ m_A^2 + m_Z^2 \pm
      \sqrt{\LT( m_A^2 + m_Z^2 \RT)^2 - 4 m_A^2 m_Z^2 \cos^2 2\beta} \RT].
\end{align}
Now, looking at Eq.~\eqref{eq:DEF.820n} it is easy to see that the lightest
higgs boson, whatever the values of $m_A$ and $\tan\beta$ are, is always
lighter than $m_Z$. This result is only valid at tree level, and radiative
corrections affect it by increasing the upper limit; however, precise
calculations show that $m_h$ cannot in general exceed $120 \div 130\GeV$,
and this unique signature of SUSY models is probably the best test of
low-energy supersymmetry which will be achieved by the next generation of
accelerator experiments.

When building the most general SUSY-invariant Lagrangean extending in a
minimal way the particle contents of the Standard Model, one quickly
realizes that due to the many new particles present in the model the
conservation of baryon and lepton numbers no longer occurs automatically.
Since phenomenologically no evidence of $B$ or $L$ violation has been found,
it is necessary to find a mechanism to suppress $B$ and $L$ violating terms,
and in the MSSM this is achieved by introducing a new discrete symmetry
called \emph{R-parity}. Formally, one can define the $R$-parity of any
particle of spin $J$, baryon number $B$ and lepton number $L$ to be $R =
(-1)^{2J + 3B + L}$, and it is immediate to see that all SM particles have
$R=+1$, while all their superpartners have $R=-1$. The conservation of this
quantities has two straightforward consequences: first, in laboratory
experiments supersymmetric particles are produced in pairs; second, the
lightest supersymmetric particle (LSP) \emph{is stable}. The prediction of
SUSY models with unbroken $R$-parity of a new stable particle is very
interesting in an astrophysical contest, since it offers a chance to solve
the problem of Dark Matter.

\begin{figure}[!t] \centering
    \includegraphics[width=0.9\textwidth]{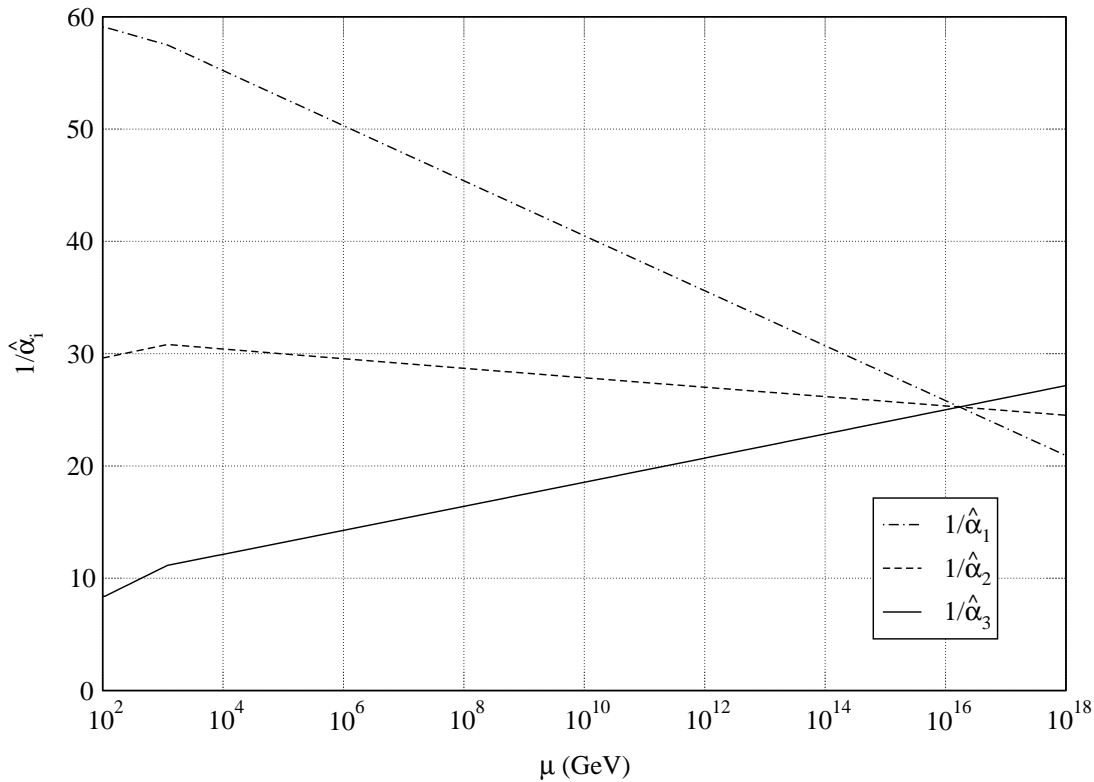}
    \caption[Running of $\hat{\alpha}_i$ in the Minimal Supersymmetric
    Standard Model]{
      Running of the $\hat{\alpha}_1$, $\hat{\alpha}_2$ and $\hat{\alpha}_3$
      coupling constants in the framework of the Minimal Supersymmetric
      Standard Model, as a function of the mass scale parameter $\mu$. The
      starting point is $m_Z=91.1867\GeV$, $\hat\alpha(m_Z)=1/128.1$,
      $\hat{s}^2=0.23147$, $\hat{\alpha}_3(m_Z)=0.119$,
      $\hat{m}_t(m_Z)=175\GeV$.}
    \label{fig:DEF.20}
\end{figure}

We already noted that in unbroken supersymmetric models all the particles
within the same supermultiplet have the same mass. Since this is not
observed to be the case in nature, we conclude that SUSY must be broken at
some scale. The mechanism adopted in the MSSM consists in an \emph{explicit}
breaking through the inclusion into the Lagrangean of the so-called ``soft''
terms, whose defining property is that they do not spoil the cancellation of
quadratic divergencies produced by supersymmetry. The origin of these terms
is usually ascribed to the spontaneous breaking of some larger symmetry, for
example \emph{supergravity} (i.e.\ \emph{local} supersymmetry), at the GUT
scale. In this framework, the MSSM is not a fundamental theory, but like the
SM it is only a step towards it; however, the hypothesis of gauge coupling
unification at a common mass scale, which is unrealizable in the Standard
Model, is instead in excellent agreement with experimental data in the MSSM
(see Fig.~\ref{fig:DEF.20} and Chapter~\ref{sec:RUN}), and this represents
nowadays a strong indirect hint in favor of supersymmetry.

\end{fmffile}

\clearemptydoublepage

\begin{fmffile}{Feynman/angles}

\chapter{The numerical closeness between $\theta$ and the $\MS$ parameter
  $\hat\theta$}
\label{sec:ANG}

As we have seen in the previous chapter, the phenomenological angle $\theta$
plays a central role in the description of the electroweak sector of the
Standard Model. This angle is defined in Eq.~\eqref{eq:DEF.590} through the
best measured quantities $G_F$, $m_Z$ and $\albar$, and is therefore very
simple to deal with and is suitable for describing the electroweak precision
measurements in a natural way. According to the last data from
Ref.~\cite{Erler98}, we have:
\begin{equation} \label{eq:ANG.10p}
    s^2 = 0.23116(22).
\end{equation}

However, as everybody knows, it is the value of $\hat s^2 \equiv \sin^2
\hat\theta$ which is used to study gauge couplings unification in the
framework of GUT models. The corresponding angle is calculated in the
modified minimal subtraction scheme~($\MS$), with $\mu = m_Z$. From
Ref.~\cite{Erler98} one can see that this quantity appears to be numerically
very close to the phenomenological parameter $s^2$:
\begin{equation} \label{eq:ANG.10g}
    \hat s^2 = 0.23144(24).
\end{equation}
It can be seen immediately that $s^2$ and $\hat s^2$ are equal with an
accuracy better that $0.1\%$. However, the splitting between them is generated
by one-loop radiative corrections, and a natural estimate for their difference
is about $1\%$. As an example, let us consider another possible definition of
the electroweak angle, $s_W^2$, which is related to the ratio between the $W$
and $Z$ boson masses:
\begin{equation} \label{eq:ANG.10m}
    s_W^2 = 0.22346(107).
\end{equation}
The splitting between Eq.~\eqref{eq:ANG.10m} and Eqs.~\eqref{eq:ANG.10p},
\eqref{eq:ANG.10g} is more than $3\%$, which is what can be expected for
one-loop radiative corrections; therefore, the numerical closeness of $s^2$
and $\hat s^2$ is completely unnatural, and represents a puzzle of the
Standard Model.

At this point it is useful to remind that there is one more coincidence in
the Standard Model: the parameter $s_l^2 \equiv \sin^2 \theta_l$, which
describes asymmetries in $Z$ boson decays, also happens to be very close to
$s^2$. Numerically we have:~\cite{NORV99}
\begin{equation} \label{eq:ANG.10l}
    s_l^2 = 0.23157(19).
\end{equation}
This coincidence is known to be accidental, since it occurs only for the top
quark mass $m_t$ close to $170\GeV$.

The aim of the present chapter is to present an explicit formula which
provides the relation between $s^2$ and $\hat s^2$. Such a relation is
relevant since it provides a simple connection between \emph{theoretical GUT
models} (which are usually described in terms of $\hat s^2$) and
\emph{experimental results} (from which $s^2$ is defined), and will be
widely used in Chapter~\ref{sec:RUN}. Analyzing this formula, we will see
that the numerical closeness between $s^2$ and $\hat s^2$ occurs only for
$m_t$ close to $170\GeV$, as for the case of $s_l^2$, so it is really a
coincidence without any physical reason. However, writing the expression for
$\hat s^2$ through $s_l^2$ it will become clear that these two angles are
naturally close, and their coincidence does not depend on the top mass and
has a straightforward physical explanation. In this way we will see that,
instead of two accidental coincidences between three mixing angles, we have
only one.

In Sec.~\ref{sec:ANG.10}, we will quickly introduce the fundamental
relations between bare quantities and physical observables. The rest of this
chapter is based on the two published papers~\cite{Maltoni98b}
and~\cite{Maltoni99a}.

\fmfset{arrow_len}{3mm}

\section{The bare quantities and the tree approximation} \label{sec:ANG.10}

Let us start our discussion analyzing some relations among \emph{bare
quantities}. As was shown in Chapter~\ref{sec:DEF}, the electroweak mixing
angle $\theta_0$ (or, equivalently, its sine $s_0$ or cosine $c_0$) is
defined from the ratio between the $W$ and $Z$ boson coupling constants
$g_0$ and $f_0$:
\begin{equation} \label{eq:ANG.20g}
    c_0 = \frac{g_0}{f_0}.
\end{equation}
However, using formulas from the previous chapter it is possible to rewrite
this expression in many different ways, so to relate $\theta_0$ to other
parameters of the model. In particular, from Eq.~\eqref{eq:DEF.230} we can
express it through the ratio between the $W$ and $Z$ boson masses:
\begin{equation} \label{eq:ANG.20m}
    c_0 = \frac{m_{W0}}{m_{Z0}}.
\end{equation}
Another interesting relation can be obtained by means of
Eqs.~\eqref{eq:DEF.230}, \eqref{eq:DEF.410} and~\eqref{eq:DEF.420}:
\begin{equation} \label{eq:ANG.20p}
    c_0^2 s_0^2 = \frac{\pi \alpha_0}{\sqrt{2} G_{F0} m_{Z0}^2}.
\end{equation}
Finally, for the present discussion it will be useful to rewrite $\theta_0$
in terms of the vector and axial neutral current coupling constants,
$g_{V0}$ and $g_{A0}$, describing the $Zee$ vertex:
\begin{equation} \label{eq:ANG.20l}
    s_0^2 = \frac{1}{4} \LT( 1 - \frac{g_{V0}}{g_{A0}} \RT).
\end{equation}
Of course, many other definitions are possible, and all of them are
equivalent to one another, but for our purposes these expressions are enough.

What we want to do now is to find the best way to relate these theoretical
expressions to the experimental measurements. Let us start neglecting
radiative corrections, so to check whether the tree approximation is
sufficient to describe the SM phenomenology at the present experimental
accuracy.

At tree level, the bare quantities appearing on the \rhs\ of
Eqs.~(\ref{eq:ANG.20g}-\ref{eq:ANG.20l}) have a straightforward physical
meaning, and all what we have to do is just to replace them with the
corresponding experimental numbers. However, if we do this, we find that the
values for the electroweak angle $\theta_0$ which follow from these
equations are in general in disagreement with one another. More exactly, the
corresponding values for the parameter $s_0^2$ are just those reported in
Eqs.~(\ref{eq:ANG.10p}-\ref{eq:ANG.10l}), provided that the numerical value
of $\albar$, rather than $\alpha$, is used for $\alpha_0$ in
Eq.~\eqref{eq:ANG.20p}. The reason for this is that Eqs.~\eqref{eq:ANG.20g},
\eqref{eq:ANG.20m}, \eqref{eq:ANG.20p} and~\eqref{eq:ANG.20l}, when
considered as relations between physical quantities, are nothing but the
definitions of the four different angles~\eqref{eq:ANG.10g},
\eqref{eq:ANG.10m}, \eqref{eq:ANG.10p} and~\eqref{eq:ANG.10l}, respectively.
In symbols:
\begin{align}
    \label{eq:ANG.30g} \hat s^2 & \equiv \frac{{\hat e}^2}{{\hat g}^2} 
      \quad \text{(in the $\MS$ scheme, with $\mu = m_Z$)}; \\
    \label{eq:ANG.30m} s_W^2 & \equiv 1 - \frac{m_W^2}{m_Z^2}; \\
    \label{eq:ANG.30p} s^2 & \equiv \frac{1}{2} \LT( 1 - \sqrt{1 -
      \frac{4\pi \albar}{\sqrt{2} G_F m_Z^2}} \RT); \\
    \label{eq:ANG.30l} s_l^2 & \equiv \frac{1}{4}
      \LT( 1 - \frac{g_V}{g_A} \RT).
\end{align}

The disagreement among the numerical
values~(\ref{eq:ANG.10p}-\ref{eq:ANG.10l}) clearly proves that the tree
approximation is not adequate to explain the experimental data, and
radiative corrections should be taken into account. To do this, we need
first of all to write down the relations between bare quantities and
physical observables; therefore, in the rest of this section we will quickly
review these formulas, addressing the reader to Ref.~\cite{NOV93} for a more
accurate analysis.

\subsection{The fine structure constant}

\renewenvironment{TempGraph}
{\begin{fmfgraph*}(30,20)
  \fmfleft{i1}
  \fmfright{o2,o1}
  \fmflabel{$\bar e$}{o1}
  \fmflabel{$e$}{o2}}
{\end{fmfgraph*}}

The fine structure constant $\alpha$ is defined from the $\gamma ee$ vertex.
The relevant Feynman diagrams, at one-loop level, are:
\begin{equation}
    \begin{split} \fmfframe(0,3)(0,3){
	  \begin{TempGraph}
	      \fmf{photon,label=$\gamma$}{i1,v}
	      \fmf{fermion}{o1,v,o2}
	      \fmfdot{v}
	  \end{TempGraph} }
    \end{split}
    \qquad\qquad
    \begin{split} \fmfframe(0,3)(0,3){
	  \begin{TempGraph}
	      \fmf{photon,tension=2,label=$\gamma$}{i1,v1}
	      \fmf{photon,tension=2,label=$\gamma$,tension=2}{v1,v}
	      \fmf{fermion}{o1,v,o2}
	      \fmfblob{.20w}{v1}
	      \fmfdot{v}
	  \end{TempGraph} }
    \end{split}
    \qquad\qquad
    \begin{split} \fmfframe(0,3)(0,3){
	  \begin{TempGraph}
	      \fmf{photon,tension=2,label=$\gamma$}{i1,v1}
	      \fmf{photon,tension=2,label=$Z$}{v1,v}
	      \fmf{fermion}{o1,v,o2}
	      \fmfblob{.20w}{v1}
	      \fmfdot{v}
	  \end{TempGraph} }
    \end{split}
\end{equation}
From these graphs, it is easy to derive a relation between $\alpha$ and
$\alpha_0$:
\begin{equation}\label{eq:ANG.40a}
    \alpha = \alpha_0 \LT[ 1
    - \Sigma'_\gamma(0) - 2\frac{s}{c} \Pi_\GZ(0) \RT].
\end{equation}
Note that in Eq.~\eqref{eq:ANG.40a} only photon self-energy and $\gamma Z$
mixing contribute. In principle, vertex corrections and electron
self-energies should also be taken into account; however, due to Ward
identities, the sum of these contributions vanishes:
\begin{equation}
    \begin{split} \fmfframe(0,3)(0,3){
	  \begin{TempGraph}
	      \fmf{photon,label=$\gamma$}{i1,v}
	      \fmf{fermion}{o1,v,o2}
	      \fmfblob{.20w}{v}
	  \end{TempGraph} }
    \end{split}
    \quad + \quad
    \begin{split} \fmfframe(0,3)(0,3){
	  \begin{TempGraph}
	      \fmf{photon,label=$\gamma$}{i1,v}
	      \fmf{fermion}{o1,v}
	      \fmf{fermion,tension=2}{v,v1,o2}
	      \fmfblob{.15w}{v1}
	      \fmfdot{v}
	  \end{TempGraph} }
    \end{split}
    \quad + \quad
    \begin{split} \fmfframe(0,3)(0,3){
	  \begin{TempGraph}
	      \fmf{photon,label=$\gamma$}{i1,v}
	      \fmf{fermion,tension=2}{o1,v1,v}
	      \fmf{fermion}{v,o2}
	      \fmfblob{.15w}{v1}
	      \fmfdot{v}
	  \end{TempGraph} }
    \end{split} 
    \quad = 0
\end{equation}
For the present discussion, it is more convenient to relate $\alpha_0$ to
$\albar$, rather than to $\alpha$. Making use of Eqs.~\eqref{eq:DEF.570}
and~\eqref{eq:DEF.580}, we can rewrite Eq.~\eqref{eq:ANG.40a} as:
\begin{equation}\label{eq:ANG.40b}
    \albar = \alpha_0 \LT[ 1 - 2 \frac{s}{c} \Pi_\GZ(0)
      - \Pi_\gamma(m_Z^2) - \altw \RT].
\end{equation}

\subsection{The $W$ and $Z$ pole masses}

The relation between physical and bare $W$ and $Z$ masses are particularly
simple. At the one-loop level, the relevant Feynman diagrams are:
\begin{equation}
    \begin{split} \fmfframe(0,0)(0,3){
	  \begin{fmfgraph*}(25,5)
	      \fmfleft{i}
	      \fmfright{o}
	      \fmf{boson,label=$W,,Z$}{i,o}
	  \end{fmfgraph*} }
    \end{split}
    \qquad\qquad    
    \begin{split} \fmfframe(0,0)(0,3){
	  \begin{fmfgraph}(25,5)
	      \fmfleft{i}
	      \fmfright{o}
	      \fmf{boson}{i,v,o}
	      \fmfblob{h}{v}
	  \end{fmfgraph} }
    \end{split}
\end{equation}
Therefore:
\begin{align}
    \label{eq:ANG.40mw} m_W^2 & = m_{W0}^2 \LT[ 1 - \Pi_W(m_W^2) \RT], \\
    \label{eq:ANG.40mz} m_Z^2 & = m_{Z0}^2 \LT[ 1 - \Pi_Z(m_Z^2) \RT].
\end{align}

\subsection{The Fermi coupling constant}    

The Fermi coupling constant $G_F$ is defined from the process $\mu \to e
\nu_\mu \bar\nu_e$. The Feynman diagrams are:
\begin{equation} \begin{gathered}
    \begin{gathered} \fmfframe(0,5)(0,5){
	  \begin{fmfgraph*}(45,20)
	      \fmfleft{i2,i1}
	      \fmfright{o2,o1}
	      \fmflabel{$\mu$}{i1}
	      \fmflabel{$e$}{i2}
	      \fmflabel{$\nu_\mu$}{o1}
	      \fmflabel{$\bar\nu_e$}{o2}
	      \fmf{fermion}{i1,v1,o1}
	      \fmf{fermion}{o2,v2,i2}
	      \fmffreeze
	      \fmf{boson,label=$W$}{v1,v2}
	      \fmfdot{v1,v2}
	  \end{fmfgraph*} } \\ \mathrm{(a)}
    \end{gathered}
    \qquad\qquad
    \begin{gathered} \fmfframe(0,5)(0,5){
	  \begin{fmfgraph}(45,20)
	      \fmfleft{i2,i1}
	      \fmfright{o2,o1}
	      \fmf{fermion}{i1,v1,o1}
	      \fmf{fermion}{o2,v2,i2}
	      \fmffreeze
	      \fmf{boson}{v1,v3,v2}
	      \fmfblob{.20w}{v3}
	      \fmfdot{v1,v2}
	  \end{fmfgraph} } \\ \mathrm{(b)}
    \end{gathered} \\[1mm]
    \begin{gathered} \fmfframe(-3,5)(-5,5){
	  \begin{fmfgraph}(45,20)
	      \fmfleft{i2,i1}
	      \fmfright{o2,o1}
	      \fmf{fermion}{i1,v1,o1}
	      \fmf{fermion}{o2,v2,i2}
	      \fmffreeze
	      \fmf{boson}{v1,v2}
	      \fmfblob{.20w}{v1}
	      \fmfdot{v2}
	  \end{fmfgraph} } \\ \mathrm{(c)}
    \end{gathered}
    \qquad
    \begin{gathered} \fmfframe(-3,5)(-5,5){
	  \begin{fmfgraph}(45,20)
	      \fmfleft{i2,i1}
	      \fmfright{o2,o1}
	      \fmf{fermion}{i1,v1,o1}
	      \fmf{fermion}{o2,v2,i2}
	      \fmffreeze
	      \fmf{boson}{v1,v2}
	      \fmfdot{v1}
	      \fmfblob{.20w}{v2}
	  \end{fmfgraph} } \\ \mathrm{(d)}
    \end{gathered}
    \qquad
    \begin{gathered} \fmfframe(-3,5)(-5,5){
	  \begin{fmfgraph}(45,20)
	      \fmfleft{i2,i1}
	      \fmfright{o2,o1}
	      \fmf{fermion}{i1,v1,v3,o1}
	      \fmf{fermion}{o2,v4,v2,i2}
	      \fmffreeze
	      \fmf{boson}{v1,v2}
	      \fmf{boson}{v3,v4}
	      \fmfdot{v1,v2,v3,v4}
	  \end{fmfgraph} } \\ \mathrm{(e)}
    \end{gathered}
\end{gathered} \end{equation}
Therefore, the relation between $G_F$ and $G_{F0}$ is:
\begin{equation} \label{eq:ANG.40g}
    G_F = G_{F0} \LT[ 1 + \Pi_W(0) + D \RT],
\end{equation}
where the term $D$ denote the sum of the last three diagrams~(c)-(e).

\subsection{The $g_V/g_A$ ratio} \label{sec:ANG.10.40}

\renewenvironment{TempGraph}
{\begin{fmfgraph*}(35,20)
  \fmfleft{i1}
  \fmfright{o2,o1}
  \fmflabel{$\bar e$}{o1}
  \fmflabel{$e$}{o2}}
{\end{fmfgraph*}}

The last expressions we are interested in relate the bare parameters
$g_{V0}$ and $g_{A0}$ to the physical observables $g_V$ and $g_A$. These
quantities describe the coupling of the $Z$ boson field with the
\emph{vector} and \emph{axial} part of the charged lepton current,
respectively, and therefore the Feynman diagrams which are relevant for the
present case are those affecting the $Zll$ vertex:
\begin{equation} \begin{gathered}
    \begin{gathered} \fmfframe(0,3)(-1,3){
	  \begin{TempGraph}
	      \fmf{boson,label=$Z$}{i1,v}
	      \fmf{fermion}{o1,v,o2}
	      \fmfblob{.20w}{v}
	  \end{TempGraph} } \\ \mathrm{(a)}
    \end{gathered}
    \qquad\qquad
    \begin{gathered} \fmfframe(0,3)(-1,3){
	  \begin{TempGraph}
	      \fmf{boson,label=$Z$}{i1,v}
	      \fmf{fermion}{o1,v}
	      \fmf{fermion,tension=2}{v,v1,o2}
	      \fmfblob{.15w}{v1}
	      \fmfdot{v}
	  \end{TempGraph} } \\ \mathrm{(b)}
    \end{gathered} \\[4mm]
    \begin{gathered} \fmfframe(0,3)(-1,3){
	  \begin{TempGraph}
	      \fmf{boson,label=$Z$}{i1,v}
	      \fmf{fermion,tension=2}{o1,v1,v}
	      \fmf{fermion}{v,o2}
	      \fmfblob{.15w}{v1}
	      \fmfdot{v}
	  \end{TempGraph} } \\ \mathrm{(c)}
    \end{gathered}
    \qquad\quad
    \begin{gathered} \fmfframe(0,3)(-1,3){
	  \begin{TempGraph}
	      \fmf{boson,label=$Z$,tension=2}{i1,v1,v}
	      \fmf{fermion}{o1,v,o2}
	      \fmfblob{.20w}{v1}
	      \fmfdot{v}
	  \end{TempGraph} } \\ \mathrm{(d)}
    \end{gathered}
    \qquad\quad
    \begin{gathered} \fmfframe(0,3)(-1,3){
	  \begin{TempGraph}
	      \fmf{boson,label=$Z$,tension=2}{i1,v1}
	      \fmf{boson,label=$\gamma$,tension=2}{v1,v}
	      \fmf{fermion}{o1,v,o2}
	      \fmfblob{.20w}{v1}
	      \fmfdot{v}
	  \end{TempGraph} } \\ \mathrm{(e)}
    \end{gathered}
\end{gathered} \end{equation}
In Ref.~\cite{NOV93}, accurate expressions relating $g_{V0}$ to $g_V$ and
$g_{A0}$ to $g_A$ are derived. However, since in the present chapter the
only place where $g_V$ and $g_A$ are used is the definition of the
electroweak angle $\theta_l$, it is enough for our purposes to consider just
the \emph{ratio} between them; therefore the only expression we report here
is:
\begin{equation} \label{eq:ANG.40l}
    \frac{g_V}{g_A} = \frac{g_{V0}}{g_{A0}} \LT( 1 - \frac{4cs}{1 - 4s^2}
    \LT[ F_V^{Ze} - \LT( 1 - 4s^2 \RT) F_A^{Ze} + \Pi_\GZ(m_Z^2) \RT] \RT).
\end{equation}
It is easy to understand \emph{qualitatively} Eq.~\eqref{eq:ANG.40l} by
comparing it with the Feynman diagrams drawn above. The terms $F_V^{Ze}$ and
$F_A^{Ze}$ describe the vector and axial part of the $Zll$ vertex, and
include both the contribution from the irreducible proper vertex~(a) and the
external lepton self-energies~(b) and~(c); the term $\Pi_\GZ(m_Z^2)$ comes
from diagram~(e). It is interesting to note that diagram~(d), which is
related to the wave function renormalization of the $Z$ boson field, gives
the same contribution both to $g_V$ and to $g_A$, and therefore it cancels
out (at one-loop level) from the ratio~\eqref{eq:ANG.40l}.

\section{The one-loop approximation}

Having proven that the tree approximation is not adequate to explain
experimental data, let us now take into consideration the effects of
radiative corrections. In the present section, we will derive relations
which are correct up to the one-loop level, leaving for the next section the
discussion of some higher-order contributions.

\subsection{$\hat s^2$ versus $s^2$}

The simplest way to relate $\hat s^2$ to $s^2$ is to start from
Eq.~\eqref{eq:ANG.20p} and substitute the bare parameters appearing on its
\rhs\ with physical quantities, by means of Eqs.~\eqref{eq:ANG.40b},
\eqref{eq:ANG.40mz} and~\eqref{eq:ANG.40g}. The final result is:
\begin{equation}
    c_0^2 s_0^2 = \frac{\pi \albar}{\sqrt{2} G_F m_Z^2}
    \frac{\LT[ 1 + \Pi_W(0) + D \RT] \LT[ 1 - \Pi_Z(m_Z^2) \RT]}
    {\LT[ 1 - 2\frac{s}{c} \Pi_\GZ(0) - \Pi_\gamma(m_Z^2) -
      \altw \RT]}.
\end{equation}
Comparing this expression with Eq.~\eqref{eq:DEF.590}, we see that the
overall factor ${\pi \albar}/{\sqrt{2} G_F m_Z^2}$ is nothing but $c^2 s^2$.
Moreover, since for the moment we are interested only in the one-loop
approximation, we can expand the involved ratio of self-energies into an
expression linear in the $\Pi_i$. In this way, we have:
\begin{equation} \label{eq:ANG.50}
    c_0^2 s_0^2 = c^2 s^2
    \LT( 1 + 2 \frac{s}{c} \Pi_\GZ(0) + \Pi_\gamma(m_Z^2) - \Pi_Z(m_Z^2)
    + \Pi_W(0) + \altw + D \RT).
\end{equation}
It is straightforward to see that this angle $\theta_0$ will coincide with
$\hat\theta$ if $D$ and $\Pi_i$ are calculated in $\MS$ framework with $\mu
= m_Z$. Therefore, from~\eqref{eq:ANG.50} we easily get:
\begin{equation} \label{eq:ANG.70}
    \boxed{
      \hat s^2 = s^2 + \frac{c^2 s^2}{c^2 - s^2}
      \LT( 2 \frac{s}{c} \hat\Pi_\GZ(0) + \hat\Pi_\gamma(m_Z^2)
      - \hat\Pi_Z(m_Z^2) + \hat\Pi_W(0) + \altw + \hat D \RT)
      }
\end{equation}
which is the relation we are interested in. Since the last equation is
central for the present discussion, let us derive it in a different way.
Substituting Eqs.~\eqref{eq:ANG.40mw} and~\eqref{eq:ANG.40mz} into
Eq.~\eqref{eq:ANG.20m} and assuming that $\Pi_W$ and $\Pi_Z$ are evaluated
in the $\MS$ scheme, we get (see also Ref.~\cite{Sirlin89}):
\begin{equation} \begin{split} \label{eq:ANG.80}
    \hat s^2 = 1 - \LT[ \frac{m_{W0}^2}{m_{Z0}^2} \RT]_\Ms
      & = 1 - \frac{m_W^2 \LT( 1 - \hat\Pi_Z(m_Z^2) \RT)}
        {m_Z^2 \LT( 1 - \hat\Pi_W(m_W^2) \RT)} \\
      & = 1 - \frac{m_W^2}{m_Z^2} + c^2
      \LT[ \hat\Pi_Z(m_Z^2) - \hat\Pi_W(m_W^2) \RT],
\end{split} \end{equation}
where in the last expression we have substituted $\LT( m_W/m_Z \RT)^2$ with
$c^2$ in the factor which multiplies $\Pi_i$, which is correct at one-loop
level. Now for the ratio $m_W/m_Z$ we should use a formula which takes
radiative corrections into account. Since we are following the general
approach to the electroweak radiative corrections which is presented partly
in Ref.~\cite{NOV93}, we use Eq.~(38) from that paper:
\begin{equation} \begin{split} \label{eq:ANG.90}
    \frac{m_W^2}{m_Z^2} = c^2 + \frac{c^2 s^2}{c^2 - s^2} \bigg( 
    & \frac{c^2}{s^2} \LT[ \Pi_Z(m_Z^2) - \Pi_W(m_W^2) \RT] + \Pi_W(m_W^2) \\
    & - \Pi_W(0) - \Pi_\gamma(m_Z^2) - 2 \frac{s}{c} \Pi_\GZ(0) - \altw - D
    \bigg).
\end{split} \end{equation}
The term $\altw$, which is absent in the original formula, has been added
here to take explicitly into account the \emph{top} and $W$ loops, which
contribute to $\Pi_\gamma$ but are not included in the definition of
$\albar$ (in Ref.~\cite{NOV93}, these contributions were accounted for by
replacing $s^2$ with $s^2+0.00015$: see Ref.~\cite{NOV94a} for a detailed
discussion on this subject).

Since both $m_W/m_Z$ and $c$ are finite, the expression for the sum of all
the radiative corrections is finite as well and we can use $\MS$ quantities
in it:
\begin{equation} \begin{split}
    \frac{m_W^2}{m_Z^2} = c^2 + \frac{c^2 s^2}{c^2 - s^2} \bigg( 
    & \frac{c^2}{s^2} \LT[ \hat\Pi_Z(m_Z^2) - \hat\Pi_W(m_W^2) \RT]
      + \hat\Pi_W(m_W^2) \\
    & - \hat\Pi_W(0) - \hat\Pi_\gamma(m_Z^2) - 2 \frac{s}{c} \hat\Pi_\GZ(0) 
      - \altw - \hat D 
    \bigg).
\end{split} \end{equation}
Substituting the last equation in~\eqref{eq:ANG.80}, we obtain once more
Eq.~\eqref{eq:ANG.70}.

\begin{table}[!t] \centering
    \begin{tabular}{>{\rule[-9pt]{0pt}{24pt}}r!{\hspace{15mm}}rrr!{\hspace{15mm}}r}
	\hline\hline
	$\hat D \approx 96.083$       &    Heavy &    Light &   Bosons &    Total \\
	\hline
	$\altw$                       &   -0.623 &        0 &    5.326 &    4.703 \\
	$\hat\Pi_W(0)$                &  149.559 &        0 &  -18.891 &  130.669 \\
	$2\dfrac{s}{c}\hat\Pi_\GZ(0)$ &        0 &        0 &   -6.493 &   -6.493 \\
	$\hat\Pi_\gamma(m_Z^2)$       &   -8.964 &   86.902 &  -14.312 &   63.627 \\
	$- \hat\Pi_Z(m_Z^2)$          & -244.399 & -119.530 &   71.806 & -292.123 \\
	\hline
	Total                         & -104.427 &  -32.628 &   37.438 &  -99.617 \\
	\hline\hline
    \end{tabular}
    \caption[Contributions of self-energies to $\hat s^2 - s^2$]{
      Contributions of the various self-energies appearing in
      Eq.~\eqref{eq:ANG.70}, in units of $10^{-4}$. The first column refers
      to \emph{top} and \emph{bottom}, the second to leptons and \emph{up},
      \emph{down}, \emph{charm}, \emph{strange} quarks. The third column
      contains vector bosons, higgs and ghost loops, and the last one is the
      sum of the first three. We assumed $m_t = 173.8\GeV$ and $m_H =
      120\GeV$. The vertex-box contribution $\hat D$ is reported in the
      upper-left corner.}
    \label{tab:ANG.10}
\end{table}

In Table~\ref{tab:ANG.10} we provide some numerical estimate for all the
polarization operators appearing in Eq.~\eqref{eq:ANG.70}, together with the
vertex-box term $\hat D$, for $m_t = 173.8\GeV$ and $m_H = 120\GeV$. It is
immediate to see that the contribution coming from top quark loops dominates
over light fermions and gauge bosons, and that the overall result for all
self-energies cancels quite exactly the vertex-box diagrams $\hat D$.
The final expression for the difference $\hat s^2 - s^2$ at one-loop level
is therefore:
\begin{equation} \label{eq:ANG.100}
    \LT. \hat s^2 - s^2 \RT|_{\substack{m_t = 173.8\GeV \\ m_H = 120\GeV}} =
    -0.00329 + 0.00318 = -0.00011,
\end{equation}
where the first term comes from polarization operators and the second from
vertices and boxes.

\begin{figure}[!t] \centering
    \includegraphics[width=0.9\textwidth]{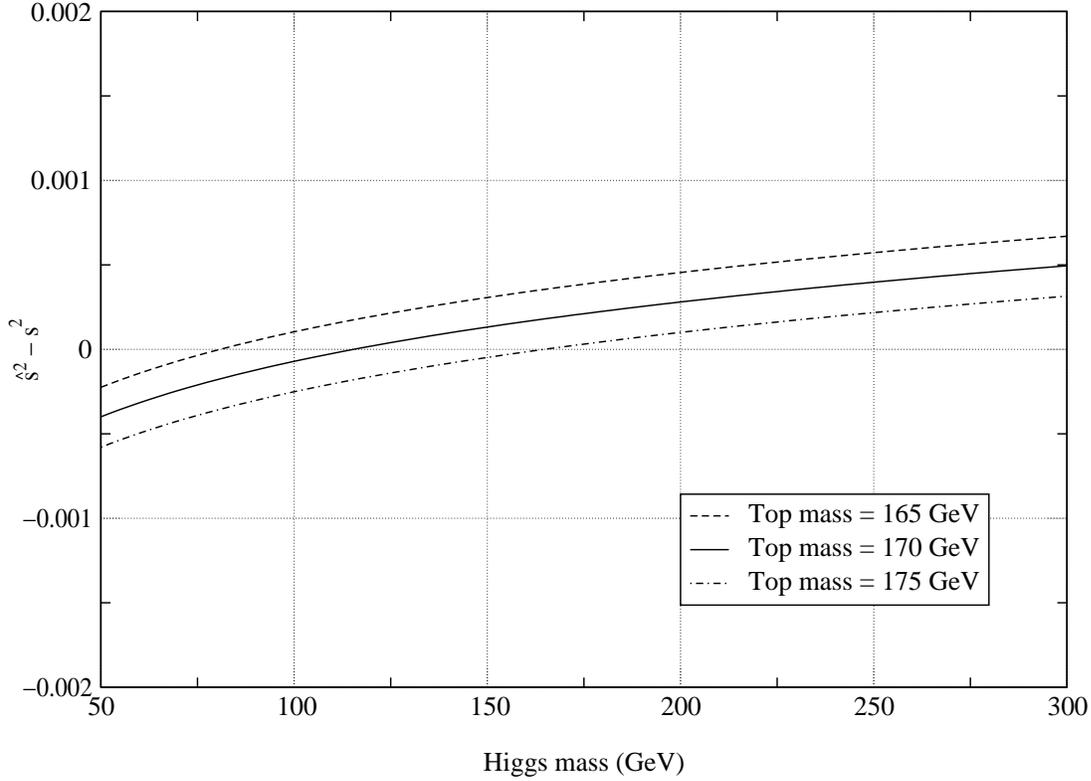}
    \caption[Plot of $\hat s^2 - s^2$ as a function of $m_H$]{
      Numerical value of the difference $\hat s^2 - s^2$, at one-loop, as a
      function of the higgs mass $m_H$. Note that the dependence on $m_H$ is
      only logarithmic.}
    \label{fig:ANG.10}
\end{figure}

\begin{figure}[!t] \centering
    \includegraphics[width=0.9\textwidth]{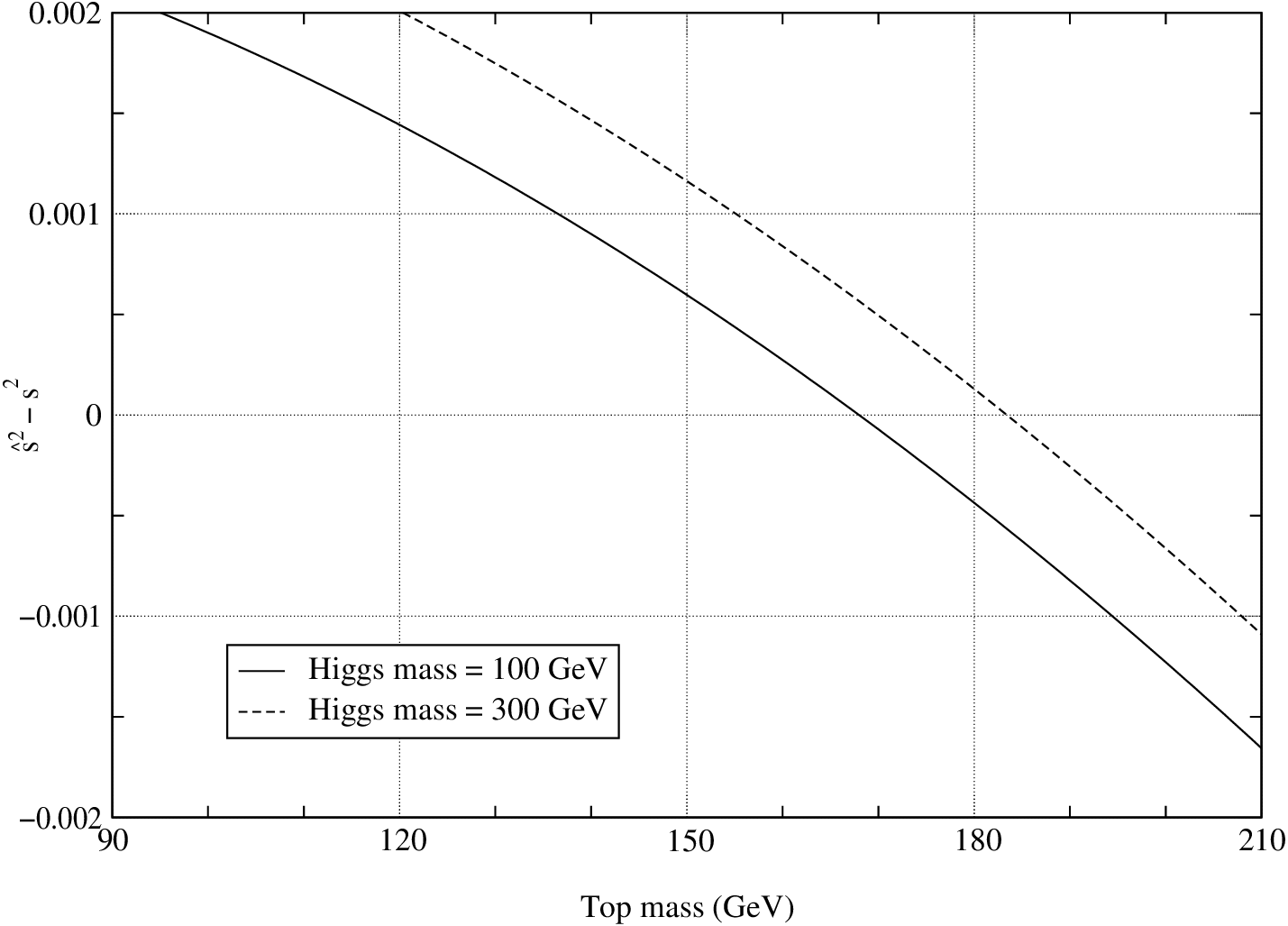}
    \caption[Plot of $\hat s^2 - s^2$ as a function of $m_t$]{
      Numerical value of the difference $\hat s^2 - s^2$, at one-loop, as a
      function of the top mass $m_t$. The strong dependence on $m_t$ is
      clearly evident.}
    \label{fig:ANG.20}
\end{figure}

To understand whether this strong cancellation has some physical meaning or
is purely accidental, let's show in Figs.~\ref{fig:ANG.10}
and~\ref{fig:ANG.20} the $\hat s^2 - s^2$ dependence on $m_H$ and $m_t$,
respectively. From Fig.~\ref{fig:ANG.10}, we see that the higgs mass play
only a marginal role, since effects of varying it are in general rather
small. This can be verified straightforwardly by looking at the leading term
in $m_H$ which enters Eq.~\eqref{eq:ANG.70}:
\begin{equation} \label{eq:ANG.120}
    \LT. \hat s^2 - s^2 \RT|_{m_H \gg m_Z} \approx
    \frac{\bar\alpha \LT( 1 + 9s^2 \RT)}{48 \pi \LT( c^2 - s^2 \RT)}
    \ln \LT( \frac{m_H}{m_Z} \RT)^2.
\end{equation}
From this formula one can see that the dependence of $\hat s^2 - s^2$ on the
higgs mass is only logarithmic.

The situation completely changes if one consider the dependence on the top
mass. From Fig.~\ref{fig:ANG.20}, it is clear that $\hat s^2$ is close to
$s^2$ only for $m_t$ around $170\GeV$, so one cannot find any physical
reason for the closeness of these two angles. The fact that $\hat s^2 - s^2$
rapidly varies with $m_t$ can be figured out from the large $m_t$
approximation:
\begin{equation} \label{eq:ANG.130}
    \LT. \hat s^2 - s^2 \RT|_{m_t \gg m_Z} \approx
    -\frac{3 \bar\alpha}{16 \pi \LT( c^2 - s^2 \RT)}
    \LT( \frac{m_t}{m_Z} \RT)^2.
\end{equation}
At this point we conclude that the numerical closeness of $\hat s^2$ and
$s^2$ is a mere coincidence without any deep physical reason.

\subsection{$s_l^2$ versus $s^2$}

\begin{figure}[!t] \centering
    \includegraphics[width=0.9\textwidth]{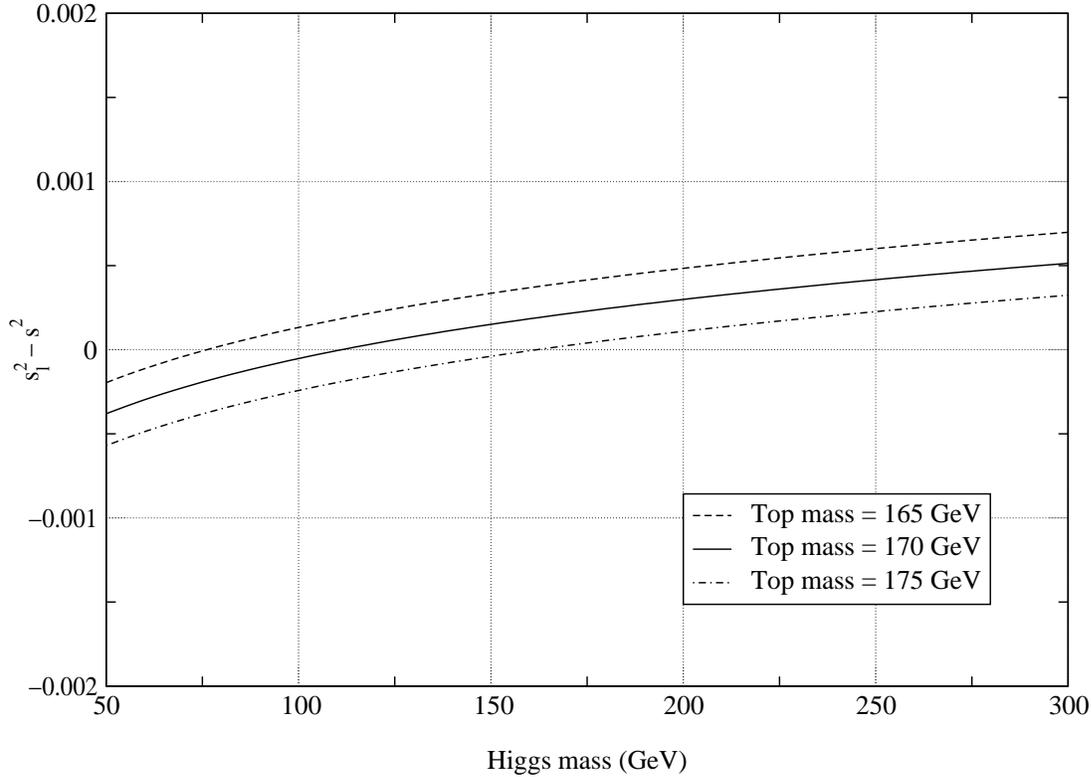}
    \caption[Plot of $s_l^2 - s^2$ as a function of $m_H$]{
      Numerical value of the difference $s_l^2 - s^2$, at one-loop, as a
      function of the higgs mass $m_H$. The dependence on $m_H$ is only
      logarithmic.} 
    \label{fig:ANG.30}
\end{figure}

\begin{figure}[!t] \centering
    \includegraphics[width=0.9\textwidth]{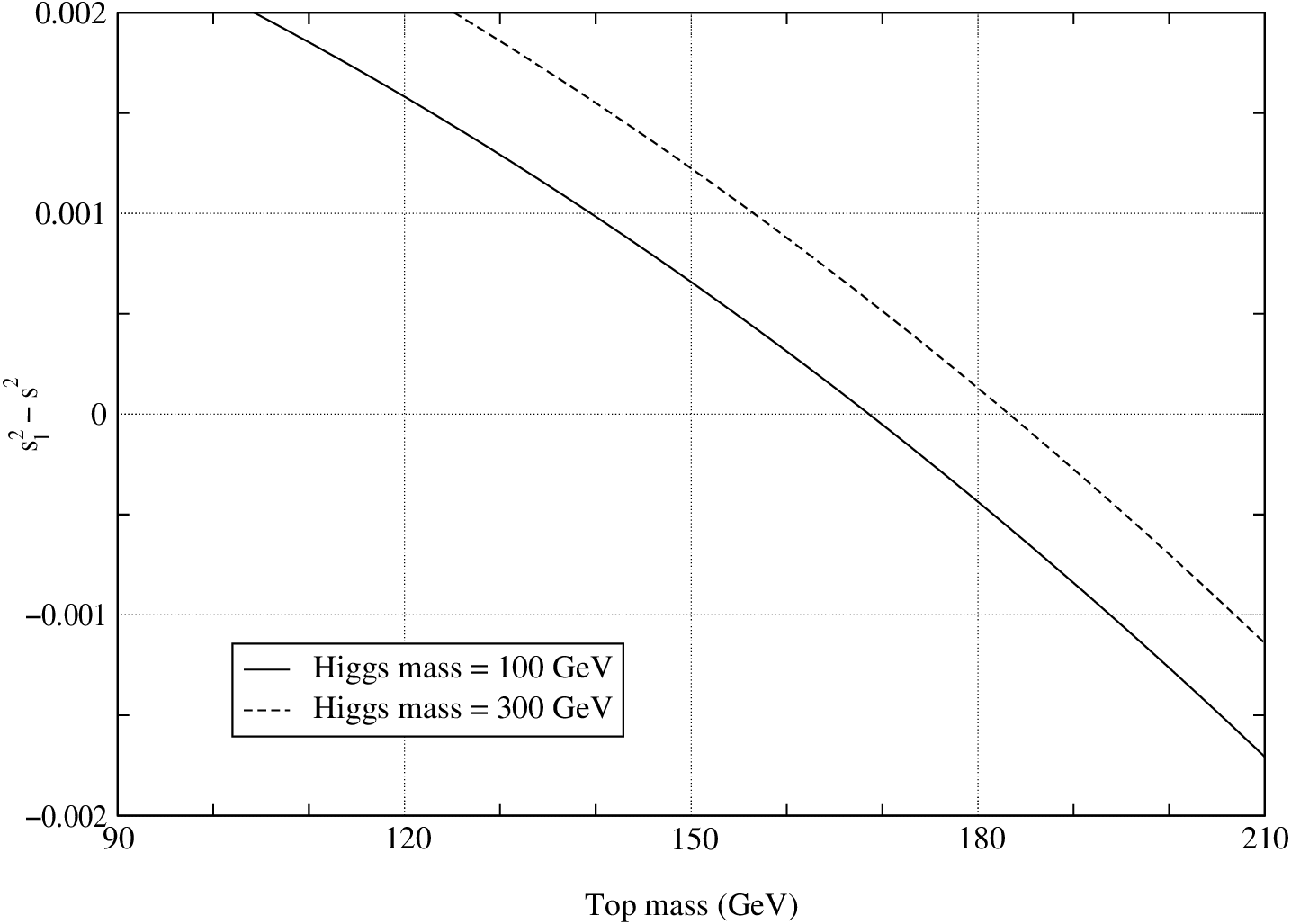}
    \caption[Plot of $s_l^2 - s^2$ as a function of $m_t$]{
      Numerical value of the difference $s_l^2 - s^2$, at one-loop, as a
      function of the top mass $m_t$. The strong dependence on $m_t$ is
      clearly evident.}
    \label{fig:ANG.40}
\end{figure}

Before going further, let us quickly review the relation between $s_l^2$ and
$s^2$. As stated in the introduction, and as can be seen directly comparing
Eqs.~\eqref{eq:ANG.10p} and~\eqref{eq:ANG.10l}, it turns out that these two
angles are numerically very close; however, this degeneracy is known to be
accidental. To check this, we need to derive an analytic expression for
their difference, and this can be done in two steps by first substituting
Eq.~\eqref{eq:ANG.40l} into~\eqref{eq:ANG.20l} and using
Eq.~\eqref{eq:ANG.30l}:
\begin{equation} \label{eq:ANG.160}
    s_0^2 = s_l^2 - cs \LT[ F_V^{Ze} - \LT( 1 - 4 s^2 \RT) F_A^{Ze}
      + \Pi_\GZ(m_Z^2) \RT],
\end{equation}
and then eliminating $s_0^2$ in favor of $s^2$ by means of
Eq.~\eqref{eq:ANG.50}:
\begin{equation} \begin{split} \label{eq:ANG.170}
    s_l^2 = s^2 
    & + cs \LT[ F_V^{Ze} - \LT( 1 - 4 s^2 \RT) F_A^{Ze}
      + cs \Pi_\GZ(m_Z^2) \RT] \\
    & + \frac{c^2 s^2}{c^2 - s^2} \LT[ \Pi_\gamma(m_Z^2) - \Pi_Z(m_Z^2)
      + \Pi_W(0) + 2 \frac{s}{c} \Pi_\GZ(0) + \altw + D \RT].
\end{split} \end{equation}
Having this formula at our disposal, we can now investigate numerically the
dependence of $s_l^2 - s^2$ on $m_H$ and $m_t$, and the corresponding plots
are shown in Fig.~\ref{fig:ANG.30} and~\ref{fig:ANG.40}. As in the previous
case, once again this difference depends logarithmically on $m_H$ but
linearly on $(m_t/m_Z)^2$, and so the coincidence occurs only for $m_t$
close to $170\GeV$. However, if now we evaluate the large $m_H$ and $m_t$
approximations and write down the corresponding asymptotic expressions, we
find an interesting result:
\begin{align}
    \LT. s_l^2 - s^2 \RT|_{m_H \gg m_Z} & \approx
      \frac{\bar\alpha \LT( 1 + 9s^2 \RT)}{48 \pi \LT( c^2 - s^2 \RT)}
      \ln \LT( \frac{m_H}{m_Z} \RT)^2, \\
    \LT. s_l^2 - s^2 \RT|_{m_t \gg m_Z} & \approx
      -\frac{3 \bar\alpha}{16 \pi \LT( c^2 - s^2 \RT)}
      \LT( \frac{m_t}{m_Z} \RT)^2.
\end{align}
which coincide with Eqs~\eqref{eq:ANG.120} and~\eqref{eq:ANG.130}! This
immediately lead us to conclude that in the difference $\LT( \hat s^2 - s^2
\RT) - \LT( s_l^2 - s^2 \RT)$ the terms of order $\ln(m_H/m_Z)^2$ and
$(m_t/m_Z)^2$ cancel, i.e.\ the dependence of $\hat s^2 - s_l^2$ on $m_t$ is
\emph{weaker} than in the two cases analyzed so far. This is just the kind
of relation we were looking for, and will be discussed widely in the next
section.

\subsection{$\hat s^2$ versus $s_l^2$}

As explained in Sec.~\ref{sec:ANG.10.40}, the quantity $s_l$ describes the
asymmetries in $Z$ boson decay; $s_l$, $s_u$ and $s_d$ refers to decays into
pairs of leptons, up-quarks and down-quarks, respectively. Let us discuss
$s_l$. Rewriting Eq.~\eqref{eq:ANG.160} in the $\MS$ scheme with $\mu =
m_Z$, we immediately get (see also Ref.~\cite{Gambino94}):
\begin{equation} \label{eq:ANG.190}
    \boxed{
      \hat s^2 = s_l^2 - cs \LT[ \hat F_V^{Ze} 
      - \LT( 1 - 4 s^2 \RT) \hat F_A^{Ze} + \hat\Pi_\GZ(m_Z^2) \RT]
      }
\end{equation}
The form of the last equation can be foreseen without any calculation. The
point is that both $\hat\theta$ and $\theta_l$ are defined by the ratio of
bare gauge coupling constants; the difference between them arises since
$\theta_l$ describes $Z \to e^+ e^-$ decays and in this case the additional
vertex radiative corrections as well as $Z \to \gamma \to e^+ e^-$
transition contribute to $\theta_l$. In~\eqref{eq:ANG.190} these additional
terms are subtracted from $s_l^2$ in order to get $\hat s^2$. The vertex
terms in~\eqref{eq:ANG.190} are mere numbers, while $\hat\Pi_\GZ$ depends on
$m_t$ only logarithmically due to the non-decoupling property of $\MS$
scheme (since a diagonal vector current is conserved, there is no $m_t^2$
term in $\hat\Pi_\GZ$, that is why $\hat\Pi_\GZ$ is numerically small).
There is also no $m_H$ dependence in the difference $\hat s^2 - s_l^2$.

From the $\hat\Pi_\GZ(m_Z^2)$ term we get the following expression for the
logarithmically enhanced contribution for $m_t \gg m_Z$:
\begin{equation} \label{eq:ANG.195}
    \LT. \hat s^2 - s_l^2 \RT|_{m_t \gg m_Z} \approx
    \frac{\albar}{\pi} \LT( \frac16 - \frac49 s^2 \RT)
    \ln \LT( \frac{m_t}{m_Z} \RT)^2.
\end{equation}

Having all the necessary formulas in our disposal, we are ready to make
numerical estimates. Using expressions~(93),~(94) from Ref.~\cite{NOV96} and
formulas from Appendix G of Ref.~\cite{NOV93}, we get:
\begin{align}
    \label{eq:ANG.200a} \hat F_V^{Ze} = 0.00197 + \frac{\albar}{8 \pi}
      \frac{c}{s^3} \ln \LT( \frac{m_W}{m_Z} \RT)^2 = 0.00133, \\
    \label{eq:ANG.200b} \hat F_A^{Ze} = 0.00186 + \frac{\albar}{8 \pi}
      \frac{c}{s^3} \ln \LT( \frac{m_W}{m_Z} \RT)^2 = 0.00122,
\end{align}
where the logarithmic terms arise from the divergent parts of vertex
functions after imposing $\MS$ renormalization conditions with $\mu = m_Z$.
Note that in numerical calculations we substituted $c^2$ for $\LT( m_W/m_Z
\RT)^2$.

To calculate $\hat\Pi_\GZ(m_Z^2)$ we use formulas from Appendices of
Ref.~\cite{NOV93}, which take into account $W^+ W^-$, light fermions and
$(t,b)$ doublet contributions. For $m_t = 173.8\GeV$ we obtain:
\begin{equation} \label{eq:ANG.210}
    \hat\Pi_\GZ(m_Z^2) = -0.00121.
\end{equation}

\begin{figure}[!t] \centering
    \includegraphics[width=0.9\textwidth]{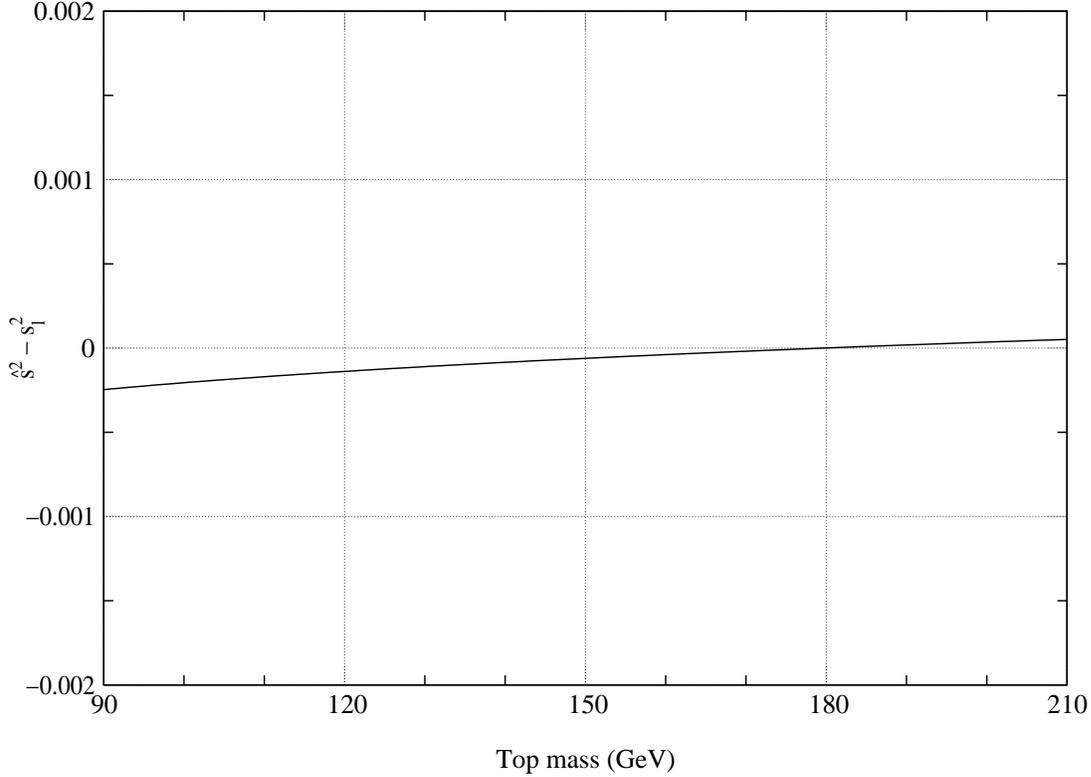}
    \caption[Plot of $\hat s^2 - s_l^2$ as a function of $m_t$]{
      Numerical value of the difference $\hat s^2 - s_l^2$, at one-loop, as
      a function of the top mass $m_t$. This quantity does not depend on the
      higgs mass, and the dependence on $m_t$ is only logarithmic.}
    \label{fig:ANG.50}
\end{figure}

Substituting~\eqref{eq:ANG.200a},~\eqref{eq:ANG.200b} and~\eqref{eq:ANG.210}
into~\eqref{eq:ANG.190}, we finally obtain:
\begin{equation} \label{eq:ANG.220}
    \LT. \hat s^2 - s_l^2 \RT|_{m_t = 173.8\GeV} =
    -0.00052 + 0.00051 = -0.00001,
\end{equation}
where the first term comes from vertex diagrams and the second from
$\hat\Pi_\GZ(m_Z^2)$. This expression should be compared with
Eq.~\eqref{eq:ANG.100}: once again, we see that a very small numerical value
arises from two large contributions of opposite sign. Also for
Eq.~\eqref{eq:ANG.220} we must conclude that this cancellation, which is
peculiar to $s_l$ and does not occur for $s_u$ or $s_d$, is merely
accidental (see also Ref.~\cite{Gambino94}); but it is essential to stress
that, unlike the case of Eq.~\eqref{eq:ANG.100}, here there is no dependence
on the higgs mass, and yet more important the dependence on the top quark
mass is now only logarithmic. As a consequence, even if the top mass would
have been quite different from its actual value $173.8(5.2)\GeV$, a
numerical closeness between $\hat s^2$ and $s_l^2$ would still have
occurred, while no coincidence between $\hat s^2$ (or $s_l^2$) and $s^2$
would have appeared. This situation is clearly depicted in
Fig.~\ref{fig:ANG.50}, where $\hat s^2 - s_l^2$ is plotted against $m_t$:
one can easily see that, unlike the case of $\hat s^2 - s^2$ difference,
here the dependence on $m_t$ is really small for a large $m_t$ values
interval.

\section{Higher-order corrections and numerical estimates}
\label{sec:ANG.30}

Due to the strong cancellation which occurs among one-loop diagrams in
Eqs.~\eqref{eq:ANG.100} and~\eqref{eq:ANG.220}, an accurate analysis of the
differences $\hat s^2 - s^2$ and $\hat s^2 - s_l^2$ requires to take into
consideration also the leading two-loop contributions, which can now be
comparable or even larger than one-loop ones.

\subsection{$\hat s^2$ versus $s_l^2$}

\renewenvironment{TempGraph}
{\begin{fmfgraph*}(50,20)
  \fmfleft{i}
  \fmfright{o}
  \fmf{boson,label=$\gamma$}{i,v1}
  \fmf{boson,label=$Z$}{v2,o}
  \fmffixed{(h,0)}{v1,v2}}
{\end{fmfgraph*}}

For the case of $\hat s^2 - s_l^2$, the leading two-loop corrections are of
the order of $\alpha \alpha_s$ and come from the insertion of a gluon into
quark loops which contribute to $\hat\Pi_\GZ(m_Z^2)$:
\begin{equation}
    \begin{split} \fmfframe(0,0.5)(0,0){
	  \begin{TempGraph}
	      \fmf{plain,left=1/2}{v1,v3,v2,v4,v1}
	      \fmf{gluon}{v3,v4}
	      \fmffixed{(0,h)}{v4,v3}
	      \fmfdot{v1,v2,v3,v4}
	  \end{TempGraph} }
    \end{split}
    \qquad\qquad
    \begin{split} \fmfframe(0,0.5)(0,0){
	  \begin{TempGraph}
	      \fmf{plain,left=1/3}{v1,v3,v4,v2}
	      \fmf{plain,left}{v2,v1}
	      \fmf{gluon,right=1/3}{v3,v4}
	      \fmffixed{(0.25h,+0.433h)}{v1,v3}
	      \fmffixed{(0.25h,-0.433h)}{v4,v2}
	      \fmfdot{v1,v2,v3,v4}
	  \end{TempGraph} }
    \end{split}
\end{equation}
There are two types of one-loop diagrams which we should consider: those
with light quarks ($u,d,c,s,b$), and those with heavy top ($t$). The
necessary two-loop formulas are widely discussed in Ref.~\cite{Kniehl90}.
However, in that article all the calculations were made with ultraviolet
cutoff $\Lambda$. To rewrite Kniehl formulas in the usual dimensional
regularization approach, in order to derive expressions for the boson
self-energies in the $\MS$ renormalization scheme, we compare them with the
calculations of Djouadi and Gambino reported in Ref.~\cite{Djouadi87}. In
this way we find the following replacement rule:
\begin{equation}
    \ln \LT( \frac{\Lambda^2}{m_Z^2} \RT)
    \longrightarrow \Delta_Z + \frac{55}{12} - 4 \zeta(3) 
    \longrightarrow \frac{55}{12} - 4 \zeta(3)
    \approx -0.225.
\end{equation}

For the case of light quarks contribution ($u,d,c,s,b$), we get:
\begin{equation} \label{eq:ANG.240}
    \delta_{\text{light}}^{\alpha \alpha_s} \hat\Pi_\GZ (m_Z^2) =
    \frac{\hat\alpha_s(m_Z)}{\pi} \frac{\albar}{\pi cs}
    \LT(\frac{7}{12} - \frac{11}{9} s^2 \RT)
    \LT[ \frac{55}{12} - 4 \zeta(3) \RT] \approx -0.00002,
\end{equation}
where we used $\hat\alpha_s(m_Z) = 0.119$ for numerical estimate. For the
contribution of the top quark we obtain:
\begin{equation} \begin{split} \label{eq:ANG.250}
    \delta_t^{\alpha \alpha_s} \hat\Pi_\GZ (m_Z^2) & =
    \frac{\hat\alpha_s(m_t)}{\pi} \frac{\albar}{\pi cs}
    \LT(\frac{1}{6} - \frac{4}{9} s^2 \RT)
    \LT[ \frac{55}{12} - 4 \zeta(3) - \ln{t}
    + 4 t V_1\LT(\frac{1}{4 t}\RT) \RT] \\
    & \approx 0.00004,
\end{split} \end{equation}
where $t \equiv (m_t/m_Z)^2$ and (see Refs.~\cite{NOV94b,Kniehl90}):
\begin{gather}
    \hat\alpha_s(m_t) = \frac{\hat\alpha_s(m_Z)}
      {1 + \frac{23}{12\pi} \hat\alpha_s(m_Z) \ln{t}} \approx 0.109, \\[5pt]
     \label{eq:ANG.260b} V_1(x) = \LT[ 4 \zeta(3) - \frac{5}{6} \RT]x
       + \frac{328}{81} x^2 + \frac{1796}{675} x^3 + \ldots, \\[5pt]
    \zeta(3) = 1.2020569\ldots \; .
\end{gather}
Substituting~\eqref{eq:ANG.240} and~\eqref{eq:ANG.250}
into~\eqref{eq:ANG.190}, we have the overall $\alpha \alpha_s$ corrections:
\begin{equation} \label{eq:ANG.270}
    \delta^{\alpha \alpha_s} \LT( \hat s^2 - s_l^2 \RT) = -0.00001.
\end{equation}

Since the leading $\sim \alpha$ correction cancel almost completely
in~\eqref{eq:ANG.220}, one start to worry about significance of two-loop
$\alpha^2$ corrections. Enhanced $\alpha^2 t$ corrections to
Eq.~\eqref{eq:ANG.190} were calculated in Ref.~\cite{Gambino96}, where it is
stated that they are numerically negligible; $\alpha^2$ corrections have not
been calculated yet. However, according to Ref.~\cite{Gambino96} there exist
enhanced two-loop $\alpha^2 \pi^2$ corrections, which come from the
interference of the imaginary parts of $\Pi_\GZ$ and $\Pi_\gamma$:
\begin{equation} 
    \begin{split} \fmfframe(0,0.5)(0,0){
	  \begin{fmfgraph*}(70,20)
	      \fmfleft{i}
	      \fmfright{o1,o2}
	      \fmf{boson,label=$Z$,label.side=right}{i,v1}
	      \fmf{boson,label=$\gamma$,label.side=right,tension=2/3}{v1,v2}
	      \fmf{boson,label=$\gamma$,label.side=right}{v2,v}
	      \fmf{plain}{o1,v,o2}
	      \fmfblob{0.5h}{v1,v2}
	      \fmfdot{v}
	  \end{fmfgraph*} }
    \end{split}
\end{equation}
Numerically they give:
\begin{equation} \label{eq:ANG.275}
    \delta_{\text{int}}^{\alpha^2} \LT( \hat s^2 - s_l^2 \RT) 
    = -0.00004.
\end{equation}

Adding Eqs.~\eqref{eq:ANG.270} and~\eqref{eq:ANG.275} to the one-loop term
given in Eq.~\eqref{eq:ANG.220}, we finally get:
\begin{gather} 
    \label{eq:ANG.280} \delta\! \LT( \hat s^2 - s_l^2 \RT) = -0.00005, \\
    \label{eq:ANG.290} \LT. \hat s^2 - s_l^2 \RT|_{m_t = 173.8\GeV} =
    -0.00001 - 0.00005 = -0.00006,
\end{gather}
in good agreement with Eqs.~\eqref{eq:ANG.10g} and~\eqref{eq:ANG.10l}. It is
instructive to compare the last formula with the corresponding numbers in
Tables~1 and~2 from Ref.~\cite{Degrassi97}, as well as the last formula in
Ref.~\cite{Gambino96}.

\subsection{$\hat s^2$ versus $s^2$}

Coming back to the aim of the present chapter, we should study
Eq.~\eqref{eq:ANG.70} in more details. Comparing Eq.~\eqref{eq:ANG.100},
which is valid at the one-loop level, with Eqs.~\eqref{eq:ANG.10p}
and~\eqref{eq:ANG.10g}, we observe an evident inconsistency. To overcome it
small higher loop corrections to Eq.~\eqref{eq:ANG.70} should be accounted
for, in analogy with what was done for the case $\hat s^2 - s_l^2$. One can
act straightforwardly, taking into account corrections to polarization
operators entering~\eqref{eq:ANG.70}. Another possible way is to rewrite the
difference $\hat s^2 - s^2$ through $s_l^2$, so to split the contributions
of the higher order radiative corrections into two parts:
\begin{equation} \label{eq:ANG.300}
    \delta\! \LT( \hat s^2 - s^2 \RT) =
    \delta\! \LT( \hat s^2 - s_l^2 \RT) + \delta\! \LT( s_l^2 - s^2 \RT).
\end{equation}
The first term was discussed in the previous section, and the leading
two-loop contributions are given in Eq.~\eqref{eq:ANG.280}. Concerning the
second term, we follow the analysis presented in
Refs.~\cite{NOV96,NOV95,NOV95b}:
\begin{equation} \label{eq:ANG.310}
    \delta\! \LT( s_l^2 - s^2 \RT) =
    -\frac{3 \albar}{16\pi \LT( c^2 - s^2 \RT)}
    \LT( \delta_2 V_R + \delta_3 V_R + \delta_4 V_R + \delta'_4 V_R\RT)
\end{equation}
where $\delta_2 V_R$ comes from gluon insertions into quark loops, $\delta_3
V_R$ accounts for three-loop $\alpha \alpha_s^2 t$ terms, and $\delta_4 V_R$
and $\delta'_4 V_R$ include the leading two-loop electroweak corrections of
order $\alpha^2 t^2$ ($\delta_4$ comes from irreducible diagrams and
$\delta'_4$ from reducible ones). Expressions for these functions are
reported in Refs.~\cite{NOV96,NOV95,NOV95b}:
\begin{align}
    \label{eq:ANG.320a} \delta_2 V_R(t,h) & = \frac{4}{3}
      \frac{\hat\alpha_s(m_t)}{\pi} \LT[ t A_1\LT(\frac{1}{4t}\RT)
      - \frac{5}{3} t V_1\LT(\frac{1}{4t}\RT)
      - 4 t F_1(0) + \frac{1}{6} \ln t \RT], \\ 
    \delta_3 V_R(t,h) & = -14.594 \frac{\hat\alpha_s^2(m_t)}{\pi^2} t, \\
    \delta_4 V_R(t,h) & = -\frac{\albar}{16 \pi s^2 c^2}
      A\LT(\frac{m_H}{m_t}\RT) t^2, \\
    \label{eq:ANG.320d} \delta'_4 V_R(t,h) & = 
      -\frac{3 \albar}{16 \pi \LT( c^2 - s^2 \RT)^2} t^2,
\end{align}
where $V_1$ is given in~\eqref{eq:ANG.260b}, and $A_1$, $F_1$ and $A$ can be
found in Refs.~\cite{NOV96,NOV95,NOV95b}.

Substituting Eqs.~(\ref{eq:ANG.320a}-\ref{eq:ANG.320d})
into~\eqref{eq:ANG.310}, we have:
\begin{equation}
    \delta\! \LT( s_l^2 - s^2 \RT) = 0.00047,
\end{equation}
which can be added to Eq.~\eqref{eq:ANG.280} into Eq.~\eqref{eq:ANG.300},
and then to the one-loop term given in Eq.~\eqref{eq:ANG.100}, to find the
numerical value:
\begin{gather}
    \delta\! \LT( \hat s^2 - s^2 \RT) = 0.00042, \\
    \label{eq:ANG.350b} \LT. \hat s^2 - s^2 
      \RT|_{\substack{m_t = 173.8\GeV \\ m_H = 120\GeV}}
      = -0.00011 + 0.00042 = 0.00031,
\end{gather}
in good agreement with Eqs.~\eqref{eq:ANG.10p} and~\eqref{eq:ANG.10g}.

\section{Conclusions}

In this chapter we have investigated the origin of the numerical closeness
among three different definitions of the electroweak mixing angle: $s^2$,
$\hat s^2$, and $s_l^2$. We have found that the degeneracy between $\hat
s^2$ and $s^2$, as well as between $s_l^2$ and $s^2$, is merely accidental,
while for the case of $\hat s^2$ and $s_l^2$ the dependence of their
difference on the $m_t$ in only logarithmic: therefore their closeness would
have occurred even if the top quark mass would have been quite different
from its actual value.

Moreover, Eq.~\eqref{eq:ANG.70} provides an explicit and very simple
relation between the phenomenological quantity $s^2$ and the $\MS$ parameter
$\hat s^2$, and this relation will be very useful in the following chapter.
The numerical estimates we did in Sec.~\ref{sec:ANG.30} proves that our
formulas are in good agreement with experimental data (or, for the case of
$\hat s^2$, with the theoretical calculations of other groups), provided
that the leading two-loop contributions are taken into account.

Effects of new physics will be studied in the next chapter. However, let us
mention since now that in generalizations of Standard Model a lot of new
heavy particles occur, and all of them contribute to $\hat s^2$ due to the
non-decoupling property of $\MS$ renormalization. To avoid this
non-universality of $\MS$ quantities, it was suggested to subtract from
vector boson self-energies $\Pi_i$ the contributions of all the particles
heavier than $m_Z$. In this approach, not only new particles, but also the
top quark is decoupled from $\hat s^2$, and in particular the logarithmic
term shown in~\eqref{eq:ANG.195} should not be included (see
Refs.~\cite{Marciano90,Erler98}). According to the definition accepted in
Ref.~\cite{Erler98}, the quantity $\hat s^2$ which has been discussed up to
now is called $\hat s_{\text{ND}}^2$, while a new ``decoupled'' $\MS$
parameter $\hat s_Z^2$ is introduced:
\begin{equation} \label{eq:ANG.400}
    \hat s_Z^2 = \hat s^2 - \frac{\albar}{\pi} \LT( \frac{1}{6}
    - \frac{4}{9} s^2 \RT) \ln \LT( \frac{m_t}{m_Z} \RT)^2
    = \hat s^2 - 0.00020.
\end{equation}
From~\eqref{eq:ANG.10g} and~\eqref{eq:ANG.400} we get:
\begin{equation}
    \hat s_Z^2 = 0.23124(24),
\end{equation}
where, unlike $\hat s^2$, $\hat s_Z^2$ is uniquely defined both in the
Standard Model and in its extensions.

\end{fmffile}

\clearemptydoublepage

\chapter{$\hat s_\SM^2$, $\hat s_\ND^2$ and the value of
  $\hat\alpha_s(m_Z)$ from SUSY Grand Unification} \label{sec:RUN}

Although Eq.~\eqref{eq:ANG.70} is very effective in providing a bridge
between the experimentally measured quantity $s^2$ and the $\MS$ parameter
$\hat s^2$, a special care is required when leaving the SM domain to study
the effects of New Physics. If this equation is used in the most
straightforward and trivial way, allowing \emph{all} particles of the model
to participate to the vector boson self-energies $\hat\Pi_i$ as it seems
natural, then a lot of new contributions appear and the numerical
estimate~\eqref{eq:ANG.350b} is no longer correct. On the other hand, it is
clear that, as long as one takes care to artificially \emph{not} include new
particles into Eq.~\eqref{eq:ANG.70} and to consider only SM contributions,
all the results found in the previous chapter remain valid. This apparent
ambiguity suggests us to to replace the $\MS$ parameter $\hat s$ with two
better defined quantities sharing its properties: $\hat s_\SM^2$, which
lives within the SM and and does not depend on New Physics, and $\hat
s_\ND^2$, which also account for all new particles present in the model.
However, having now at our disposal two different $\MS$ angles, one may
wonder which one should be used when writing down the Renormalization Group
equations, and how the Grand Unification mass scale $m_\GUT$ is affected by
this choice.

In order to investigate the running of the $SU(3)$, $SU(2)$ and $U(1)$ gauge
coupling constants, it is essential as a first step to implement a unified
method for calculating their initial values. In the literature, the $\MS$
renormalization scheme is usually used. Since in this prescription heavy
particles do not decouple, it happens that even at low $\mu$ the value of
the coupling constants depends on the (unknown) contents of the theory at
high energies. To avoid this evident inconvenience Marciano and Rosner
introduced the so-called ``$\MS$ with heavy particles decoupling
procedure'': when performing calculations at some value of $\mu$ the
contribution of particles with masses larger than $\mu$ must be
neglected~\cite{Marciano90}. In this approach, low-energy data and particle
spectrum are all what is needed to determine initial values for coupling
constants. This procedure is very reasonable: using $\MS$ without decoupling
in GUT models will lead to the inconvenient result that all gauge couplings
remain equal even \emph{below} $m_\GUT$, as far as they unify above $m_\GUT$
into unique $\alpha_5$ value. This is why evolving below $m_\GUT$ it is
desirable to decouple heavy degrees of freedom.

Going down in energy, we cross thresholds associated with superpartners and
finally reach the domain $\mu \sim m_Z$. Instead of this theoretically
preferable up-down running, in practice down-up running is performed because
what are experimentally measured are the values of gauge coupling constants
at low energies, $\mu \sim m_Z$. Starting from this domain one has two
different options: to use $\hat s_\SM^2$, where SUSY partners are decoupled,
or $\hat s_\ND^2$, where they are taken into account and not decoupled. The
same ambiguity take place for electromagnetic coupling $\albar$:
$\hat\alpha_\SM(m_Z)$ or $\hat\alpha_\ND(m_Z)$. These two procedures lead to
different initial values of coupling constants $\hat g_1$ and $\hat g_2$ and
a natural question arise: how it is possible to get one and the same value
of $m_\GUT$ (where $\hat g_1$ and $\hat g_2$ become equal) starting from
different sets of initial values.

We will provide an answer to this question in Sec.~\ref{sec:RUN.10}, showing
how calculations based on $\hat s_\SM^2$ and $\hat s_\ND^2$ produce just the
same numerical value for the Grand Unification mass $m_\GUT$. As an
application of the techniques described there, in Sec.~\ref{sec:RUN.20} we
will quickly discuss what is the value of $\hat\alpha_s(m_Z)$ which can be
predicted from the demand of SUSY Grand Unification.

\section{RGE in the $\MS$ scheme and the concept of threshold}
\label{sec:RUN.10}

One of the most remarkable properties of the $\MS$ renormalization
prescription is that the Renormalization Group equations for the gauge
coupling constants do not contain explicitly the mass scale $\mu$. As a
consequence, in this scheme the RGE can be easily written and solved, at
least within perturbation theory, and this is why it is very well suited for
the study of Grand Unified theories.

On the other hand, it is well known that $\MS$ quantities have no simple
interpretations in terms of physical quantities, so special care is required
when setting the \emph{initial value} (i.e., the value at some fixed scale
$\mu_0$) of $\MS$ parameters which will enter RGE. In particular, as already
stated at the beginning of this chapter, $\MS$ quantities get radiative
corrections from \emph{any} particle in the model, no matter how light or
heavy it is, and so heavy boson and higgs multiplets, with masses $\sim
10^{16}\GeV$, which are usually present in GUT models should be taken into
account even when considering the low-scale ($\sim m_Z$) behavior of the
theory.

Although not theoretically unacceptable (we remind that $\MS$ quantities
have no physical meaning), this \emph{non-decoupling} property is clearly
unpractical, since it forces us to set different low-scale initial
conditions for models which differ \emph{only} in the high-energy particle
contents and have the same low-energy limit (the $SU(3) \times SU(2) \times
U(1)$ gauge group of the Standard Model and its particle content). Clearly,
it would be much more attractive to split the study of GUT into two
independent parts, namely a model-independent relation between physical
observables and some variant of $\MS$ quantities (taking into consideration
only the low-energy spectrum, for example the Standard Model or the MSSM)
and a model-dependent way to run these quantities up to the GUT scale.

A way to realize this project, which has been widely used in literature for
the accurate study of Grand Unified theories, make use of the concept of
\emph{threshold}. The purpose of the first part of this chapter is to review
in an extensive but simple way how this idea arises.

\subsection{Effects of new physics on the initial values}

Let's start with the SM. In terms of the physical parameters $G_F$, $\albar$
and $s^2$, the $\MS$ quantities $\hat s^2$ and $\hat\alpha^2$ at the scale
$\mu = \mu_0 = m_Z$ can be written (at one-loop level) as:
\begin{align}
    \begin{split}
	\hat s^2 = s^2 & + \frac{c^2 s^2}{c^2 - s^2} A, \\
	A & = \LT. 2 \frac{s}{c} \hat\Pi_\GZ(0) + \hat\Pi_\gamma(m_Z^2)
          - \hat\Pi_Z(m_Z^2) + \hat\Pi_W(0) + \altw
	  + \hat D \RT|_{\mu = \mu_0};
    \end{split} \\
    \begin{split} \label{eq:RUN.30b}
	\hat\alpha = \albar & + \albar B, \\
	B & = \LT. 2 \frac{s}{c} \hat\Pi_\GZ(0) + \hat\Pi_\gamma(m_Z^2) 
	  + \altw \RT|_{\mu = \mu_0}.
    \end{split}
\end{align}

Now, let's introduce some set of new particles, all with almost degenerate
masses $m$, and let's assume that the new model obtained is still
renormalizable. Due to their definitions, the value of $\albar$ and $s^2$ is
clearly unaffected by the introduction of the new particles (they are
defined directly from experimentally measured quantities), but the $\MS$
parameters $\hat\alpha$ and $\hat s^2$ will get new contributions from
self-energies, vertices and boxes coming from new physics. If we denote with
$\hat s_\SM^2$ and $\hat\alpha_\SM$ their old value, and with $\hat s_\ND^2$
and $\hat\alpha_\ND$ the new one, we can write:
\begin{align}
    \hat s_\ND^2 & = \hat s_\SM^2 + \delta \hat s^2, &
    \delta \hat s^2 & = \frac{c^2 s^2}{c^2 - s^2} \delta A; \\
    \hat\alpha_\ND & = \hat\alpha_\SM + \delta \alpha, &
    \delta \hat\alpha & = \albar \delta B;
\end{align}
where:
\begin{align} 
    \begin{split} \label{eq:RUN.50a}
      \delta A & = \LT. 2 \frac{s}{c} \hat\Pi_\GZ(0) + \hat\Pi_\gamma(m_Z^2)
        - \hat\Pi_Z(m_Z^2) + \hat\Pi_W(0) + \altwx
	+ \hat D \RT|_{\mu = \mu_0}^{\text{New Physics}} \\
      & = 2 \frac{s}{c} \delta\hat\Pi_\GZ(0) + \delta \hat\Sigma'_\gamma(0) 
        - \delta\hat\Pi_Z(m_Z^2) + \delta\hat\Pi_W(0) + \delta\hat D;
      \end{split} \\
    \label{eq:RUN.50b} \delta B & = \delta \hat\Sigma'_\gamma(0)
      + 2 \frac{s}{c} \delta\hat\Pi_\GZ(0);
\end{align}

Since we are interested in application to GUT models, it is more convenient
to deal with $\hat\alpha_1$ and $\hat\alpha_2$, rather than $\hat\alpha$ and
$\hat s^2$. We have:
\begin{align}
    \hat\alpha_1 & = \frac{5}{3} \frac{\hat\alpha}{\hat c^2}; 
    & \delta \hat\alpha_1 & = \hat\alpha_1 \LT( \frac{\delta \hat\alpha}
      {\hat\alpha} - \frac{\delta \hat c^2}{\hat c^2} \RT); \\
    \hat\alpha_2 & = \frac{\hat\alpha}{\hat s^2}; 
    & \delta \hat\alpha_2 & = \hat\alpha_2 \LT( \frac{\delta \hat\alpha}
      {\hat\alpha} - \frac{\delta \hat s^2}{\hat s^2} \RT).
\end{align}
After some simple calculations, one finds:
\begin{align}
    \label{eq:RUN.100a} \frac{\delta \hat\alpha_1}{\hat\alpha_1} 
    & = \frac{1}{c^2-s^2} \LT\{ c^2 \LT( \delta \hat\Sigma'_\gamma(0) 
      + 2 \frac{s}{c} \delta \hat\Pi_\GZ(0) \RT) 
      + s^2 \LT( \delta\hat\Pi_W(0) - \delta\hat\Pi_Z(m_Z^2) 
        + \delta \hat D \RT) \RT\}; \\
    \label{eq:RUN.100b} \frac{\delta \hat\alpha_2}{\hat\alpha_2} 
    & = - \frac{1}{c^2-s^2} \LT\{ s^2 \LT( \delta \hat\Sigma'_\gamma(0) 
      + 2 \frac{s}{c} \delta \hat\Pi_\GZ(0) \RT) 
      + c^2 \LT( \delta\hat\Pi_W(0) - \delta\hat\Pi_Z(m_Z^2) 
        + \delta \hat D \RT) \RT\}.
\end{align}
Although these formulae has been derived with the assumption that $\mu =
\mu_0 \equiv m_Z$ (cfr.~Eqs.~\eqref{eq:RUN.50a} and~\eqref{eq:RUN.50b}), it is
straightforward to note that the condition $\mu_0 = m_Z$ has never been used
for deriving them, and so they are correct for \emph{any} possible value of
$\mu_0$; therefore, we can conclude they provide a relation between
$\hat\alpha_i^\SM$ and $\hat\alpha_i^\ND$ for \emph{any} value of $\mu$. As
a consequence, there is no need for writing down RGE for $\hat\alpha_i^\ND$,
since we already know their solution; the only thing which we should do is
to write down explicitly the dependence of $\hat\alpha_i^\ND$ on the mass
parameter $\mu$. Although this would be the simplest approach, since the
goal of this article is to give a simple presentation of how the concept of
threshold arises, in the following sections we will forget about this and
keep on assuming that $\mu = \mu_0$ in Eqs.~\eqref{eq:RUN.100a}
and~\eqref{eq:RUN.100b}, so that their meaning is simply to set the
\emph{shift} in the \emph{initial value} of $\hat\alpha_i^\ND$; instead, the
running of the coupling constants will be derived from the RGE.

\subsection{Writing and solving the RGE} \label{sec:RUN.10.20}

Now let's consider in more detail which kind of contribution the new set of
particles gives to the initial value of the coupling constants and to their
running. It is well known that any unrenormalized self-energy, vertex or box
diagram can be written as a sum of three parts:

\begin{enumerate}[(1)]
  \item\label{misc:RUN.a} a purely divergent contribution, proportional to
    $\frac{1}{\epsilon}$ and independent of $\mu$, coming from dimensional
    regularization ($\delta \hat Z$);
  \item\label{misc:RUN.b} a finite $\mu$-dependent part, also arising from
    dimensional regularization ($\delta \tilde Z$);
  \item\label{misc:RUN.c} a finite and $\mu$-independent part.
\end{enumerate}

Only the third term is physically relevant and contributes to physical
observables; the second one has no physical meaning, but is present in the
definition of $\MS$ quantities and contributes to $\delta \hat\alpha_i$. On
the other hand, the first term ($\delta \hat Z$) is directly involved when
writing the RG equations. In general, $\delta \hat Z$ and $\delta \tilde Z$
are not independent, but (at one-loop level) can be written as:
\begin{align}
    \label{eq:RUN.190a} \delta \hat Z & = 4\pi \hat\alpha \delta C
      \frac{1}{\epsilon}, \\
    \label{eq:RUN.190b} \delta \tilde Z & = -4\pi \hat\alpha \delta C
      \ln\frac{m^2}{\mu_0^2}.
\end{align}
where $\delta C$ is a constant which depends on the particular Feynman
diagram.

Since in Eqs.~\eqref{eq:RUN.100a} and~\eqref{eq:RUN.100b} the contributions
of the new particles enter through the \emph{$\MS$ renormalized} vertex-box
and self-energies, the term~\enref{misc:RUN.a} is absent by definition. On
the other hand, the term~\enref{misc:RUN.c} directly affects physical
observables, so a large value (larger for example than the experimental
error of some measured quantity) will result either in the discovery of new
physics or in the rejection of the model; therefore, we will assume that
this term is small, and neglect it.

Now, let $2 \delta \tilde Z_i$ be the sum of the $\mu$-dependent
part~\enref{misc:RUN.b} coming from all the renormalized self-energies and
vertex-box contributions appearing in Eqs.~\eqref{eq:RUN.100a}
and~\eqref{eq:RUN.100b}, and let $2\delta \hat Z_i$ be the sum of all the
divergent part~\enref{misc:RUN.a} of the corresponding unrenormalized
diagrams. Using Eqs.~\eqref{eq:RUN.190a} and~\eqref{eq:RUN.190b}, we have:
\begin{align}
    \label{eq:RUN.220a} \frac{\delta \hat\alpha_i}{\hat\alpha_i} & \equiv
      2 \delta \tilde Z_i, \\
    \label{eq:RUN.220b} \delta \tilde Z_i & \equiv -4\pi \hat\alpha_i \delta C_i
      \ln\frac{m^2}{\mu_0^2}, \\
    \label{eq:RUN.220c} \delta \hat Z_i & \equiv 4\pi \hat\alpha_i \delta C_i
      \frac{1}{\epsilon}.
\end{align}
Substituting Eq.~\eqref{eq:RUN.220b} into Eq.~\eqref{eq:RUN.220a}, we have
that the shift in the starting point induced by the new particles in the
coupling constants is:
\begin{equation} \begin{split} \label{eq:RUN.250}
    \delta \LT( \frac{1}{\hat\alpha_i} \RT) & = - \frac{1}{\hat\alpha_i} 
      \LT( \frac{\delta \hat\alpha_i}{\hat\alpha_i} \RT) =
      - \frac{2}{\hat\alpha_i} \delta \tilde Z_i = 
      - \frac{2}{\hat\alpha_i} 4 \pi \hat\alpha_i
        \delta C_i \LT( - \ln\frac{m^2}{\mu_0^2} \RT) \\
    & = 8\pi \delta C_i \ln\frac{m^2}{\mu_0^2}.
\end{split} \end{equation}

On the other hand, due to the presence of the $\delta \hat Z_i$
contribution, the RG equation changes, and the running of the $\MS$
quantities $\hat\alpha_i$ is consequently modified. At one-loop:
\begin{equation}
    \mu^2 \frac{d \hat\alpha_i}{d \mu^2} = -\hat\alpha_i \epsilon
    - 2 \frac{\mu^2}{Z_i} \frac{d \hat Z_i}{d \mu^2} \hat\alpha_i.
\end{equation}
The solution of this equation, both in the case when only SM is considered
and when new particles are included, is:
\begin{align}
    \label{eq:RUN.280a} \frac{1}{\hat\alpha_i^\SM(\mu^2)} & =
      8\pi C_i^\SM \ln\frac{K_i^\SM}{\mu^2}, \\
    \label{eq:RUN.280b} \frac{1}{\hat\alpha_i^\ND(\mu^2)} & =
      8\pi C_i^\ND \ln\frac{K_i^\ND}{\mu^2}.
\end{align}
where $K_i^\SM$ and $K_i^\ND$ are integration constants, and $C_i$ have the
same meaning as $\delta C_i$ appearing in Eq.~\eqref{eq:RUN.220c}
and~\eqref{eq:RUN.220b} (and, in particular, $\delta C_i = C_i^\ND -
C_i^\SM$). Subtracting Eq.~\eqref{eq:RUN.280a} from Eq.~\eqref{eq:RUN.280b}
and after some algebra, we find:
\begin{gather}
    \label{eq:RUN.300a} \frac{1}{\hat\alpha_i^\ND(\mu^2)} -
      \frac{1}{\hat\alpha_i^\SM(\mu^2)} = 
      8\pi \delta C_i \ln\frac{K_i}{\mu^2}, \\
    K_i = K_i^\SM \LT( \frac{K_i^\ND}{K_i^\SM} 
      \RT)^{C_i^\SM / \delta C_i}.
\end{gather}

The constants $K_i$ can be determined by remembering that, for $\mu=\mu_0$,
the value of $1/\hat\alpha_i^\ND(\mu_0) - 1/\hat\alpha_i^\SM(\mu_0)$ is
given by Eq.~\eqref{eq:RUN.250}. The solution is then:
\begin{equation}
    \frac{1}{\hat\alpha_i^\ND(\mu)} = \frac{1}{\hat\alpha_i^\SM(\mu)} +
      8 \pi \delta C_i \ln \frac{m^2}{\mu^2}.
\end{equation}

\begin{figure}[!t] \centering
    \includegraphics[width=0.9\textwidth]{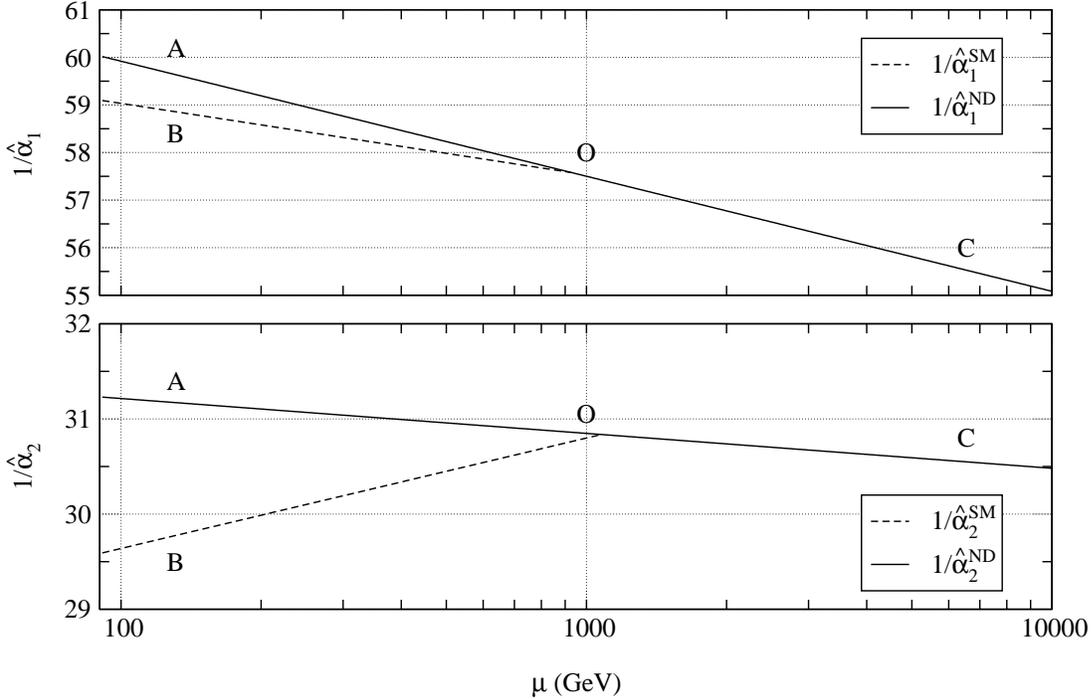}
    \caption[Running of $\hat{\alpha}_i$ in the \emph{decoupling} and
    \emph{non-decoupling} approaches]{
      Running of $\hat\alpha_1$ and $\hat\alpha_2$ coupling constants, at
      one-loop, as a function of the mass scale parameter $\mu$, both in the
      \emph{non-decoupling} (paths $\mathsc{\overline{AOC}}$) and
      \emph{decoupling} (paths $\mathsc{\overline{BOC}}$) approaches. As SM
      model extension, MSSM with $m_\SUSY = 1000\GeV$ is used. The small
      shift of the $\hat\alpha_i^\SM$~-~$\hat\alpha_i^\ND$ crossing point
      with respect to $m_\SUSY$ is due to the term~\enref{misc:RUN.c} which
      was neglected in equations.}
    \label{fig:RUN.10}
\end{figure}

This result, although mathematically trivial, is nevertheless rather
interesting, since it put into clear light the effects of the introduction
of new particles on the running of the $\MS$ coupling constants. First of
all, as expected, the coefficients of the logarithmic term are changed, so
the running occurs with a different slope. But it can be seen that
absolutely nothing happens when crossing the new particles mass threshold:
the running of $\hat \alpha_i$ is always the same, both below and above $m$.
Instead, an extra contribution to the \emph{initial value}
$\hat\alpha_i(\mu_0)$ should be accounted for, due to the non-decoupling
property of $\MS$ scheme. This behavior corresponds to paths
$\mathsc{\overline{AOC}}$ in Fig.~\ref{fig:RUN.10}. It is straightforward to
see that, for $\mu = m$, we have $\alpha_i^\ND = \alpha_i^\SM$: here, as
expected, the contribution of the new particles to the value of
$\hat\alpha_i$ vanishes.

\subsection{Running in the decoupling approach} \label{sec:RUN.10.30}

We started this chapter stating that we would have shown how it could be
possible to split the study of GUT models into two independent parts, namely
a model-independent relation between physical quantities and some variant of
$\MS$ quantities, and a model-dependent way to run these quantities up to
the GUT scale. We have seen in the previous section that new particles
contributes both to the setting of the initial point (the relation between
physical quantities and $\MS$ coupling constants) and to the running of
$\hat\alpha_i$, but we also noted that $\hat\alpha_i^\SM$ and
$\hat\alpha_i^\ND$ become equals when $\mu = m$. This provides us with the
solution to our problem: up to small $\mu$-independent contributions (see
beginning of Sec.~\ref{sec:RUN.10.20}, term~\enref{misc:RUN.c} in the list),
it is possible to use \emph{ordinary} (i.e.\ accounting only for the
contribution of SM particles) RGE, without introducing any extra initial
shift, to run coupling constants up to the scale $\mu = m$, and then to run
them up to the GUT scale by means of the \emph{full} (i.e.\ accounting also
for the new particles) RG equations. This behavior corresponds to paths
$\mathsc{\overline{BOC}}$ in Fig.~\ref{fig:RUN.10}.

This approach has also a very simple physical interpretation: it corresponds
to working with a variant of $\MS$ scheme, in which the contribution of
particles heavier than the considered mass scale $\mu$ is decoupled from the
very beginning. In other words, Eqs.~\eqref{eq:RUN.190a}
and~\eqref{eq:RUN.190b} are replaced by:
\begin{align}
    \delta \hat Z & = 4\pi \hat \alpha C \LT[ \frac{1}{\epsilon}
      - \theta(m - \mu) \ln\frac{m^2}{\mu^2} \RT], \\
    \delta \tilde Z & = -4\pi \hat \alpha \delta C \theta(\mu - m)
      \ln\frac{m^2}{\mu^2}. 
\end{align}

Solving the RGE, it is easy to see that, for $\mu < m$, the one-loop
contribution of the new particles is now canceled by the logarithmic
$\mu$-dependent term in $\delta \hat Z_i$, so that the total result is now
zero; for $\mu > m$, nothing is changed and the running of the coupling
constants simply reflects the usual behavior.

\section{The numerical value of $\hat\alpha_s(m_Z)$ from SUSY GUT}
\label{sec:RUN.20}

As an application of the results found in the previous section, we will now
present a short overview of the prediction on $\hat\alpha_s(m_Z)$ from the
demand of SUSY Grand Unification. Since our purpose is simplicity, rather
than completeness, we will consider in some detail only effects related to
low energy physics, and in particular we will analyze the dependence of
$\hat\alpha_s(m_Z)$ on the Standard Model parameters $\hat\alpha_\SM$, $\hat
s_\SM^2$, and $\hat m_t^\SM$. Instead, we will not consider the
contributions of heavy vector bosons and higgs multiplets which usually
occur close to the unification scale in any reasonable GUT model; effects of
unknown non-renormalizable operators which may be induced by (possible) new
physics above $m_\GUT$ will be neglected as well. Also, we will assume the
complete degeneracy of all SUSY particles at a common mass scale $m_\SUSY$;
therefore, in our approach (which is based on Ref.~\cite{Barger93}) the
superpartners enter the RGE and affect $\hat\alpha_s(m_Z)$ only through the
two parameters $m_\SUSY$ and $\tan\beta$. Although these assumption are very
restrictive and physically unrealistic, they will allow us to keep our
discussion at a very simple level; for a detailed analysis of all possible
contributions which may affect the prediction of $\hat\alpha_s(m_Z)$, we
address the reader to the literature (see for example
Refs.~\cite{Barger93,Langacker93,Barbieri95}).

In our approach, we start from the value of the gauge coupling constants
$\hat\alpha_1^\SM$ and $\hat\alpha_2^\SM$, the Yukawa couplings
$\hat\lambda_t^\SM$, $\hat\lambda_b^\SM$ and $\hat\lambda_\tau^\SM$, and the
quartic higgs coupling $\hat\lambda^\SM$, at the $\mu=m_Z$ scale; then we
run them up to $\mu=m_\SUSY$ using the SM two-loop RGE (Eqs.~(35-43),
(A17-A20) and~(A25-A26) of Ref.~\cite{Barger93}), evaluate the corresponding
ND quantities, and finally run them up to $\mu=m_\GUT$ using the MSSM
two-loop RGE (Eqs.~(1-4) and~(A5-A11) of Ref.~\cite{Barger93}). The mass
scale $m_\GUT$ is defined be the condition $\hat\alpha_1^\ND(m_\GUT) =
\hat\alpha_2^\ND(m_\GUT)$; imposing the requirement
$\hat\alpha_3^\ND(m_\GUT) = \hat\alpha_{1,2}^\ND(m_\GUT)$ and running down
all quantities back to $\mu=m_Z$, we obtain a prediction for
$\hat\alpha_3^\SM(m_Z)$. When crossing the SUSY threshold, both upwards and
downwards, the following conditions between SM and ND quantities are imposed
(see Eqs.~(34), (44) and~(45-49) from Ref.~\cite{Barger93}):
\begin{align}
    \hat\alpha_i^\SM(m_\SUSY) & = \hat\alpha_i^\ND(m_\SUSY), \\
    \hat m_j^\SM(m_\SUSY) & = \hat m_j^\ND(m_\SUSY), \\
    \hat\lambda^\SM(m_\SUSY) & = \pi \LT( \frac{3}{5}
      \hat\alpha_1^\ND(m_\SUSY) + \hat\alpha_2^\ND(m_\SUSY) \RT) 
      \cos^2 2\beta.
\end{align}
Let us add that, for small values of $\tan\beta$, the contributions of
$\hat\lambda_b$ and $\hat\lambda_\tau$ to the running of $\hat\alpha_i$ are
usually very small and could safely be neglected; however, since they are
already included in formulas given in Ref.~\cite{Barger93}, we decided to
take them into account as well.

While we are using explicitly the decoupling procedure described in
Sec.~\ref{sec:RUN.10.30} for taking SUSY particles into account, our
approach is completely non-decoupling for what concerns the top quark: for
example, the quantity $\hat s^2$ discussed in Chapter~\ref{sec:ANG} depends
strongly on $m_t$. On the other hand, the $\MS$ parameter
$\hat\alpha_s(m_Z)$ used in literature accounts only for quarks lighter than
$m_Z$. To allow comparison of our numerical calculations with experimental
data, we need to subtract $t$ contributions from $\hat\alpha_3^\SM(m_Z)$,
and this can be done by means of the following relation:
\begin{equation}
    \hat\alpha_s(m_Z) = \frac{\hat\alpha_3^\SM(m_Z)}{1 -
      \frac{1}{3\pi} \hat\alpha_3^\SM(m_Z) \ln \frac{m_t}{m_Z}},
\end{equation}
which is valid at the one-loop level.

\subsection{The SUSY parameters $\tan\beta$ and $m_\SUSY$}

\begin{figure}[!t] \centering
    \includegraphics[width=0.9\textwidth]{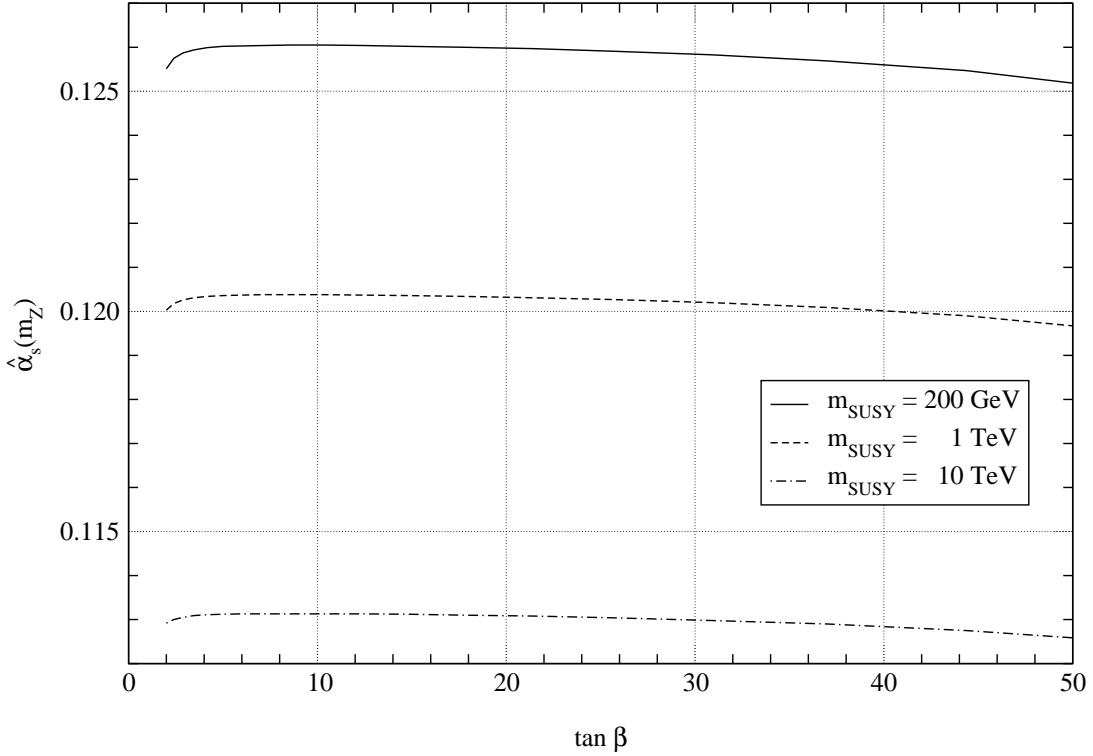}
    \caption[Plot of $\hat\alpha_s$ as a function of $\tan\beta$]{
      Dependence of $\hat\alpha_s$ on $\tan\beta$, for $\albar = 1/128.878$,
      $s^2 = 0.23116$, $m_Z = 91.1867\GeV$, $m_H = 120\GeV$, $m_t =
      173.8\GeV$.}
    \label{fig:RUN.20}
\end{figure}

\begin{figure}[!t] \centering
    \includegraphics[width=0.9\textwidth]{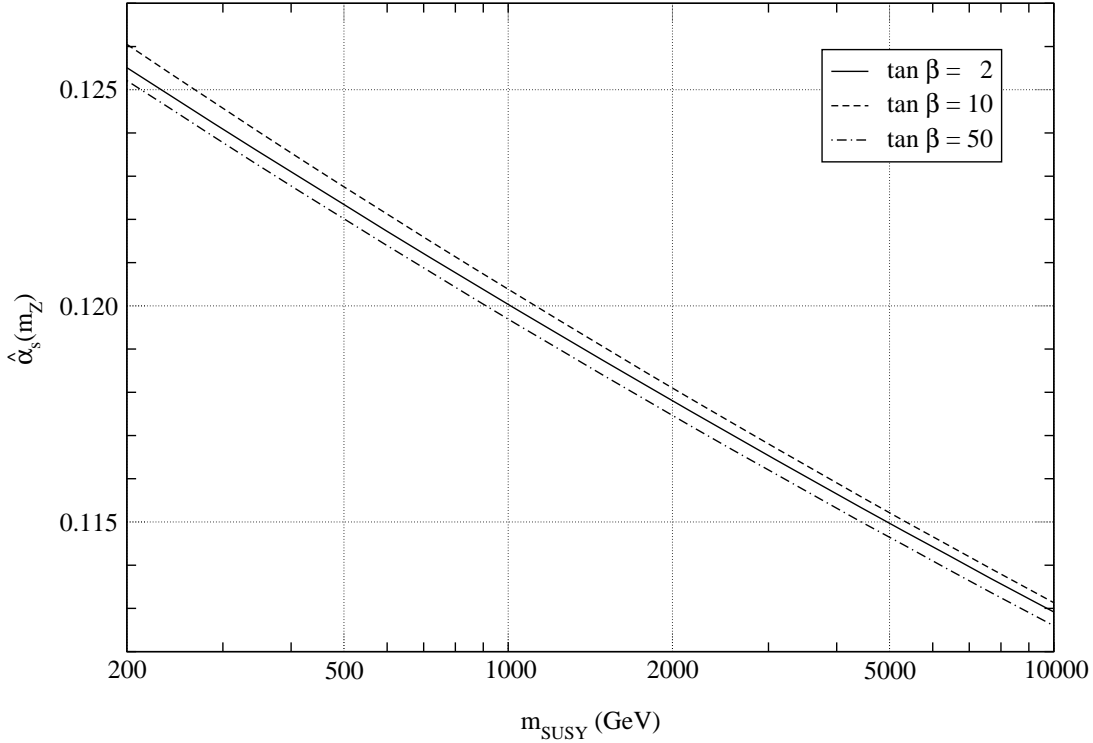}
    \caption[Plot of $\hat\alpha_s$ as a function of $m_\SUSY$]{
      Dependence of $\hat\alpha_s$ on $m_\SUSY$, for $\albar = 1/128.878$,
      $s^2 = 0.23116$, $m_Z = 91.1867\GeV$, $m_H = 120\GeV$, $m_t =
      173.8\GeV$.}
    \label{fig:RUN.30}
\end{figure}

We are now ready to investigate the dependence of $\hat\alpha_s(m_Z)$ on the
only two SUSY parameters, $m_\SUSY$ and $\tan\beta$, which are relevant in
our approximation. From Fig.~\ref{fig:RUN.20}, we immediately see that
varying $\tan\beta$ do not introduce any relevant effect, and the value of
$\hat\alpha_s$ remains almost the same for a large window of $\tan\beta$
values.

The situation is rather different if one considers the dependence of
$\hat\alpha_s$ on $m_\SUSY$, which is shown in Fig.~\ref{fig:RUN.30}. One
can see that, when changing $m_\SUSY$ from $200\GeV$ to $10\TeV$, the value
of $\hat\alpha_s$ varies of about $0.012$, which is much more than its
experimental error. Since the mass scale at which supersymmetry occurs is
unknown, this shift should be regarded as a theoretical uncertainty in the
prediction of $\hat\alpha_s$. Therefore, we must conclude that, up to a
concrete discovery of superpartners, it will not be possible to reduce the
error on the \emph{predicted} value of $\hat\alpha_s$ to less than $\sim
10\%$. However, we should note that SUSY Grand Unification is not in
disagreement with the experimental value $0.119(2)$, and in particular
$m_\SUSY \sim 1\TeV$ reproduce nicely this number (although it is well known
that the effects of \emph{mass splitting} among superpartners can change
drastically this picture, and so $m_\SUSY$ should be regarded only as an
\emph{effective} parameter without any physical meaning).

Fig.~\ref{fig:RUN.30} also shows us that the dependence of $\hat\alpha_s$ on
$m_\SUSY$ is almost logarithmic. From this plot we can derive an useful
approximate relation:
\begin{equation}
    \hat\alpha_s(m_Z) = 
    \begin{cases}
	0.12014 - 0.0073934 \, \log_{10} \LT( m_\SUSY / \mathrm{TeV} \RT)
	  & \text{for $\tan\beta =  2$}, \\
	0.12051 - 0.0075748 \, \log_{10} \LT( m_\SUSY / \mathrm{TeV} \RT)
	  & \text{for $\tan\beta = 10$}, \\
	0.11981 - 0.0074000 \, \log_{10} \LT( m_\SUSY / \mathrm{TeV} \RT)
	  & \text{for $\tan\beta = 50$}.
    \end{cases}
\end{equation}

\subsection{The SM parameters $\hat\alpha$, $\hat s^2$ and $\hat m_t$}

\begin{figure}[!t] \centering
    \includegraphics[width=0.9\textwidth]{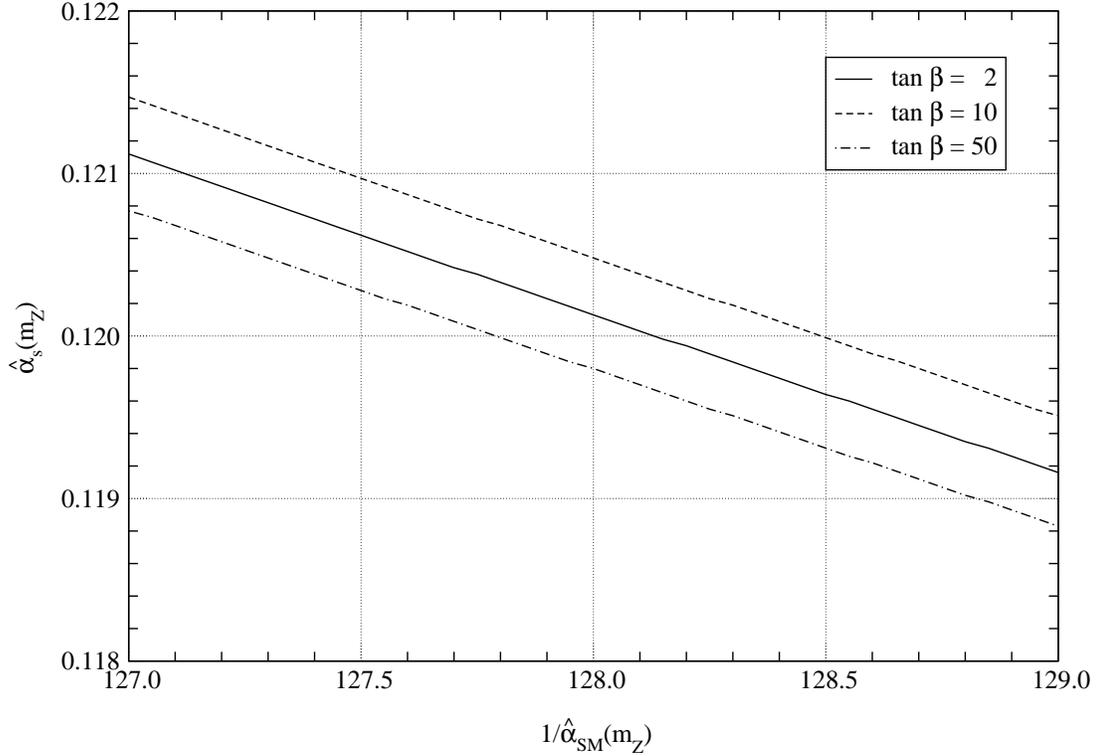}
    \caption[Plot of $\hat{\alpha}_s$ as a function of $1/\hat{\alpha}$]{
      Dependence of $\hat{\alpha}_s$ on $1/\hat{\alpha}$, for $m_\SUSY =
      1\TeV$, $\hat{s}^2(m_Z) = 0.23147$, $\hat{m}_t(m_Z) = 175\GeV$.}
    \label{fig:RUN.40}
\end{figure}

\begin{figure}[!t] \centering
    \includegraphics[width=0.9\textwidth]{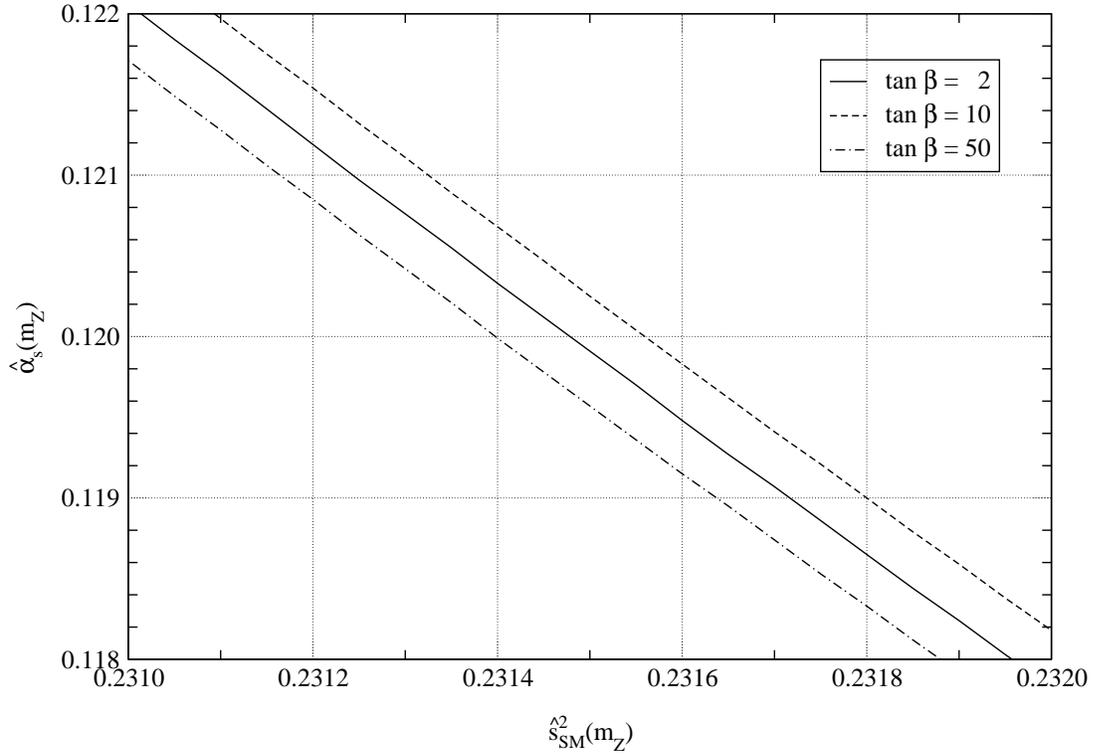}
    \caption[Plot of $\hat\alpha_s$ as a function of $\hat s^2$]{
      Dependence of $\hat\alpha_s$ on $\hat s^2$, for $m_\SUSY = 1\TeV$,
      $\hat\alpha(m_Z) = 1/128.1$, $\hat m_t(m_Z) = 175\GeV$.}
    \label{fig:RUN.50}
\end{figure}

\begin{figure}[!t] \centering
    \includegraphics[width=0.9\textwidth]{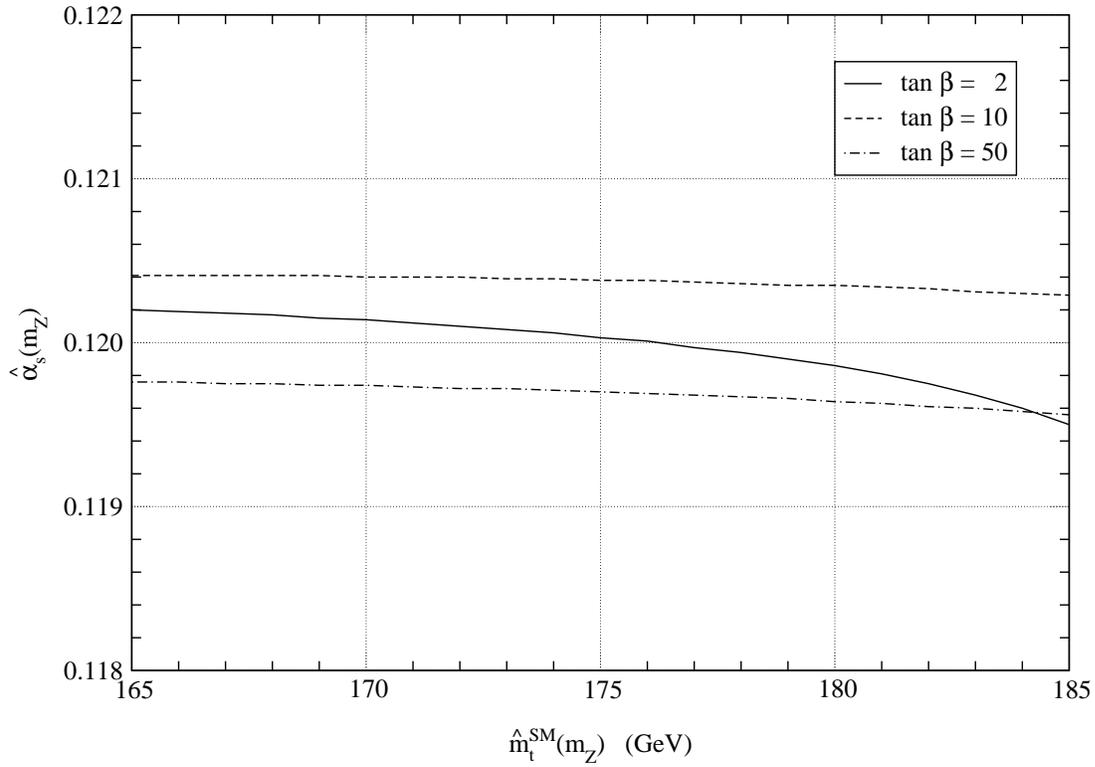}
    \caption[Plot of $\hat\alpha_s$ as a function of $\hat m_t$]{
      Dependence of $\hat\alpha_s$ on $\hat m_t$, for $m_\SUSY = 1\TeV$,
      $\hat\alpha(m_Z) = 1/128.1$, $\hat s^2(m_Z) = 0.23147$.}
    \label{fig:RUN.60}
\end{figure}

As already stated, before dealing with the running of $\MS$ coupling
constants up to the GUT scale we need first of all to set their initial
values. Since our original project of splitting the study of SUSY GUT into
two independent parts, one dealing with ``low-energy'' SM quantities and one
accounting for ``high-energy'' MSSM effects, has been successfully realized
by the introduction of one SUSY threshold, we can now forget about
superpartners when setting the starting point. As a consequence, all
formulas derived in Chapter~\ref{sec:ANG} for the case of SM are now
perfectly adequate for our purposes, and we can use them.

As reference values for the relevant physical observables, we use $\albar =
1/128.878$, $s^2 = 0.23116$, $m_Z = 91.1867\GeV$, $m_H = 120\GeV$ and $m_t =
173.8\GeV$. From Eqs.~\eqref{eq:ANG.10p} and~\eqref{eq:ANG.350b} given in
the previous chapter, we get $\hat s^2(m_Z) = 0.23147$, while the value of
$\hat\alpha$ at the one-loop order can be derived by Eq.~\eqref{eq:RUN.30b}
leading to $\hat\alpha(m_Z) = 1/128.1$. Concerning $\hat m_t$ the situation
is slightly more difficult: Eq.~(17) from Ref.~\cite{Barger93} is valid at
the scale $\mu = m_t$, and to get an estimation for $\hat m_t(m_Z)$ we need
to run it down to the $Z$ scale. However, doing this one finds that the big
splitting between $m_t$ and $\hat m_t(m_t)$ which occurs at the top scale
due to the large value of $\hat\alpha_s$ is completely compensated by the
running effects, and for $m_t$ in the range $165 \div 185\GeV$ it follows
that $\hat m_t(m_Z)$ is about $1\GeV$ larger than $m_t$. This complete the
analysis of the initial conditions.

Now let's consider the prediction of $\hat\alpha_s$ coming from the demand
of gauge coupling unification at the GUT scale. In Figs.~\ref{fig:RUN.40},
\ref{fig:RUN.50} and~\ref{fig:RUN.60} we show the dependence of
$\hat\alpha_s$ on the $\MS$ quantities $\hat\alpha$, $\hat s^2$ and $\hat
m_t$, respectively. One can see immediately that, unlike the case of the
completely unbounded parameter $m_\SUSY$, here the small theoretical
uncertainty on these quantities induces a correspondingly small error on
$\hat\alpha_s$. The largest effect is due to $\hat s^2$, and in particular
from Eq.~\eqref{eq:ANG.350b} and Fig.~\ref{fig:RUN.50} one can see that
inclusion of the leading two-loop contributions in the relation between
$\hat s^2$ and $s^2$ produce a sensible change in $\hat\alpha_s$ (about half
of the experimental error). On the contrary, from Fig.~\ref{fig:RUN.40} and
Eq.~(43) of Ref.~\cite{NOV94a} we see that contributions of order
$\alpha\alpha_s$ into $\hat\alpha$ have negligible consequences on
$\hat\alpha_s$, and this is why we used the one-loop
relation~\eqref{eq:RUN.30b} to set the initial value of $\hat\alpha$.
Finally, Fig.~\ref{fig:RUN.60} proves that changing $\hat m_t(m_Z)$ does not
induce any relevant effect on $\hat\alpha_s$.

Figs.~\ref{fig:RUN.40} and~\ref{fig:RUN.50} also show that $\hat\alpha_s$
depends almost linearly on $1 / \hat\alpha$ and $\hat s^2$; therefore, we
can derive the following approximate relations:
\begin{align}
    \hat\alpha_s(m_Z) & =
    \begin{cases}
	0.12003 - 0.00097812 \LT[ 1/\hat\alpha(m_Z) - 128.1 \RT]
	  & \text{for $\tan\beta =  2$}, \\
	0.12038 - 0.00098094 \LT[ 1/\hat\alpha(m_Z) - 128.1 \RT]
	  & \text{for $\tan\beta = 10$}, \\
	0.11971 - 0.00097153 \LT[ 1/\hat\alpha(m_Z) - 128.1 \RT]
	  & \text{for $\tan\beta = 50$};
    \end{cases} \\[2mm]
    \hat\alpha_s(m_Z) & = 
    \begin{cases}
	0.12006 - 4.2340 \LT[ \hat s^2(m_Z) - 0.23147 \RT]
	  & \text{for $\tan\beta =  2$}, \\
	0.12044 - 4.2306 \LT[ \hat s^2(m_Z) - 0.23147 \RT]
	  & \text{for $\tan\beta = 10$}, \\
	0.11973 - 4.2000 \LT[ \hat s^2(m_Z) - 0.23147 \RT]
	  & \text{for $\tan\beta = 50$}.
    \end{cases}
\end{align}

It is interesting to discuss explicitly how effects of varying the
\emph{physical observables} $\albar$, $s^2$ and $m_t$ on $\hat\alpha_s$ may
be derived from comparing Figs.~\ref{fig:RUN.40}, \ref{fig:RUN.50}
and~\ref{fig:RUN.60} with Figs.~\ref{fig:ANG.10} and~\ref{fig:ANG.20} from
the previous chapter. As an example, let us consider $m_t$. Changing it
induces shifts on the initial values of $\hat\alpha$, $\hat s^2$ and $\hat
m_t$, and these shifts will result in a different prediction for
$\hat\alpha_s$. However, from Fig.~\ref{fig:RUN.60} we immediately see that
the variation of $\hat m_t$ does not produce any sensible effect on
$\hat\alpha_s$, while since $\hat\alpha$ depends on $m_t$ only
logarithmically its variation itself is very small. Therefore, in our
approach the only relevant contribution of $m_t$ into $\hat\alpha_s$ happens
\emph{through} $\hat s^2$, and can easily be estimated from
Figs.~\ref{fig:ANG.20} and~\ref{fig:RUN.50}. An analogous situation holds
for the higgs mass $m_H$.

\section{Conclusions} \label{sec:RUN.30}

In this chapter we have discussed in detail how both the decoupling and the
non-decoupling approach to the running of coupling constants in the $\MS$
scheme produce the same numerical value of $m_\GUT$ despite of the different
initial conditions. The concept of threshold has been introduced in
Sec.~\ref{sec:RUN.10} and used in Sec.~\ref{sec:RUN.20}, where the
dependence of $\hat\alpha_s$ on the SM parameters $\hat\alpha$, $\hat s^2$
and $\hat m_t$ and the MSSM quantities $m_\SUSY$ and $\tan\beta$ has been
studied. Combining the results obtained in this chapter with analysis of
$\hat s^2$ carried out in Chapter~\ref{sec:ANG} it is straightforward to
evaluate the impact of the numerical value of the physical observables
$\albar$, $s^2$ and $m_t$ on the prediction of $\hat\alpha_s$ from the
demand of SUSY Grand Unification.

\clearemptydoublepage

\begin{fmffile}{Feynman/gauginos}

\chapter{Charginos nearly degenerate with the lightest neutralino}
\label{sec:GAU}

Despite of its success in predicting the correct value of
$\hat{\alpha}_s(m_Z)$ from the hypothesis of gauge coupling unification at
the GUT scale, the MSSM still lacks a direct confirmation from experimental
search. None of the many new particles expected in SUSY models have yet been
observed, and the high energy and luminosity reached at LEP~II and Tevatron
in the last years allow now to put stringent bounds on the masses of most of
the superpartners. However, to further constrain SUSY parameters, or
alternatively to determine them through the discovery of new particles, more
powerful accelerators are required, and they won't be available till the
construction of LHC~- i.e.\ many years.

To overcome this limitation of direct search experiments, one may try to
find \emph{indirect} evidence of the superpartners analyzing their impact on
precision measurements through radiative corrections. Since no deviation
from the predictions of the SM has so far been observed, nowadays precision
data can only provide further bounds on MSSM parameters. Unfortunately,
unlike other kind of New Physics (for example, an extra generation of
fermions, which we will discuss in the next chapter), SUSY particles have
the remarkable property to \emph{decouple} from low-energy physical
observables when their masses are pushed to a high energy scale. This means
that loop effects may be appreciable only if superpartners are light, but
this possibility is strongly constrained by the results of direct
observations. Therefore, one is led to answer the following question: is
there still something to learn about supersymmetry from precision
measurements \emph{after} having imposed all bounds required by direct
search?

In most of the cases, the answer is negative. Concerning sfermions,
radiative corrections could be large if the mass splitting between the
up-left and down-left superpartners were big enough, but this situation
never occurs in the MSSM (except for the third generations of squarks,
see~\cite{Gaidaenko98}). Concerning gauginos, radiative corrections are
\emph{always} small, at least in the region still allowed by LEP~II bounds.

However, there is still one class of situations where precision measurements
can be useful, and it occurs when direct search analysis, for some
particular reason, fails, leaving a hole in a domain which would otherwise
have been closed. This is just what happens in the gaugino sector when the
neutralino is the LSP and the lightest chargino is only $\sim 1\GeV$
heavier: in this case, the chargino is too short-lived to be identified as a
stable heavy particle, and its decay products are too soft to be
distinguished from the background, so it escapes detection and no bound
other than LEP~I can be applied.

In the present chapter, we will discuss in detail the impact on precision
measurements of a chargino almost degenerate in mass with the lightest
neutralino. In particular, we will show that the bound $m_\tdC \gtrsim
45\GeV$ which follows from direct searches can be improved up to $51\GeV$
for the higgsino-dominated case, and $56\GeV$ for the wino-dominated case,
by analyzing the precision data. Also, we will derive explicit formulas for
the oblique electroweak radiative corrections.

In Sec.~\ref{sec:GAU.10}, we will quickly review the present experimental
situation, showing the results found by the DELPHI
collaboration~\cite{DELPHI99}. Secs.~\ref{sec:GAU.20} and~\ref{sec:GAU.30}
are based on the published paper~\cite{Maltoni99b}. In Sec.~\ref{sec:GAU.40}
we will discuss the latest (preliminary) results presented by A.~Perrotta
(DELPHI \mbox{Collab.}) at the conference PASCOS99~\cite{Perrotta99},
together with an improved analysis of $\chi^2$ that we also presented
there~\cite{Maltoni99d}.

\section{Experimental bounds} \label{sec:GAU.10}

LEP~II is very effective in bounding from below the masses of charginos
which, if kinematically allowed, should be produced in a pair in $e^+ e^-
\to \chi^+ \chi^-$ annihilation. This process can be mediated by either a
gauge boson or a sneutrino, and the relevant Feynman diagrams fall into two
classes:
\begin{equation} \label{eq:GAU.5}
    \begin{gathered} \fmfframe(2,5)(0,5){ 
	  \begin{fmfgraph*}(45,20)
	      \fmfleft{i2,i1}
	      \fmfright{o2,o1}
	      \fmflabel{$e^+$}{i1}
	      \fmflabel{$e^-$}{i2}
	      \fmflabel{$\chi^+$}{o1}
	      \fmflabel{$\chi^-$}{o2}
	      \fmf{fermion}{i2,v1,i1}
	      \fmf{fermion}{o1,v2,o2}
	      \fmf{boson,label=$\gamma,,Z$}{v1,v2}
	      \fmfdot{v1,v2}
	  \end{fmfgraph*}} \\ \mathrm{(a)} 
    \end{gathered}
    \qquad\qquad
    \begin{gathered} \fmfframe(2,5)(0,5){
	  \begin{fmfgraph*}(45,20)
	      \fmfleft{i2,i1}
	      \fmfright{o2,o1}
	      \fmflabel{$e^+$}{i1}
	      \fmflabel{$e^-$}{i2}
	      \fmflabel{$\chi^+$}{o1}
	      \fmflabel{$\chi^-$}{o2}
	      \fmf{fermion}{o1,v1,i1}
	      \fmf{fermion}{i2,v2,o2}
	      \fmf{dashes,label=$\tilde{\nu}_e$}{v1,v2}
	      \fmfdot{v1,v2}
	  \end{fmfgraph*}} \\ \mathrm{(b)}
    \end{gathered}
\end{equation}
The relevance of each graph changes depending on the nature of the outgoing
chargino. In the higgsino-dominated case, diagram~(\ref{eq:GAU.5}b) gives
negligible contribution, and only diagram~(\ref{eq:GAU.5}a) should be
considered. On the other hand, in the gaugino-dominated case both these
graphs contribute, but the second is suppressed if the sneutrino is heavy.
Since the interference term between~(\ref{eq:GAU.5}a) and~(\ref{eq:GAU.5}b)
is negative, the total cross section in the gaugino-dominated scenario will
be smaller if the sneutrino is light, and the corresponding lower limit on
$m_\tdC$ will be weaker. According to Refs.~\cite{DELPHI98,DELPHI99}, the
present bound is $m_{\tdC^\pm} \gtrsim 90\GeV$ for the higgsino-dominated
case or when the sneutrino is heavy, while if the sneutrino is light and the
chargino is mainly a gaugino this bound is reduced to $m_{\tdC^\pm} \gtrsim
70\GeV$.

However, when the lightest chargino and neutralino (the latter being the
LSP) are almost degenerate in mass, the charged decay products of the light
chargino are very soft, and the above quoted bounds are no longer valid. A
special search for such light chargino has been performed recently by the
DELPHI collaboration, and the case of $\Delta M^\pm \equiv m_{\tdC_1^\pm} -
m_{\tdC_1^0} \lesssim 100\MeV$ is now excluded~\cite{DELPHI99}. However,
still in the case of $\Delta M^\pm \sim 1\GeV$ LEP~II does not provide a
lower bound and charginos as light as $45\GeV$ are allowed (this bound comes
from the measurements of $Z$ decays at LEP~I and SLC). The case of almost
degenerate chargino and neutralino can be naturally realized in SUSY and the
possibilities to find such particles are discussed in
literature~\cite{Gunion99}.

\begin{figure}[!t] \centering
    \includegraphics[width=0.9\textwidth]{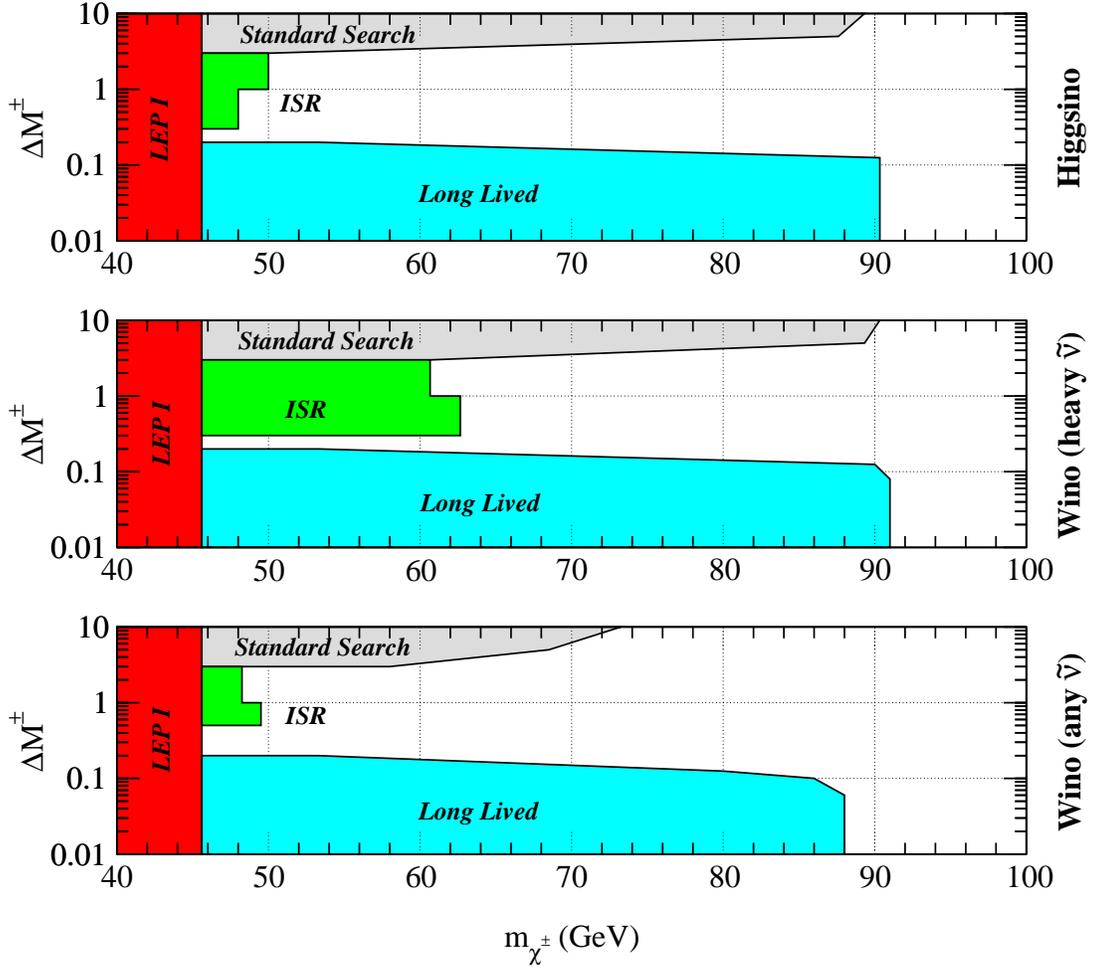}
    \caption[Exclusion plot in the plane $(m_{\tilde{\chi}}, \Delta M^\pm)$]{ 
      Regions in the plane $(m_{\tilde{\chi}}, \Delta M^\pm)$ excluded at
      $95\pCL$ by the DELPHI collaboration. The upper panel refers to the
      Higgsino-dominated case, and the lower ones to the wino-dominated
      scenario with and without assuming $m_{\tilde{\nu}_e} > 500\GeV$. This
      figure reproduces schematically Fig.~12 of Ref.~\cite{DELPHI99}.}
    \label{fig:GAU.10}
\end{figure}

In Fig.~\ref{fig:GAU.10} we summarize the present experimental bounds on the
chargino mass for different values of the chargino-neutralino mass splitting
$\Delta M^\pm$. One can see that, for $0.2\GeV \lesssim \Delta M^\pm
\lesssim 3\GeV$, not much more than $m_{\tdC^\pm} < 45\GeV$ can be excluded.
However, some interesting results can be obtained by looking at the
so-called Initial State Radiation (ISR), which consists in the emission of a
single photon from the initial $e^+$ or $e^-$. In the analysis presented in
Ref.~\cite{DELPHI99}, the use of ISR does not led to relevant improvements
of the chargino mass bounds in the region $0.2\GeV \lesssim \Delta M^\pm
\lesssim 3\GeV$, except for the case of gaugino domination with a heavy
sneutrino. However, recently many progresses have been done in this
direction, and some preliminary results found by the DELPHI
collaboration~\cite{Perrotta99} will be discussed in Sec.~\ref{sec:GAU.40}.

\section{Theoretical analysis} \label{sec:GAU.20}

In the simplest supersymmetric extensions of the Standard Model, the
chargino-neutralino sector is characterized by the numerical values of four
parameters: $M_1$, $M_2$, $\mu$ and $\tan\beta$. The particle contents of
the model includes one $SU(2)$ singlet (bino), one $SU(2)$ triplet (wino)
and two $SU(2)$ doublets (higgsinos):
\begin{equation} \label{eq:GAU.10}
    \tdB, \qquad 
    \begin{pmatrix}
	\tdW_1 \\
	\tdW_2 \\
	\tdW_3
    \end{pmatrix}, \qquad
    \begin{pmatrix} 
	\tdH_1^0 \\ 
	\tdH_1^-
    \end{pmatrix}, \qquad
    \begin{pmatrix} 
	\tdH_2^+ \\ 
	\tdH_2^0
    \end{pmatrix},
\end{equation}
where all the fields are Weyl spinors. For later convenience, let us also
define the Weyl fields $\tdW^\pm$ and the charged Dirac fields $\tdW$ and
$\tdH$:
\begin{equation}
    \tdW^\pm \equiv \frac{ \tdW_1 \mp i \tdW_2 }{\sqrt{2}}, \qquad
    \tdW \equiv
    \begin{pmatrix}
	-i \tdW^+ \\
	i \Bar{\tdW}^-
    \end{pmatrix}, \qquad
    \tdH \equiv
    \begin{pmatrix}
	\tdH_2^+ \\
	\Bar{\tdH}_1^-
    \end{pmatrix}.
\end{equation}
The terms of the SUSY Lagrangean which are relevant for our purposes are
those describing the coupling of gauginos with the electroweak gauge bosons
(which we will discuss in the following sections), and of course the mass
terms:
\begin{equation} \begin{split} \label{eq:GAU.20}
    \mathcal{L}_m & = \frac{1}{2} M_1 \tdB \tdB 
    + \frac{1}{2} M_2 \tdW_i \tdW_i + \mu 
    \LT( \tdH_1^0 \tdH_2^0 - \tdH_1^+ \tdH_2^- \RT)
    - i \frac{g'}{2} \tdB \LT( v_1 \tdH_1^0 - v_2 \tdH_2^0 \RT) \\
    & + i \frac{g}{2} \LT[ v_1 \LT( \sqrt{2} \tdW^+ \tdH_1^- 
    + \tdW_3 \tdH_1^0 \RT) + v_2 \LT( \sqrt{2} \tdW^- \tdH_2^+ 
    - \tdW_3 \tdH_2^0 \RT) \RT] + \hc
\end{split} \end{equation}
Looking at Eq.~\eqref{eq:GAU.20}, it is straightforward to see that the
states~\eqref{eq:GAU.10} are \emph{not} mass eigenstates. To find mass
eigenstates, it is convenient to introduce the following definitions:
\begin{align}
    \label{eq:GAU.30x} \mathbf{X} & =
    \begin{pmatrix}
	M_2 & \sqrt{2} \, m_W \sin\beta \\
	\sqrt{2} \, m_W \cos\beta & \mu
    \end{pmatrix}, \qquad\quad \psi^+ =
    \begin{pmatrix}
	-i \tdW^+ \\
	\tdH_2^+
    \end{pmatrix}, & \psi^- & =
    \begin{pmatrix}
	-i \tdW^- \\
	\tdH_1^-
    \end{pmatrix}, \\[1mm]
    \label{eq:GAU.30y} \mathbf{Y} & =
    \begin{pmatrix}
	M_1 & 0 & -m_Z\, s\, \cos\beta & ~m_Z\, s\, \sin\beta \\
	0 & M_2 & ~m_Z\, c\, \cos\beta & -m_Z\, c\, \sin\beta \\
	-m_Z\, s\, \cos\beta & ~m_Z\, s\, \sin\beta & 0 & -\mu \\
	~m_Z\, c\, \cos\beta & -m_Z\, c\, \sin\beta & -\mu & 0
    \end{pmatrix}, & \psi^0 & =
    \begin{pmatrix}
	-i \tdB   \\
	-i \tdW_3 \\
	\tdH_1^0  \\
	\tdH_2^0
    \end{pmatrix}.
\end{align}
Now the mass Lagrangean $\mathcal{L}_m$ can be written in a more compact
form:
\begin{equation}
    \mathcal{L}_m = -\frac{1}{2} 
    \LT( \psi^+ \; \psi^- \RT)
    \begin{pmatrix} 
	0 & \mathbf{X}^T \\
	\mathbf{X} & 0 
    \end{pmatrix}
    \begin{pmatrix}
	\psi^+ \\ \psi^-
    \end{pmatrix} - \frac{1}{2} \LT( \psi^0 \RT)^T \mathbf{Y} \psi^0
    + \hc,
\end{equation}
and the mass eigenstates can be obtained from~\eqref{eq:GAU.10} by means of
a unitary transformation, chosen in such a way that the new mass
matrices~(\ref{eq:GAU.30x},~\ref{eq:GAU.30y}) will be diagonal:
\begin{equation}
    \mathbf{U^* X V^{-1}} = \mathbf{diag}(m_{\tdC_1^\pm},
        m_{\tdC_2^\pm}), \qquad
    \mathbf{N^* Y N^{-1}} = \mathbf{diag}(m_{\tdC_1^0},
        m_{\tdC_2^0}, m_{\tdC_3^0}, m_{\tdC_4^0}).
\end{equation}
At this point, let us note that, while it is rather simple to find the
eigenstates for the $(2 \times 2)$ chargino matrix $\mathbf{X}$, the $(4
\times 4)$ neutralino matrix $\mathbf{Y}$ cannot in general be diagonalized
analytically; therefore, no results written explicitly through the SUSY
parameters $M_1$, $M_2$, $\mu$ and $\tan\beta$ can usually be obtained when
discussing the neutralino sector.

From Eqs.~\eqref{eq:GAU.30x} and~\eqref{eq:GAU.30y} it is easy to see that
the case of a nearly degenerate lightest chargino and neutralino naturally
arises under two different circumstances:
\begin{enumerate}[(1)]
  \item\label{misc:GAU.a} $M_2 \gg \mu$: in this case the particles of
    interest form an $SU(2)$ doublet of Dirac fermions, whose wave functions
    are dominated by higgsinos;
  \item\label{misc:GAU.b} $\mu \gg M_2$: in this way we get an $SU(2)$
    triplet of Majorana fermions, with the wave functions dominated by
    winos.
\end{enumerate}
Although physically different, the mathematical description of these two
cases is the same. In fact, it is simple to prove that, given an arbitrary
matrix, if a diagonal element dominates over the others then the
corresponding non-diagonal elements which lie on the same line or column may
be neglected in the first approximation (even if they are comparable in size
with the other diagonal elements). For what concerns the mass
matrices~(\ref{eq:GAU.30x},~\ref{eq:GAU.30y}), limits~\enref{misc:GAU.a}
and~\enref{misc:GAU.b} are mathematically equivalent to setting $m_{W,Z} =
0$ (or equivalently $v_1 = v_2 = 0$ in Eq.~\eqref{eq:GAU.20}), and this has
some straightforward consequences:
\begin{itemize}
  \item the SUSY parameter $\tan\beta$ disappears from
    Eqs.~\eqref{eq:GAU.30x} and~\eqref{eq:GAU.30y};

  \item the matrices $\mathbf{X}$ and $\mathbf{Y}$ are now automatically
    diagonal (except for the $(2 \times 2)$ block involving higgsinos in the
    neutralino matrix $\mathbf{Y}$), therefore the states~\eqref{eq:GAU.10}
    are now mass eigenstates;
    
  \item the two neutral Weyl higgsinos merge to form a Dirac particle with
    mass $\mu$;
    
  \item for each neutralino it exists a chargino with the same mass, except
    for the bino which has no charged counterpart.
\end{itemize}

Being $\mathbf{X}$ and $\mathbf{Y}$ in both the cases already in the desired
form, it is now possible to choose the matrices $\mathbf{U}$, $\mathbf{V}$
and $\mathbf{N}$ to be simply the identity matrix, so in these particular
limits we can derive \emph{analytical} expressions for the radiative
corrections. The coupling Lagrangean is naturally split into two parts, one
describing the interactions of the wino components (with mass $M_2$), and
the other describing the interactions of the higgsino components (with mass
$\mu$); depending on whether limit~\enref{misc:GAU.a} or~\enref{misc:GAU.b}
is considered, either the former or the latter part will describe the
low-energy spectrum of the model.

In the following sections, we will analyze in detail both these cases and
the corresponding relevant parts of the coupling Lagrangean. Let us note
since now that the bino is decoupled from gauge bosons, therefore the mass
parameter $M_1$ never enters in the expressions for the oblique corrections.

\subsection{The higgsino-dominated case}

As already noted, the two neutral Weyl higgsinos $\tdH_1^0$ and $\tdH_2^0$
can be seen as components of a neutral Dirac fermion $\tdH^0$, which
together with the charged higgsino $\tdH$ form an $SU(2)$ doublet with mass
$\mu$. The interesting parts of the mass and coupling Lagrangean are:
\begin{align}
    \label{eq:GAU.100m} \mathcal{L}_m^\tdH & =
        -\mu \Bar{\tdH} \tdH - \mu \Bar{\tdH}^0 \tdH^0, \\[1mm]
    \label{eq:GAU.100cg} \mathcal{L}_\gamma^\tdH & = 
        -e A_\mu \Bar{\tdH} \gamma^\mu \tdH, \\[1mm]
    \label{eq:GAU.100cz} \mathcal{L}_Z^\tdH & = 
        \frac{e}{cs} \LT( s^2 - \frac{1}{2} \RT) 
        Z_\mu \Bar{\tdH} \gamma^\mu \tdH
        + \frac{e}{2cs} Z_\mu \Bar{\tdH}^0 \gamma^\mu \tdH^0, \\[1mm]
    \label{eq:GAU.100cw} \mathcal{L}_W^\tdH & 
        = -\frac{e}{\sqrt{2} s} W_\mu \Bar{\tdH}^0 \gamma^\mu \tdH 
	+ \hc
\end{align}
It is immediate to note that all the couplings are vector-like. As a
consequence, all the gauge boson self-energies are simply proportional to
the photon self-energy $\Sigma_\gamma(q^2)$, and the proportionality factor
can be easily derived from the coupling constants entering
Eqs.~(\ref{eq:GAU.100cg}-\ref{eq:GAU.100cw}). We have:
\begin{align}
    \label{eq:GAU.110w} \Sigma_W^\tdH(q^2) & = 
        \frac{1}{2 s^2} \Sigma_\gamma(q^2), \\[1mm]
    \label{eq:GAU.110z} \Sigma_Z^\tdH(q^2) & = 
        \frac{1 + \LT( c^2 - s^2 \RT)^2}{4 c^2 s^2}
        \Sigma_\gamma(q^2), \\[1mm]
    \label{eq:GAU.110x} \Sigma_\GZ^\tdH(q^2) & = 
        \frac{c^2 - s^2}{2 cs} \Sigma_\gamma(q^2).
\end{align}
The expression for the photon self-energy $\Sigma_\gamma(q^2)$ due to a
charged Dirac fermion with mass $m$ is known from QED:
\begin{equation} \label{eq:GAU.120}
    \Sigma_\gamma(q^2) = \frac{\albar}{3\pi} \LT\{
      q^2 \LT( \Delta - \ln \frac{m^2}{m_Z^2} \RT)
      + \LT( q^2 + 2 m^2 \RT) F\! \LT( \frac{q^2}{m_Z^2} \RT) 
      - \frac{q^2}{3} \RT\},
\end{equation}
where the function $F(x)$ is defined in Eq.~\eqref{eq:APP.40f}. Now, looking
at Eqs.~(\ref{eq:APP.20m}-\ref{eq:APP.20r}) and using
Eqs.~(\ref{eq:GAU.110w}-\ref{eq:GAU.110x}), we see that the quantities which
are relevant for evaluating the $V_i$ functions are $\Sigma'_\gamma(0)$,
$\Sigma'_\gamma(m_Z^2)$, $\Pi_\gamma(m_Z^2)$ and $\Pi_\gamma(m_W^2)$:
\begin{align}
    \label{eq:GAU.130a} \Sigma'_\gamma(0) & = \frac{\albar}{3\pi}
      \Delta_\chi, \\
    \label{eq:GAU.130b} \Sigma'_\gamma(m_Z^2) & = \frac{\albar}{3\pi} \LT\{
      \Delta_\chi + F(\chi) + \LT( 1 + 2\chi\RT) F'(\chi)
      - \frac{1}{3} \RT\}, \\
    \label{eq:GAU.130c} \Pi_\gamma(m_Z^2) & = \frac{\albar}{3\pi} \LT\{
      \Delta_\chi + \LT( 1 + 2\chi \RT) F(\chi) - \frac{1}{3} \RT\}, \\
    \label{eq:GAU.130d} \Pi_\gamma(m_W^2) & = \frac{\albar c^2}{3\pi} \LT\{
      \Delta_\chi + \LT( 1 + 2\frac{\chi}{c^2} \RT)
      F\! \LT( \frac{\chi}{c^2} \RT) - \frac{1}{3} \RT\},
\end{align}
where $\chi \equiv \LT( m_\tdC / m_Z \RT)^2$ and $m_\tdC = \mu$ is the
common mass of the light chargino and neutralino. Finally, substituting
Eqs.~(\ref{eq:GAU.110w}-\ref{eq:GAU.110x})
and~(\ref{eq:GAU.130a}-\ref{eq:GAU.130d}) into
Eqs.~(\ref{eq:APP.20m}-\ref{eq:APP.20r}), we find:
\begin{align}
    \label{eq:GAU.140m} \delta^\tdH V_m &= \frac{16}{9} \LT[
       \LT( \frac{1}{2} - s^2 + s^4 \RT) \LT( 1 + 2\chi \RT) F(\chi) 
       - \LT( \frac{1}{2} - s^2 \RT) \LT( 1 + 2\frac{\chi}{c^2} \RT)
        F\LT( \frac{\chi}{c^2} \RT) - \frac{s^4}{3} \RT], \\[1mm]
    \label{eq:GAU.140a} \delta^\tdH V_A &= \frac{16}{9} 
      \LT( \frac{1}{2} - s^2 + s^4 \RT)
      \LT[ \frac{12 \chi^2 F(\chi) - 2\chi - 1}{4\chi - 1} \RT], \\[1mm]
    \label{eq:GAU.140r} \delta^\tdH V_R &= \frac{16}{9} c^2 s^2
      \LT[ \LT( 1 + 2\chi \RT) F(\chi) - \frac{1}{3} \RT],
\end{align}

\begin{figure}[!t] \centering
    \includegraphics[width=0.9\textwidth]{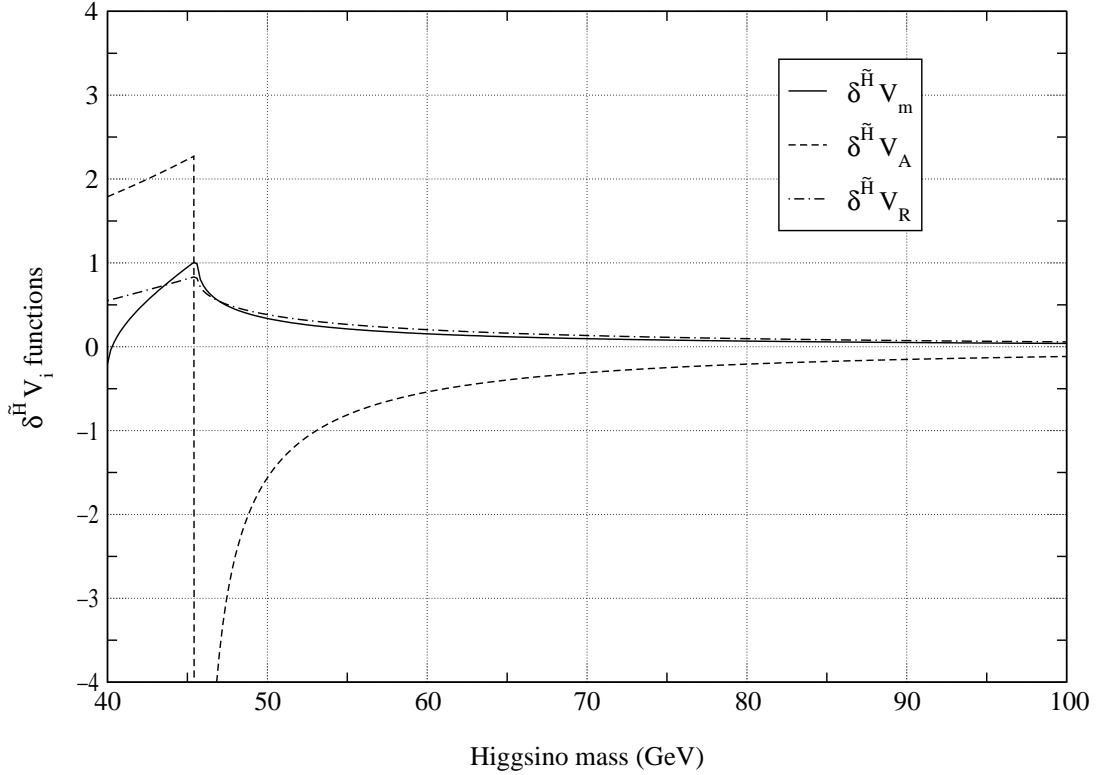}
    \caption[Plot of $\delta^{\tilde{H}} V_i$ as a function of
    $m_{\tilde{\chi}}$]{
      Dependence of the $\delta^{\tilde{H}} V_i$ functions on the light
      gaugino mass $m_{\tilde{\chi}}$, in the limit $M_{1,2} \gg \mu$
      (higgsino-dominated case).} 
    \label{fig:GAU.20}
\end{figure}

In Fig.~\ref{fig:GAU.20} the $V_i$ functions are plotted against the
chargino-neutralino mass $m_\tdC$. Looking at it we can see that the
numerical values of $\delta^\tdH V_m$ and $\delta^\tdH V_R$ are always
rather small, and the main contribution comes from $\delta^\tdH V_A$ which
is singular at $m_\tdC = m_Z/2$. This singularity is not physical and our
formulas are valid only for $2 m_\tdC \gtrsim m_Z + \Gamma_Z$; the existence
of $\chi^\pm$ with a mass closer to $m_Z/2$ will change Z-boson Breit-Wigner
curve, therefore it is also not allowed. The importance of the $Z$ wave
function renormalization for the case of light charginos was emphasized
in~\cite{Barbieri92}.

\subsection{The wino-dominated case}

Let us now discuss case~\enref{misc:GAU.b}. The charged Dirac wino $\tdW$
and the neutral Weyl zino $\tdZ$ form an $SU(2)$ triplet of Majorana
fermions, with mass $M_2$. The relevant Lagrangean in this case is:
\begin{align}
    \label{eq:GAU.200m} \mathcal{L}_m^\tdW & = 
        -M_2 \Bar{\tdW} \tdW - \frac{1}{2} M_2 \Bar{\tdZ} \tdZ, \\[1mm]
    \label{eq:GAU.200cg} \mathcal{L}_\gamma^\tdW &= 
        -e A_\mu \Bar{\tdW} \gamma^\mu \tdW, \\[1mm]
    \label{eq:GAU.200cz} \mathcal{L}_Z^\tdW & =
      -e \frac{c}{s} Z_\mu \Bar{\tdW} \gamma^\mu \tdW, \\[1mm]
    \label{eq:GAU.200cw} \mathcal{L}_W^\tdW & =
      \frac{e}{s} W_\mu \Bar{\tdZ} \gamma^\mu \tdW + \hc
\end{align}
Again, we see that in Eqs.~(\ref{eq:GAU.200cg}-\ref{eq:GAU.200cw}) we have
only vector currents. Therefore, as in the previous case all gauge boson
self-energies can be written through $\Sigma_\gamma(q^2)$, but the
coefficients are now different:
\begin{align}
    \label{eq:GAU.210w} \Sigma_W^\tdW(q^2) & =
        \frac{1}{s^2} \Sigma_\gamma(q^2), \\[1mm]
    \label{eq:GAU.210z} \Sigma_Z^\tdW(q^2) & =
        \frac{c^2}{s^2} \Sigma_\gamma(q^2), \\[1mm]
    \label{eq:GAU.210x} \Sigma_\GZ^\tdW(q^2) & =
        \frac{1}{c^2 s^2} \Sigma_\gamma(q^2).
\end{align}
Substituting Eqs.~(\ref{eq:GAU.210w}-\ref{eq:GAU.210x})
into~(\ref{eq:APP.20m}-\ref{eq:APP.20r}) and using once more
Eqs.~(\ref{eq:GAU.130a}-\ref{eq:GAU.130d}), we find the expressions for
$\delta^\tdW V_i$:
\begin{align}
    \label{eq:GAU.240m} \delta^\tdW V_m & =
      \frac{16}{9} \LT[ c^4 \LT( 1 + 2\chi \RT) F(\chi)
      - \LT( 1 - 2s^2 \RT) \LT( 1 + 2\frac{\chi}{c^2} \RT)
        F\LT( \frac{\chi}{c^2} \RT) - \frac{s^4}{3} \RT], \\[1mm]
    \label{eq:GAU.240a} \delta^\tdW V_A & = \frac{16}{9} c^4
      \LT[ \frac{12 \chi^2 F(\chi) - 2\chi - 1}{4\chi - 1} \RT], \\[1mm]
    \label{eq:GAU.240r} \delta^\tdW V_R & = \frac{16}{9} c^2 s^2
      \LT[ \LT( 1+2\chi \RT) F(\chi) - \frac{1}{3} \RT].
\end{align}

Comparing Eqs.~(\ref{eq:GAU.240m}-\ref{eq:GAU.240r}) with
Eqs.~(\ref{eq:GAU.140m}-\ref{eq:GAU.140r}), we see that the corresponding
formulas for $\delta^\tdH V_i$ and $\delta^\tdW V_i$ are very similar,
differing from each other only in some constant numerical coefficients. In
particular, $\delta^\tdH V_R$ and $\delta^\tdW V_R$ are exactly identical,
while for the case of $\delta V_A$ the ratio $\delta^\tdW V_A / \delta^\tdH
V_A$ does not depend on $\chi$ and its numerical value is $2c^4 / (c^4 +
s^4) \approx 1.8$.

\begin{figure}[!t] \centering
    \includegraphics[width=0.9\textwidth]{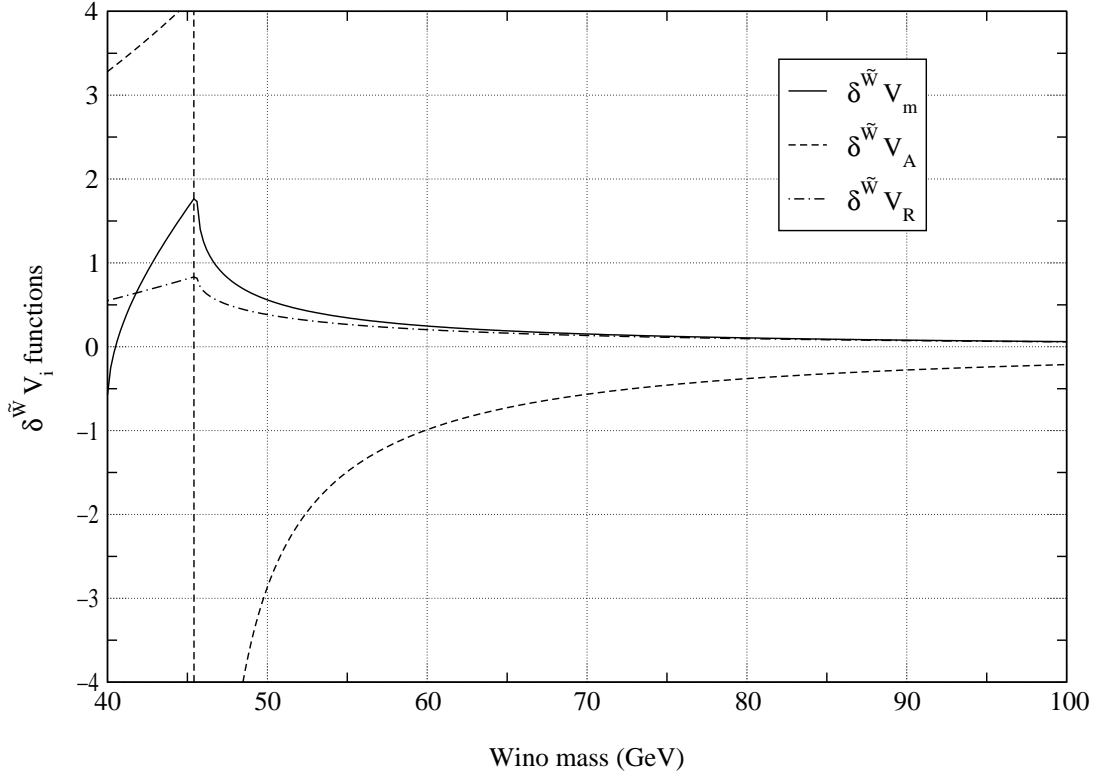}
    \caption[Plot of $\delta^{\tilde{W}} V_i$ as a function of
    $m_{\tilde{\chi}}$]{
      Dependence of the $\delta^{\tilde{W}} V_i$ functions on the light
      gaugino mass $m_{\tilde{\chi}}$, in the limit $\mu \gg M_{1,2}$
      (wino-dominated case).}
    \label{fig:GAU.30}
\end{figure}

In Fig.~\ref{fig:GAU.30} we show the dependence of $\delta^\tdW V_i$ on the
chargino-neutralino mass $m_\tdC$. Note that for $m_\chi \gtrsim 50\GeV$ we
have $\delta^\tdW V_m > \delta^\tdW V_R$, while in the previous case it was
$\delta^\tdH V_m < \delta^\tdH V_R$: this fact is irrelevant for the present
discussion, but will be useful in the next chapter to explain why higgsinos
help to improve fits for an extra generation of fermions more than winos do.
Also, note that in the limit $m_\chi \to \infty$ all $\delta V_i$ vanish, as
expected due to the \emph{decoupling} property of SUSY mentioned in the
introduction.

At this point, one may wonder whether the results found so far are really
useful for the problem under investigation or not.
Formulae~(\ref{eq:GAU.140m}-\ref{eq:GAU.140r})
and~(\ref{eq:GAU.240m}-\ref{eq:GAU.240r}) are valid in the case of
\emph{exactly} degenerate chargino and neutralino, and we know from
Sec.~\ref{sec:GAU.10} that the experimental bounds are very strong for
$\Delta M^\pm < 0.2\GeV$. However, it is important to note that, while for
the direct search experiments the crucial quantity is the \emph{mass
difference} between the lightest chargino and neutralino (because it is on
this parameter that the chargino life time and the energy of its decay
products depend), for the case of precision measurements what really matters
is the \emph{mass} itself of the particles involved. This imply that, when
the masses of chargino and neutralino are slightly moved from each other,
the consequences on the results of direct search may be drastic, but effects
on radiative corrections are very small and the overall picture does not
change. Therefore, even in the case of not-complete degeneracy
Eqs.~(\ref{eq:GAU.140m}-\ref{eq:GAU.140r})
and~(\ref{eq:GAU.240m}-\ref{eq:GAU.240r}) provide a very good approximation
to the exact numerical results and the conclusions still hold. This has been
verified numerically using equations from Ref.~\cite{Hollik99}.

\section{Data analysis and fits} \label{sec:GAU.30}

\begin{table}[!t] \centering
    \newcolumntype{d}{D{.}{.}{-1}}
    \begin{tabular}{>{\rule{0pt}{12pt}}l!{\hspace{5mm}}ddr!{\hspace{5mm}}c}
	\hline\hline
	Observable & \multicolumn{1}{r}{Experimental data}
	           & \multicolumn{1}{r}{Theoretical fit}
		   & Pull & $V_i$ function \\
	\hline
	$\Gamma_Z$ (GeV) &    2.4939(24)  &   2.4960(18)  & -0.9 & $V_A$, $V_R$ \\
	$\sigma_h$ (nb)  &   41.491(58)   &  41.472(16)   &  0.3 & $V_A$, $V_R$ \\
	$R_l$            &   20.765(26)   &  20.746(20)   &  0.7 & $V_A$, $V_R$ \\
	$R_b$            &    0.2166(7)   &   0.2158(2)   &  1.1 & $V_A$, $V_R$ \\
	$R_c$            &    0.1735(44)  &   0.1723(1)   &  0.3 & $V_A$, $V_R$ \\
	$A_{\tau}$       &    0.1431(45)  &   0.1467(16)  & -0.8 & $V_R$ \\
	$A_e$            &    0.1479(51)  &   0.1467(16)  &  0.2 & $V_R$ \\
	$A_b$            &    0.8670(350) &   0.9348(2)   & -1.9 & $V_R$ \\
	$A_c$            &    0.6470(400) &   0.6677(7)   & -0.5 & $V_R$ \\
	$A_{FB}^l$       &    0.0168(10)  &   0.0161(4)   &  0.7 & $V_R$ \\
	$A_{FB}^b$       &    0.0990(21)  &   0.1028(12)  & -1.8 & $V_R$ \\
	$A_{FB}^c$       &    0.0709(44)  &   0.0734(9)   & -0.6 & $V_R$ \\
	$A_{LR}$         &    0.1504(23)  &   0.1467(16)  &  1.6 & $V_R$ \\
	$s_l^2(Q_{FB})$  &    0.2321(10)  &   0.2316(2)   &  0.5 & $V_R$ \\
	$m_W$ (GeV)      &   80.390(64)   &  80.366(34)   &  0.4 & $V_m$ \\
	$s_W^2$          &    0.2254(21)  &   0.2233(7)   &  1.0 & $V_m$ \\
	$m_t$ (GeV)      &  173.8(5.0)    & 170.8(4.9)    &  0.6 & \\
	\hline\hline
    \end{tabular}
    \caption[Experimental value and theoretical prediction of precision
    measurements]{
      Experimental value and SM theoretical prediction of the electroweak
      precision measurements, as reported in Tab.~7 of Ref.~\cite{NORV99}.
      In the last column we list the $V_i$ functions which are relevant for
      evaluating each observable.}
    \label{tab:GAU.10}
\end{table}

Having all the necessary formulas at our disposal, we can now analyze the
impact of a chargino almost degenerate with the lightest neutralino on
precision measurements, so to derive a lower limit for its mass. In
Table~\ref{tab:GAU.10}, which we extracted from Table~7 of
Ref.~\cite{NORV99}, we report the present experimental value of many
electroweak observables, together with their SM expectation. The key point
is that the contribution of New Physics into these observables occurs mostly
\emph{through} the oblique corrections, which are fully parameterized by the
three functions $V_i$ described above. This situation is summarized in the
following diagram:
\begin{center}
    \fbox{\parbox[c][20mm]{40mm}{ \centering
	\underline{\bf New Physics} \\ \vfil
	$M_1$, $M_2$, $\mu$, $\tan\beta$, \ldots
	}}
    $\xrightarrow{~(\mathrm{A})~}$
    \fbox{\parbox[c][20mm]{40mm}{\centering
	\underline{\bf Oblique corrections} \\ \vfil
	$\delta V_m$, $\delta V_A$, $\delta V_R$
	}}
    $\xrightarrow{~(\mathrm{B})~}$
    \fbox{\parbox[c][20mm]{40mm}{\centering
	\underline{\bf Experimental data} \\ \vfil
	$\Gamma_Z$, $\sigma_h$, $A_{FB}^b$, $m_W$, \ldots
	}}
\end{center}
Step~(A) is clearly model dependent, and was performed in the previous
sections by deriving Eqs.~(\ref{eq:GAU.140m}-\ref{eq:GAU.140r})
and~(\ref{eq:GAU.240m}-\ref{eq:GAU.240r}). On the contrary, step~(B) is
model independent, and can be done once for all by summarizing all the
experimental data into three ``best fit'' values $\overline{\delta V_i}$ and
a $(3 \times 3)$ correlation matrix $C_{ij}$. Using formulas from
Ref.~\cite{NORV99} and the numerical values from Table~\ref{tab:GAU.10}, we
get:
\begin{equation} \label{eq:GAU.300}
    \begin{pmatrix}
	C_{mm} & C_{mA} & C_{mR} \\
	C_{mA} & C_{AA} & C_{AR} \\
	C_{mR} & C_{AR} & C_{RR}
    \end{pmatrix} = \begin{pmatrix}
	7.28   & 0      &  0     \\ 
	0      & 7.24   &  2.54  \\
	0      & 2.54   & 23.03
    \end{pmatrix}; \qquad \begin{pmatrix}
	\overline{\delta V_m} \\
	\overline{\delta V_A} \\
	\overline{\delta V_R}
    \end{pmatrix} = \begin{pmatrix}
	-0.07  \\ 
	-0.33  \\
	+0.01
    \end{pmatrix}.
\end{equation}
Now it is very easy to evaluate the $\chi^2$:
\begin{equation} \label{eq:GAU.310}
    \chi^2 = C_{ij} \LT( \delta^\NP V_i - \overline{\delta V_i} \RT)
    \LT( \delta^\NP V_j - \overline{\delta V_j} \RT).
\end{equation}

\begin{figure}[!t] \centering
    \includegraphics[width=0.9\textwidth]{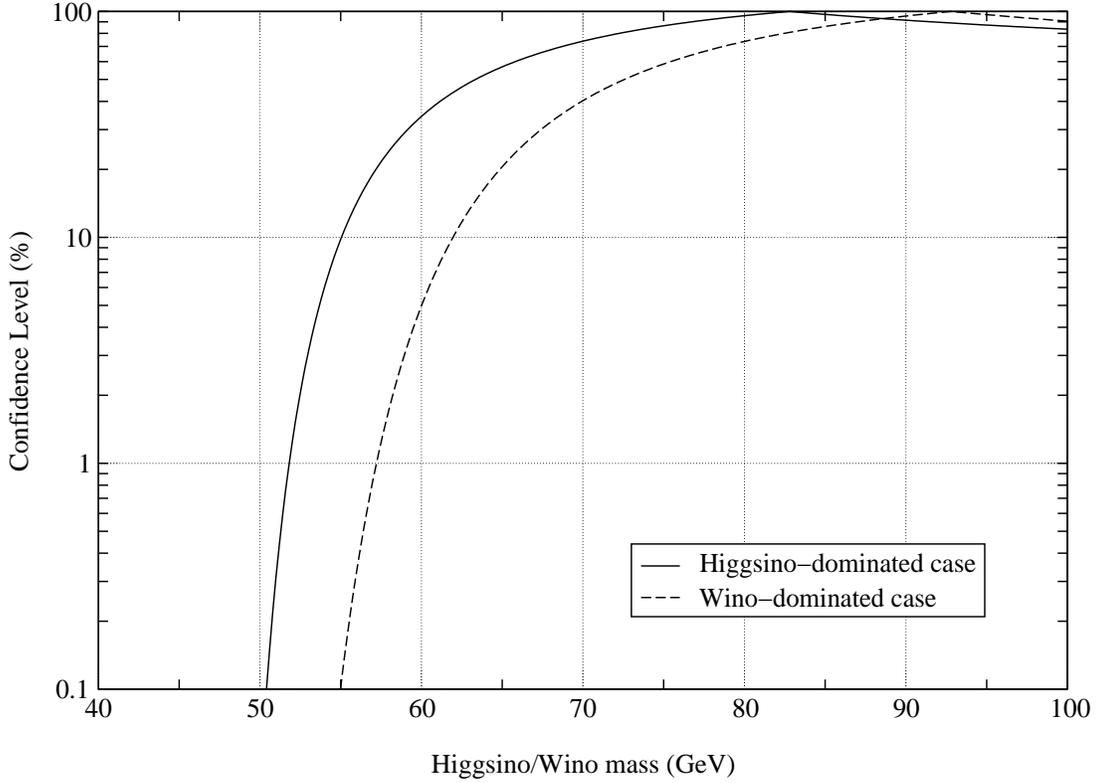}
    \caption[Plot of C.L.\ as a function of $m_{\tilde{\chi}}$]{
      Dependence of the confidence level on the light gaugino mass
      $m_{\tilde{\chi}}$, both in the limit $M_{1,2} \gg \mu$
      (higgsino-dominated case) and $\mu \gg M_{1,2}$ (wino-dominated
      case).}
    \label{fig:GAU.40}
\end{figure}

In Fig.~\ref{fig:GAU.40} we report the confidence level curve for both the
higgsino-dominated and the wino-dominated case. At $95\pCL$, we have:
\begin{align}
    \label{eq:GAU.320a} m_\chi &
        \gtrsim 54\GeV \text{ for the higgsino-dominated case}; \\
    \label{eq:GAU.320b} m_\chi &
        \gtrsim 60\GeV \text{ for the wino-dominated case}.
\end{align}

Since there are a number of new additional particles in SUSY extensions, we
will briefly discuss their contributions to the functions $V_i$. In the
considered limits the remaining charginos and neutralinos are very heavy, so
they simply decouple and produce negligible contributions. The contributions
of the three generations of sleptons (with masses larger than $90\GeV$) into
$V_A$ are smaller than $0.1$, so they can be safely neglected. The
contributions of squarks of the first two generations are also negligible
since they should be heavier than Tevatron direct search bounds; taking
$m_{\tilde{q}} \gtrsim 200\GeV$, we have $| \delta^{\tilde{q}} V_i |
\lesssim 0.1$. Concerning the contributions of the third generation squarks,
they are enhanced by the large top-bottom mass difference and are not
negligible. However, being positive and almost universal~\cite{Gaidaenko98},
they do not affect our analysis: compensating negative contributions of
chargino-neutralino into $V_A$ they will generate positive contributions to
$V_R$ and $V_m$, and $\chi^2$ will not be better. When squarks are heavy
enough (for $m_{\tilde{b}} \gtrsim 300\GeV$), they simply decouple and their
contributions become negligible as well.

The last sector of the theory to be discussed is Higgs bosons. Unlike the
case of Standard Model now we have one extra charged higgs and two extra
neutral higgses. Their contributions to radiative corrections were studied
in detail in~\cite{Chankowski94}. According to Fig.~2 from that paper it is
clear that the contributions of MSSM higgses (and $SU(2) \times U(1)$ gauge
bosons) equal with very good accuracy those of the Standard Model with the
mass of SM higgs being equal to that of the lightest neutral higgs in SUSY
generalization. That is why the contributions from the higgs sector of the
theory also cannot compensate those of the light chargino-neutralino.

In addition to SUSY particles, one may wonder about the possibility of
improving the overall $\chi^2$ by shifting a bit some of the SM parameters
from their central values. The most important contributions of this kind are
those coming from the worst-measured quantities, i.e.\ the top mass and the
higgs mass. However, concerning the top its contributions are almost
universal, just as for the case of the third generation squarks, therefore
they also cannot affect our analysis. Concerning the higgs, in our model it
is bounded to be heavier than $90\GeV$ by direct search experiments and
lighter than $120 \div 130\GeV$ by the basic assumption that SUSY exists
(otherwise, we wouldn't discuss chargino mass!), and since its contributions
to physical observables are only logarithmic, we do not expect them to be
relevant.

Apart from oblique corrections (those arising from vector bosons self
energies) which have been considered in this letter, there are process
dependent vertex and box corrections. However, due to LEP~II and Tevatron
low bounds on squarks and sleptons masses they are small.

\section{Recent developments and new fits} \label{sec:GAU.40}

The analysis of the $\chi^2$ done in the previous section is not completely
satisfactory for two main reasons:
\begin{itemize}
  \item the data reported in Table~\ref{tab:GAU.10} are not independent from
    one another, since many of them come from the same experiment and
    therefore are strongly correlated. In the analysis performed in
    Sec.~\ref{sec:GAU.30}, the cross correlations have been neglected. This
    led to a small underestimation of the errors associated to each
    $\overline{\delta V_i}$ (or equivalently to an overestimation of the
    coefficients $C_{ij}$), so bounds~(\ref{eq:GAU.320a}-\ref{eq:GAU.320b})
    may be slightly stronger than the correct ones;

  \item effects of changing the higgs mass, although known to be small, may
    affect the precise numerical value of
    bounds~(\ref{eq:GAU.320a}-\ref{eq:GAU.320b}).
\end{itemize}

\begin{figure}[!t] \centering
    \includegraphics[width=0.9\textwidth]{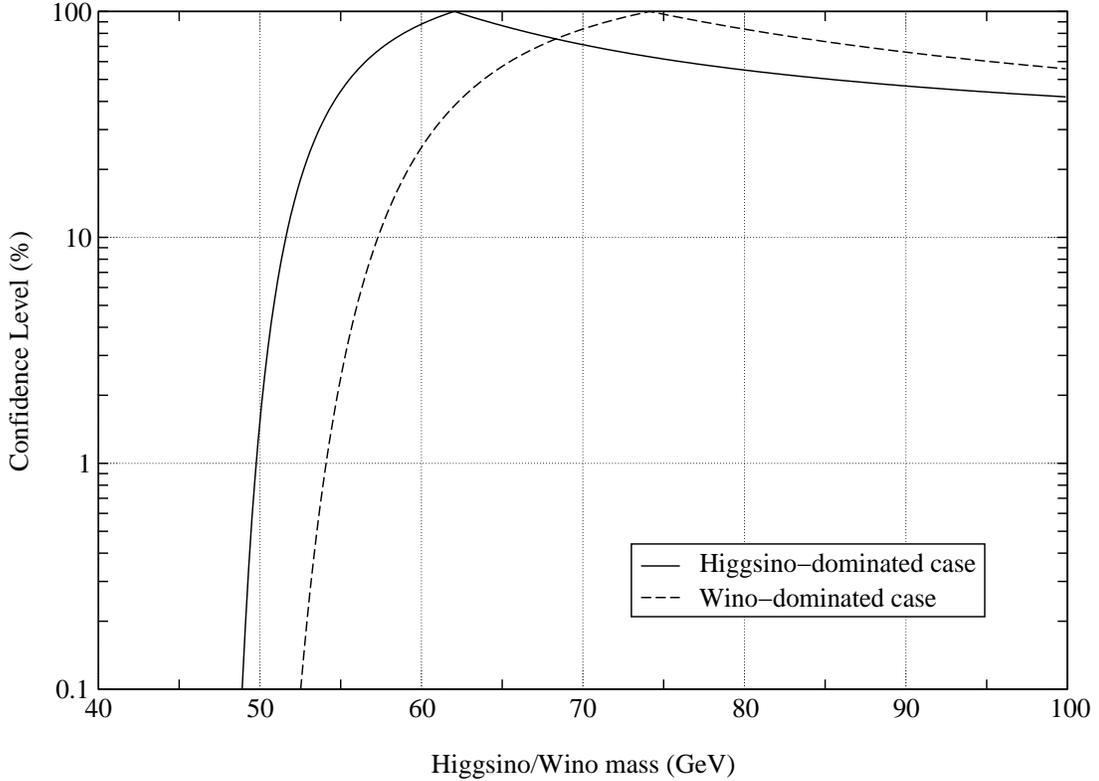}
    \caption[Improved plot of C.L.\ as a function of $m_{\tilde{\chi}}$]{
      Dependence of the confidence level, evaluated with the program
      \texttt{LEPTOP}, on the light gaugino mass $m_{\tilde{\chi}}$, both in
      the limit $M_{1,2} \gg \mu$ (higgsino-dominated case) and $\mu \gg
      M_{1,2}$ (wino-dominated case).}
    \label{fig:GAU.50}
\end{figure}

To overcome these problems, we asked to A.~Rozanov to repeat our analysis
using the program \texttt{LEPTOP}. In Fig.~\ref{fig:GAU.50} we show the new
profile for the confidence level curve, which now account both for all the
cross correlations among experimental data and for the effects of varying
the SM parameters $m_t$, $m_H$ and $\hat{\alpha}_s$. The new bounds at
$95\pCL$ for the chargino-neutralino mass are:
\begin{align}
    \label{eq:GAU.400a} m_\chi &
        \gtrsim 51\GeV \text{ for the higgsino-dominated case}; \\
    \label{eq:GAU.400b} m_\chi &
        \gtrsim 56\GeV \text{ for the wino-dominated case}.
\end{align}
Although these limits are slightly weaker than the previous one, the main
conclusion still hold.

\begin{figure}[!t] \centering
    \includegraphics[width=0.9\textwidth]{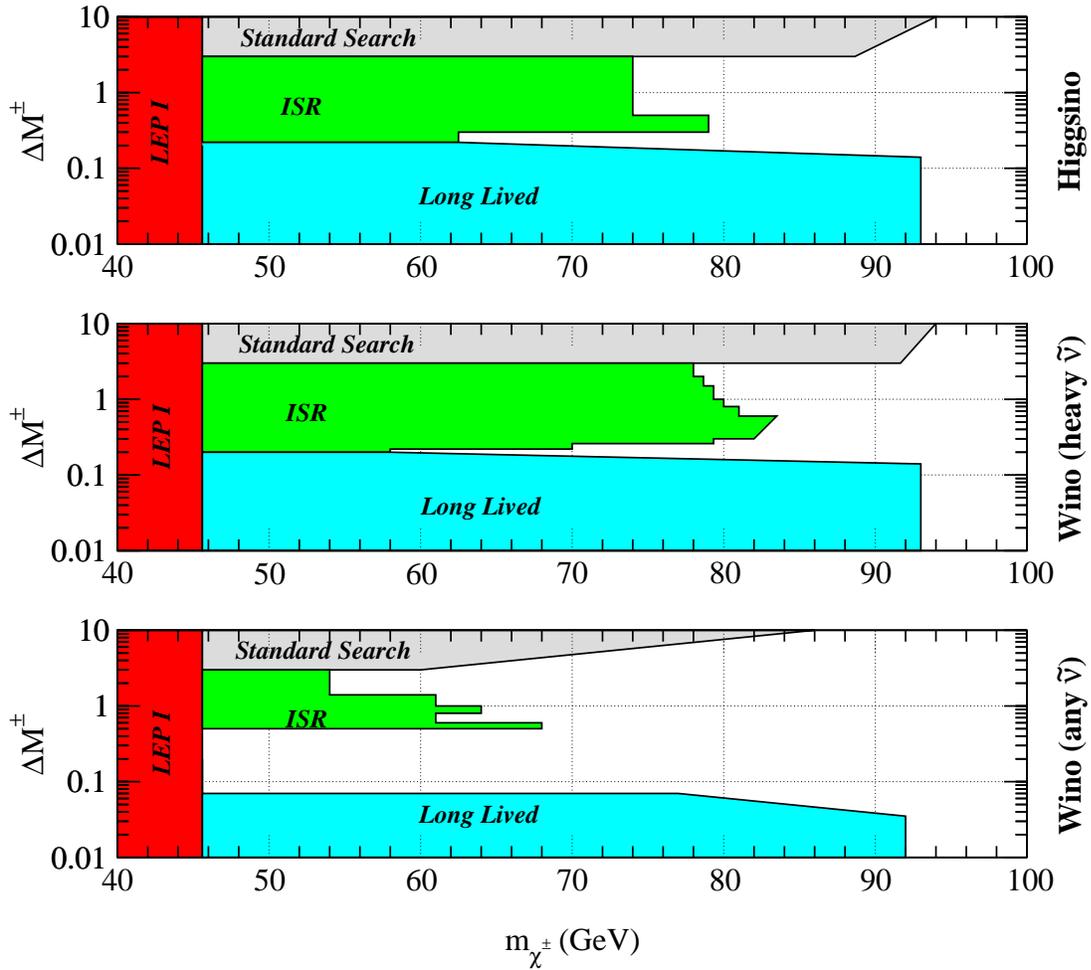}
    \caption[Improved exclusion plot in the plane $(m_{\tilde{\chi}}, \Delta
    M^\pm)$]{
      Regions in the plane $(m_{\tilde{\chi}}, \Delta M^\pm)$ excluded at
      $95\pCL$ by the DELPHI collaboration. The upper panel refers to the
      Higgsino-dominated case, and the lower ones to the wino-dominated
      scenario with and without assuming $m_{\tilde{\nu}_e} > 500\GeV$. This
      figure reproduces schematically the results presented by A.~Perrotta
      at the conference PASCOS99~\cite{Perrotta99}.}
    \label{fig:GAU.60}
\end{figure}

Recently, the DELPHI collaboration has strongly improved its analysis of the
ISR in the region $0.2\GeV \lesssim \Delta M^\pm \lesssim 3\GeV$. The
preliminary results were presented in~\cite{Perrotta99}, and are summarized in
Fig.~\ref{fig:GAU.60}. It can immediately be seen that the narrow window for
a light chargino-neutralino has now be completely closed (up to $80\GeV$)
for the higgsino-dominated case, and for the wino-dominated case with a
sneutrino heavier than $500\GeV$. However, for a light sneutrino the
negative interference between diagrams~(\ref{eq:GAU.5}a)
and~(\ref{eq:GAU.5}b) makes the total cross section very small, and analysis
based on ISR becomes impossible. Therefore, in the wino-dominated scenario
with unconstrained sneutrino mass the study of radiative corrections is
currently the only way to go beyond LEP~I bound.

\section{Conclusions} \label{sec:GAU.50}

In this chapter, we analyzed in detail the effects of radiative corrections
for the case of a chargino almost degenerate with the lightest neutralino on
the electroweak precision measurements. Both the higgsino-dominated and the
wino-dominated scenario have been studied, and for these limits simple
analytical formulae have been derived. For the case of wino domination, the
bound $m_\tdC \gtrsim 56\GeV$ is presently the strongest constraint which
can be imposed without making any assumption on the mass spectrum of other
superpartners.

\end{fmffile}

\clearemptydoublepage

\chapter{Extra quark-lepton generations and precision measurements}
\label{sec:GEN}

The analysis performed in the previous chapter has shown that, except for
special cases when direct search experiments fail for some peculiar reason,
at the present time precision measurements are not very useful to
investigate those extensions of the SM which, like SUSY, exhibit the
\emph{decoupling} property. However, the situation completely changes when
dealing with \emph{non-decoupling} New Physics, since in this framework even
particles far too heavy to be directly produced and observed at accelerators
may give relevant contributions to ``low'' ($\sim m_Z$) energy observables
through radiative corrections. The most straightforward generalization of
the Standard Model through inclusion of extra chiral generations of heavy
fermions, both quarks ($q=U,D$) and leptons ($l=N,E$), is an example of such
non decoupled New Physics.

The aim of this chapter is to understand to what extent the existing data on
the $Z$-boson parameters, the $W$-boson and the top quark masses allow to
bound effectively high energy models which do not decouple at low energies.
In particular, we will analyze in detail the case of extra generations of
fermions, showing that it is excluded by the electroweak precision
measurements if all the new particles are heavier than $m_Z$. We find
however that if masses of extra neutrinos are close to $50\GeV$, then
additional generations can exist, and in this case up to three extra
families are allowed within $2\sigma$. Finally, inclusion of new generations
in SUSY extensions of the Standard Model is discussed.

When speaking of extra generations, the first thing to bother about is the
extra neutrinos, $N$: being coupled to $Z$-boson they would increase the
invisible $Z$-width, so to avoid contradiction with experimental data their
masses should be larger than $45\GeV$~\cite{Abreu92}. In order to meet this
condition, one has to introduce not only left-handed states $N_L$ but also
right-handed states $N_R$, and supply new ``neutrinos'' with Dirac masses
analogously to the case of charged leptons and quarks. Unfortunately, gauge
symmetries do not forbid a Majorana mass term for $N_R$, and if it is large
one would get light $N_L$ through the see-saw mechanism. To avoid such a
failure we will suppose that the Majorana mass of $N_R$ is negligible,
closing eyes on the emerging (un)naturalness problem. Thus, our neutral
lepton $N$ is a heavy Dirac particle.

Contributions of extra generations to the electroweak radiative corrections
were considered in
Refs.~\cite{Evans94,Bamert95,Inami95,Masiero95,Novikov95,Erler98}. In what
follows we will assume that the mixing among new generations and the three
existing ones is small, hence new fermions contribute only to oblique
corrections (vector boson self-energies). This case is discussed in
Ref.~\cite{Erler98} and the authors come to the conclusion that one extra
generation of heavy fermions is excluded at $99.2\%$~C.L. The authors
of~\cite{Erler98} follow a two step procedure: first, they show that
experimental bounds on parameter $\rho$ exclude the existence of
non-degenerate extra generations (a degenerate generation is decoupled from
$\rho$); then, they consider the parameter $S$~\cite{Peskin90}, which is
sensitive also to degenerate families, and find that its experimental value
excludes the existence of extra degenerate generation as well. Such
procedure is not general enough: it does not use all precision data. In the
present chapter we perform global fits of all precision measurements.

We study both degenerate and non-degenerate extra generations on the equal
footing. By considering the contributions of new families into all precision
electroweak observables simultaneously we see that the fit of the data is
worsened by them if all the new particles are heavy. Taking the number of
new generations $N_g$ as a continuous parameter (just as it was done with
the determination of the number of neutrinos from invisible $Z$ width) we
get a bound on it. The best $\chi^2$ corresponds to $N_g \simeq -0.5$, while
$N_g=1$ is excluded by more than 2 standard deviations.

This chapter is based on Ref.~\cite{Maltoni99c}, currently submitted for
publication on PLB, and on the talk recently presented at the conference
PASCOS99~\cite{Maltoni99e}. Sec.~\ref{sec:GEN.10} contains general formulas
for oblique radiative corrections caused by an extra doublet of quarks or
leptons. In Sec.~\ref{sec:GEN.20} we consider the case when \emph{all} the
extra fermions are heavy ($m \gtrsim m_Z$), while in Sec.~\ref{sec:GEN.30}
we allow the extra neutrino to be ``light'' ($m_N \simeq 50\GeV$): in this
case, the contribution of $N$ compensates that of $U$, $D$, $E$. Finally, in
Sec.~\ref{sec:GEN.40} we analyze the SUSY version of four generations.

\section{General formulas} \label{sec:GEN.10}

New particles contribute to physical observables through the self-energies
of vector and axial currents: this induces extra corrections $\delta V_i$ to
the functions $V_i$ which are used to describe the theoretical relations
among precision measurements (see Sec.~\ref{sec:GAU.30} and
Ref.~\cite{NORV99}). In this section we present explicit expressions for
these corrections. Eqs.~(\ref{eq:GEN.20m}-\ref{eq:GEN.80r}) are valid both
for lepton and quark contributions, provided that the values of $Q_U$,
$Q_D$, $Y$, $N_c$, $u$ and $d$ are correctly set according to the following
table:
\begin{equation} \label{eq:GEN.5}
    \begin{tabular}{>{\rule[-4pt]{0pt}{15pt}}l!{\hspace{10mm}}cccccc}
	\hline\hline
	& $Q_U$ &  $Q_D$ &   $Y$ & $N_c$ & $u$           & $d$ \\
	\hline
	leptons  &   $0$ &   $-1$ &  $-1$ &   $1$ & $(m_N/m_Z)^2$ & $(m_E/m_Z)^2$ \\
	quarks   & $2/3$ & $-1/3$ & $1/3$ &   $3$ & $(m_U/m_Z)^2$ & $(m_D/m_Z)^2$ \\
	\hline\hline
    \end{tabular}
\end{equation}
where $Q_U$ and $Q_D$ are the electric charges of the up and down component
of the $SU(2)_L$ doublet, $Y = Q_U + Q_D$ is the doublet hypercharge (the
hypercharge of isosinglets being equal to its doubled electric charge), and
$N_c$ is the number of colors. Corrections produced by the whole extra
generation are given by the sum of lepton and quark contributions:
\begin{equation} \label{eq:GEN.10}
    \Delta V_i = \delta V_i^q +\delta V_i^l
\end{equation}

Contributions of quark or lepton doublets to $V_i$ functions are:
\begin{align}
    \begin{split} \label{eq:GEN.20m}
	\delta V_m^{q,l}
	& = \frac{2}{9} N_c \LT(1 - \frac{s^2}{c^2} \RT) \LT\{
	  -F(m_W^2, m_U^2, m_D^2) \LT[2 c^2 - u - d - \frac{(u-d)^2}{c^2} \RT]
	  + u + d - \frac{4}{3} c^2 \RT\} \\
	& - \frac{4s^2}{9} N_c Y \LT\{
	  (1+2u) F(u) - (1+2d) F(d) - \ln\LT(\frac{u}{d}\RT) \RT\} \\
	& + \frac{16}{9} N_c s^4 \LT\{
	  Q_U^2 \LT[(1+2u) F(u) - \frac{1}{3} \RT] +
	  Q_D^2 \LT[(1+2d) F(d) - \frac{1}{3} \RT] \RT\} \\
	& + \frac{2}{9} N_c \LT\{
	  (1-u) F(u) + (1-d) F(d) - \frac{2}{3} \RT\}
	  + \frac{s^2}{3c^2} N_c \LT( u+d \RT) \\
	& - \frac{4}{9} N_c s^2 \LT\{
	  (1+2u) F(u) + (1+2d) F(d) - \frac{2}{3} \RT\} \\
	& - \frac{2}{9} N_c \ln\LT(\frac{u}{d}\RT) \LT\{
	  \LT( 1 + \frac{1}{c^2} \RT) \frac{ud}{u-d} - \LT( c^2 - s^2 \RT)
	  \frac{u+d}{u-d} \RT\},
    \end{split} \\[\baselineskip]
    \begin{split} \label{eq:GEN.20a}
	\delta V_A^{q,l}
	& = \frac{1}{3} N_c \LT[ u + d - 2\frac{ud}{u-d}
	  \ln\LT(\frac{u}{d}\RT) - F'(u) - F'(d) \RT] \\
	& - N_c \LT( \frac{4}{9} s^2 + \frac{1}{9} \RT) \LT\{
	  \LT[ 2 u F(u) - (1+2u) F'(u) \RT] +
	  \LT[ 2 d F(d) - (1+2d) F'(d) \RT] \RT\} \\
	& + \frac{16}{9} N_c s^4 \LT\{
	  Q_U^2 \LT[ 2 u F(u) - (1+2u) F'(u) \RT] +
	  Q_D^2 \LT[ 2 d F(d) - (1+2d) F'(d) \RT] \RT\} \\
	& - \frac{4}{9} N_c Y s^2 \LT\{
	  \LT[ 2 u F(u) - (1+2u) F'(u) \RT] -
	  \LT[ 2 d F(d) - (1+2d) F'(d) \RT] \RT\},
    \end{split} \\[\baselineskip]
    \begin{split} \label{eq:GEN.20r}
	\delta V_R^{q,l} 
	& = -\frac{2}{3} N_c \LT[ u F(u) + d F(d) + \frac{ud}{u-d}
 	  \ln\LT(\frac{u}{d}\RT) - \frac{u+d}{2} \RT] \\
	& + \frac{16}{9} N_c s^2 c^2 \LT\{
	  Q_U^2 \LT[ (1+2u) F(u) - \frac{1}{3} \RT] +
	  Q_D^2 \LT[ (1+2d) F(d) - \frac{1}{3} \RT] \RT\} \\
	& - \frac{2 Y}{9} N_c \LT\{
	  (1+2u) F(u) - (1+2d) F(d) - \ln\LT(\frac{u}{d}\RT) \RT\},
    \end{split}
\end{align}
where the expressions for $F(m_W^2, m_U^2, m_D^2)$, $F(x)$ and $F'(x)$ are
given in Eqs.~(\ref{eq:APP.40w}-\ref{eq:APP.40p}).

In the asymptotic limit (denoted by prime) where the extra generation
particles are much heavier than $Z$-boson ($u, d\gg 1$), power suppressed
terms can be neglected and Eqs.~(\ref{eq:GEN.20m}-\ref{eq:GEN.20r}) reduce
to:
\begin{align}
    \begin{split} \label{eq:GEN.30m}
	\delta' V_m^{q,l} 
	& = \delta' V_A^{q,l} - \frac{2}{9} N_c 
	  + \frac{4}{9} N_c Y s^2 \ln\LT(\frac{u}{d}\RT) 
	  - \frac{4}{9} N_c \LT( c^2 - s^2 \RT) \cdot \\
	& \cdot \LT\{
	  \frac{1}{3} - 2 \frac{ud}{\LT( u - d \RT)^2}
	  - \frac{u^3 - 3u^2d - 3ud^2 + d^3}{2 \LT( u - d \RT)^3}
	  \ln\LT(\frac{u}{d}\RT) \RT\},
    \end{split} \\[0.5\baselineskip]
    \label{eq:GEN.30a} \delta' V_A^{q,l} & = \frac{1}{3} N_c
      \LT[ u + d - 2\frac{ud}{u-d} \ln\LT(\frac{u}{d}\RT) \RT], 
      \\[0.5\baselineskip]
    \label{eq:GEN.30r} \delta' V_R^{q,l} & = \delta' V_A^{q,l}
      - \frac{2}{9} N_c + \frac{2}{9} N_c Y
      \ln\LT(\frac{u}{d}\RT).
\end{align}

A stronger approximation can be used when the mass differences $\LT| m_N -
m_E \RT|$ and $\LT| m_U - m_D \RT|$ are small with respect to the masses of
the extra particles, i.e.\ the total amount of $SU(2)_V$ breaking is not too
large. In this case, further terms can be omitted from
Eqs.~(\ref{eq:GEN.30m}-\ref{eq:GEN.30r}) and the resulting expressions are
almost trivial:
\begin{align}
    \label{eq:GEN.40m} \delta'' V_m^{q,l} & = \delta'' V_A^{q,l}
      - \frac{4}{9} N_c s^2 \LT[ 1 - Y \ln\LT(\frac{u}{d}\RT) \RT], \\
    \label{eq:GEN.40a} \delta'' V_A^{q,l} & = 
      \frac{4}{9} N_c \LT[ \sqrt{u} - \sqrt{d} \RT]^2, \\
    \label{eq:GEN.40r} \delta'' V_R^{q,l} & = \delta' V_A^{q,l}
      - \frac{2}{9} N_c \LT[ 1 - Y \ln\LT(\frac{u}{d}\RT) \RT].
\end{align}
From these formulas we can see that the $SU(2)_V$ breaking terms are
universal, i.e.\ their contribution is the same for all the three $V_i$
functions. Also, they are even under the exchange $u \leftrightarrow d$,
while terms proportional to the doublet hypercharge $Y$ are odd.

In the opposite case, when $SU(2)_V$ is strongly violated ($m_U \gg m_D$ or
$m_U \ll m_D$), Eqs.~(\ref{eq:GEN.40m}-\ref{eq:GEN.40r}) are no longer
valid. However, also in this limit the $SU(2)_V$ breaking terms are
universal and even under $u \leftrightarrow d$:
\begin{equation} \label{eq:GEN.50}
    \delta' V_i = \frac{1}{3} N_c \LT| u - d \RT|.
\end{equation}

In order to obtain the contributions of the whole generation, one should sum
those of quarks and leptons, as shown in Eq.~\eqref{eq:GEN.10}. Looking
at~\eqref{eq:GEN.5}, it is easy to see that in the ``horizontally
degenerate'' case ($m_N = m_U$ and $m_E = m_D$) terms proportional to the
doublet hypercharge cancel between quarks and leptons and the overall result
is:
\begin{equation} \label{eq:GEN.60}
    \LT.
    \begin{aligned}
	\Delta'' V_m & = -\frac{16}{9} s^2 \\
	\Delta'' V_A & = \hspace{4mm} 0 \\
	\Delta'' V_R & = -\frac{8}{9}
    \end{aligned} \RT\} 
    + \frac{16}{9} \LT( \frac{m_{N,U} - m_{E,D}}{m_Z} \RT)^2.
\end{equation}

In the present chapter we consider the general case described \emph{not} by
asymptotic formulas~(\ref{eq:GEN.30m}-\ref{eq:GEN.30r}), but by general
formulas~(\ref{eq:GEN.20m}-\ref{eq:GEN.20r}), which will allow us to study
the case of ``light'' new neutrinos ($m_N \approx m_Z/2$). Effects of extra
generations with masses heavier than $m_Z$ were already analyzed in
Refs.~\cite{Inami95,Masiero95,Novikov95} using asymptotic
expressions~(\ref{eq:GEN.30m}-\ref{eq:GEN.30r}); however, with new
experimental data the bounds we get are much more restrictive than those
obtained in~\cite{Novikov95}. Concerning
Eqs.~(\ref{eq:GEN.40m}-\ref{eq:GEN.40r}), they are so simple that using them
it will be possible to understand qualitatively many aspects of the
numerical results produced by fits: this is the reason why we presented them
here.

\begin{figure}[!p] \centering
    \includegraphics[width=0.8\textwidth]{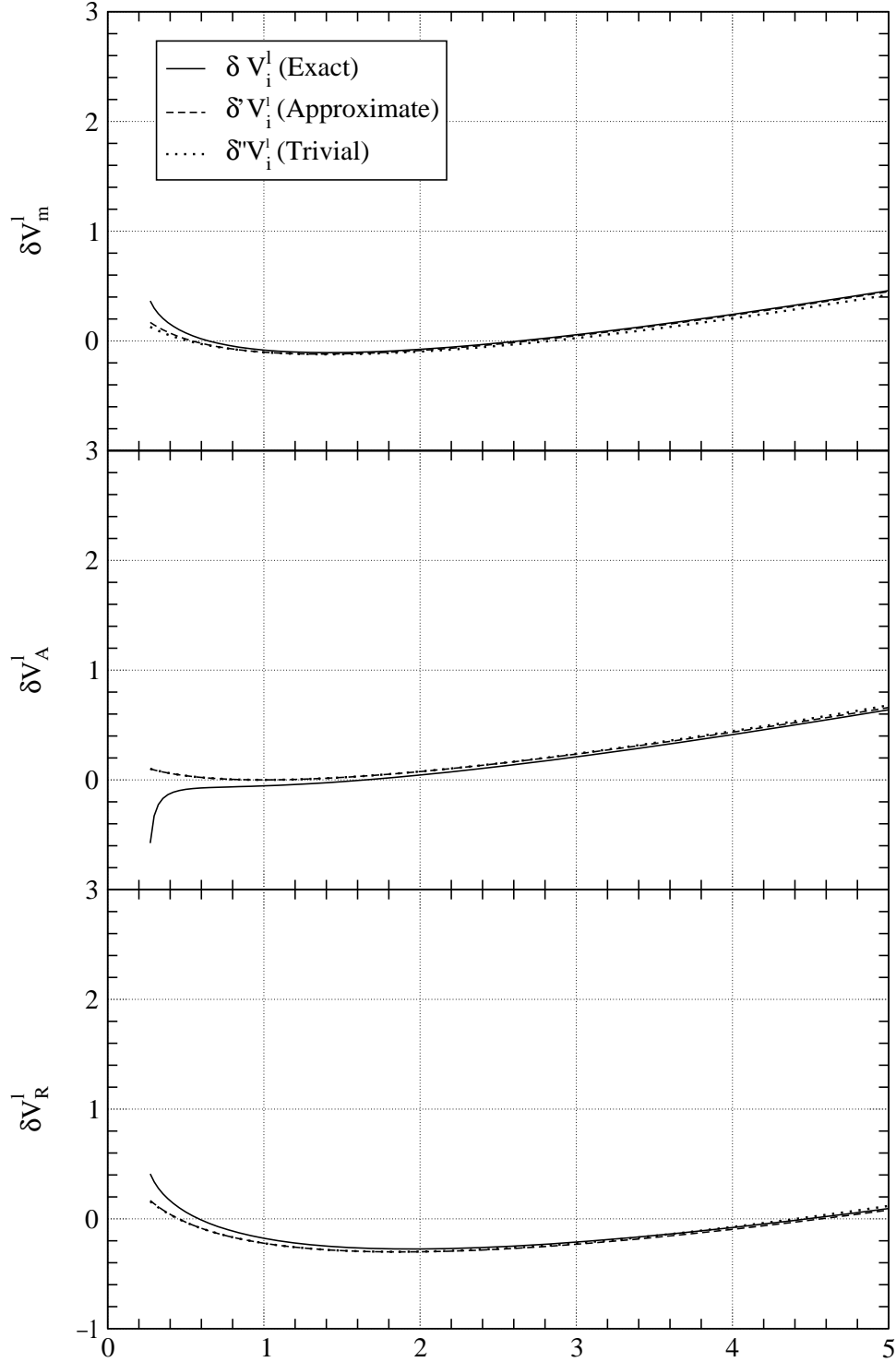}
    \caption[Plot of $\delta V_i^l$ as a function of $(m_N/m_Z)^2$]{
      Contributions of an extra lepton doublet to $\delta V_i^l$ as a
      function of $u \equiv (m_N/m_Z)^2$. We assume $m_E = m_Z$. Solid lines
      correspond to exact formulas~(\ref{eq:GEN.20m}-\ref{eq:GEN.20r}),
      dashed lines to approximated
      expressions~(\ref{eq:GEN.30m}-\ref{eq:GEN.30r}), and dotted lines to
      even more approximated relations~(\ref{eq:GEN.40m}-\ref{eq:GEN.40r}).}
    \label{fig:GEN.100}
\end{figure}

\begin{figure}[!p] \centering
    \includegraphics[width=0.8\textwidth]{Fig.generations/graph_Qrk_Y.eps}
    \caption[Plot of $\delta V_i^q$ as a function of $(m_U/m_Z)^2$]{
      Contributions of an extra quark doublet to $\delta V_i^q$ as a
      function of $u \equiv (m_U/m_Z)^2$. We assume $m_D = m_Z$. Solid lines
      correspond to exact formulas~(\ref{eq:GEN.20m}-\ref{eq:GEN.20r}),
      dashed lines to approximated
      expressions~(\ref{eq:GEN.30m}-\ref{eq:GEN.30r}), and dotted lines to
      even more approximated relations~(\ref{eq:GEN.40m}-\ref{eq:GEN.40r}).}
    \label{fig:GEN.110}
\end{figure}

\begin{figure}[!p] \centering
    \includegraphics[width=0.8\textwidth]{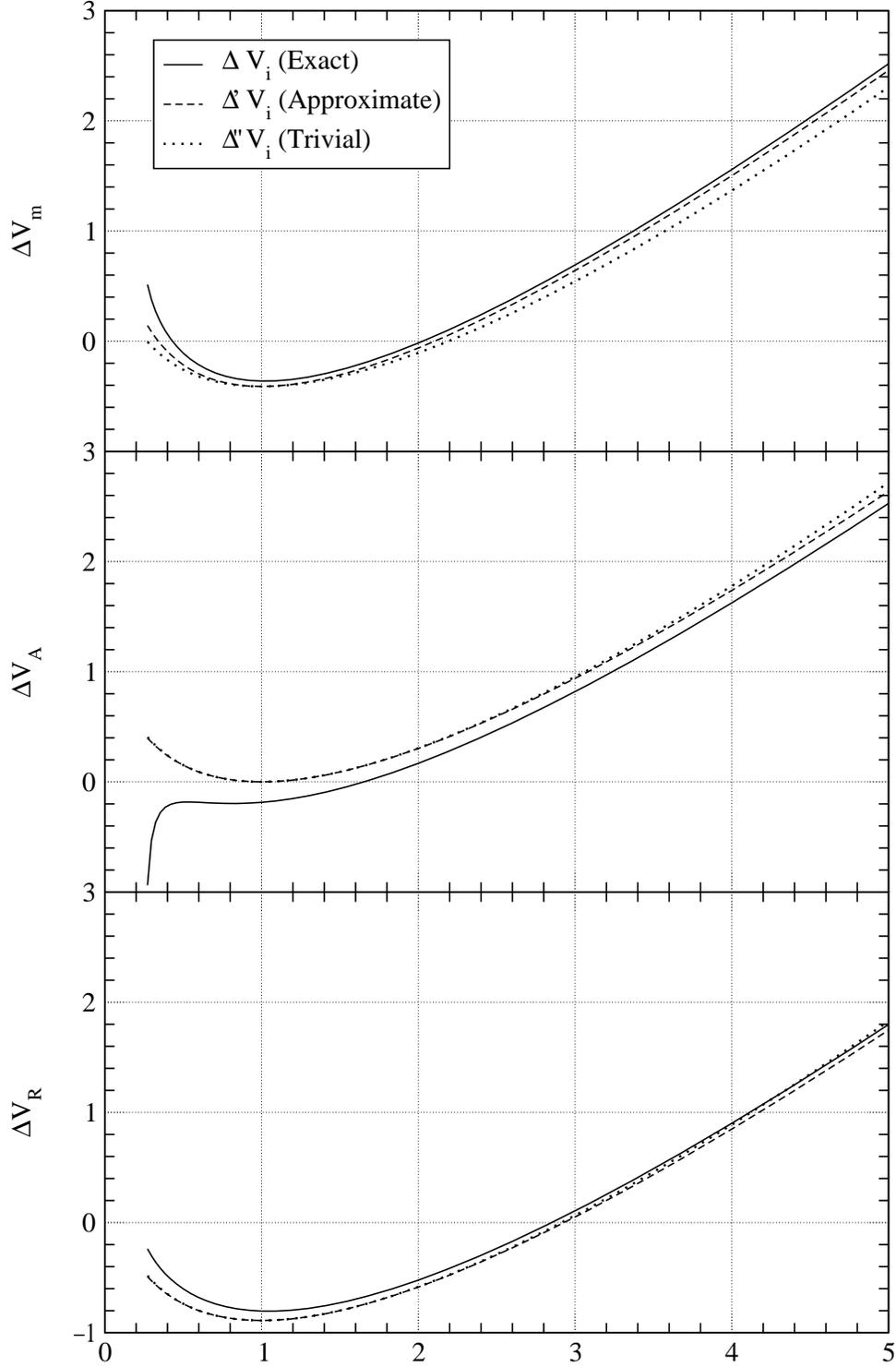}
    \caption[Plot of $\Delta V_i$ as a function of $(m_{N,U}/m_Z)^2$]{
      Contributions of an extra generation of fermions to $\Delta V_i$ as a
      function of $u \equiv (m_{N,U}/m_Z)^2$. We assume $m_N = m_U$ and $m_E
      = m_D = m_Z$. Solid lines correspond to exact
      formulas~(\ref{eq:GEN.20m}-\ref{eq:GEN.20r}), dashed lines to
      approximated expressions~(\ref{eq:GEN.30m}-\ref{eq:GEN.30r}), and
      dotted lines to even more approximated
      relations~(\ref{eq:GEN.40m}-\ref{eq:GEN.40r}). These plots help to
      study accuracy of approximations~(\ref{eq:GEN.30m}-\ref{eq:GEN.40r})
      \emph{outside} its formal domain of validity, that is why we neglect
      experimental bounds on $m_U$ and $m_D$.}
    \label{fig:GEN.120}
\end{figure}

In Figs.~\ref{fig:GEN.100}-\ref{fig:GEN.120} we show the $u$ dependence of
the functions $\delta V_i$, $\delta' V_i$ and $\delta'' V_i$ for $d=1$, for
the case of leptons alone (Fig.~\ref{fig:GEN.100}), quarks alone
(Fig.~\ref{fig:GEN.110}), and horizontally degenerate quarks and leptons
(Fig.~\ref{fig:GEN.120}, see also Eqs.~(\ref{eq:GEN.80m}-\ref{eq:GEN.80r})).
It is clear that accuracy of equations~(\ref{eq:GEN.30m}-\ref{eq:GEN.30r})
is very good as soon as new fermions are heavier than $Z$-boson. Also, we
see that Eqs.~(\ref{eq:GEN.40m}-\ref{eq:GEN.40r}) perfectly approximate the
exact functions if the value of $\LT| m_U - m_D \RT|$ or $\LT| m_N - m_E
\RT|$ is small enough. For larger $SU(2)_V$ breaking, corresponding to the
(not shown) region $u \gg 1$ in Figs.~\ref{fig:GEN.100}-\ref{fig:GEN.120},
there is a discrepancy; however, in this limit the contributions to $\delta
V_i$ which are correctly described by Eq.~\eqref{eq:GEN.50} are so large
that the quality of $\chi^2$ is awful. Therefore this region is not
interesting for our analysis.

Since in the rest of this chapter we will frequently assume horizontal
degeneracy between lepton and quark doublets, let us conclude this section
giving explicit formulas for this case:
\begin{align}
    \begin{split} \label{eq:GEN.80m}
	\Delta V_m
	& = \frac{16}{9} s^2 \LT( \frac{4}{9} s^2 - 1 \RT)
	  \LT\{ (1+2u) F(u) + (1+2d) F(d) - \frac{2}{3} \RT\} \\
	& + \frac{8}{9} \LT\{ (1-u) F(u) + (1-d) F(d) - \frac{2}{3} \RT\}
	  + \frac{4}{3} \frac{s^2}{c^2} 
	  \LT\{ u + d - 2\frac{ud}{u-d}\ln\LT(\frac{u}{d}\RT) \RT\} \\
	& + \frac{8}{9} \LT(1 - \frac{s^2}{c^2} \RT)
	  \LT\{ \frac{u-d}{2}\ln\LT(\frac{u}{d}\RT) + u + d
	  + \LT( c^2 - \frac{u+d}{2} \RT) 
	  \frac{u+d}{u-d} \ln\LT(\frac{u}{d}\RT) - \frac{4}{3} c^2 \RT\} \\
	& - \frac{8}{9} \LT(1 - \frac{s^2}{c^2} \RT)
	  \LT[ 2 c^2 - u - d - \frac{(u-d)^2}{c^2} \RT]
	  F(m_W^2, m_U^2, m_D^2),
    \end{split} \\[8pt]
    \begin{split} \label{eq:GEN.80a}
	\Delta V_A 
	& = \frac{4}{9} \LT( \frac{16}{3} s^4 - 4 s^2 - 1 \RT) \bigg\{\!
	  \LT[ 2u F(u) - (1+2u) F'(u) \RT] +
	  \LT[ 2d F(d) - (1+2d) F'(d) \RT] \!\bigg\} \\
	& + \frac{4}{3} \LT\{ u + d - 2\frac{ud}{u-d} \ln\LT(\frac{u}{d}\RT) 
	  - F'(u) - F'(d) \RT\},
    \end{split} \\[8pt]
    \begin{split} \label{eq:GEN.80r}
	\Delta V_R
	& = - \frac{8}{3} \LT\{ u F(u) + d F(d) + \frac{ud}{u-d}
	  \ln\LT(\frac{u}{d}\RT) - \frac{u+d}{2} \RT\} \\
	& + \frac{64}{27} s^2 c^2 \LT\{ (1+2u) F(u) + (1+2d) F(d)
	  - \frac{2}{3} \RT\}.
    \end{split}
\end{align}
It is worth noting that these expressions are exactly even under the
exchange $u \leftrightarrow d$. This happens since the cancellation of odd
terms proportional to the doublet hypercharge $Y$ between leptons and
quarks, which we noted when deriving the approximate
formula~\eqref{eq:GEN.60}, occurs also for the exact
relations~(\ref{eq:GEN.20m}-\ref{eq:GEN.20r}).

\section{Comparison with experimental data: heavy fermions}
\label{sec:GEN.20}

\begin{figure}[!t] \centering
    \includegraphics[width=0.9\textwidth]{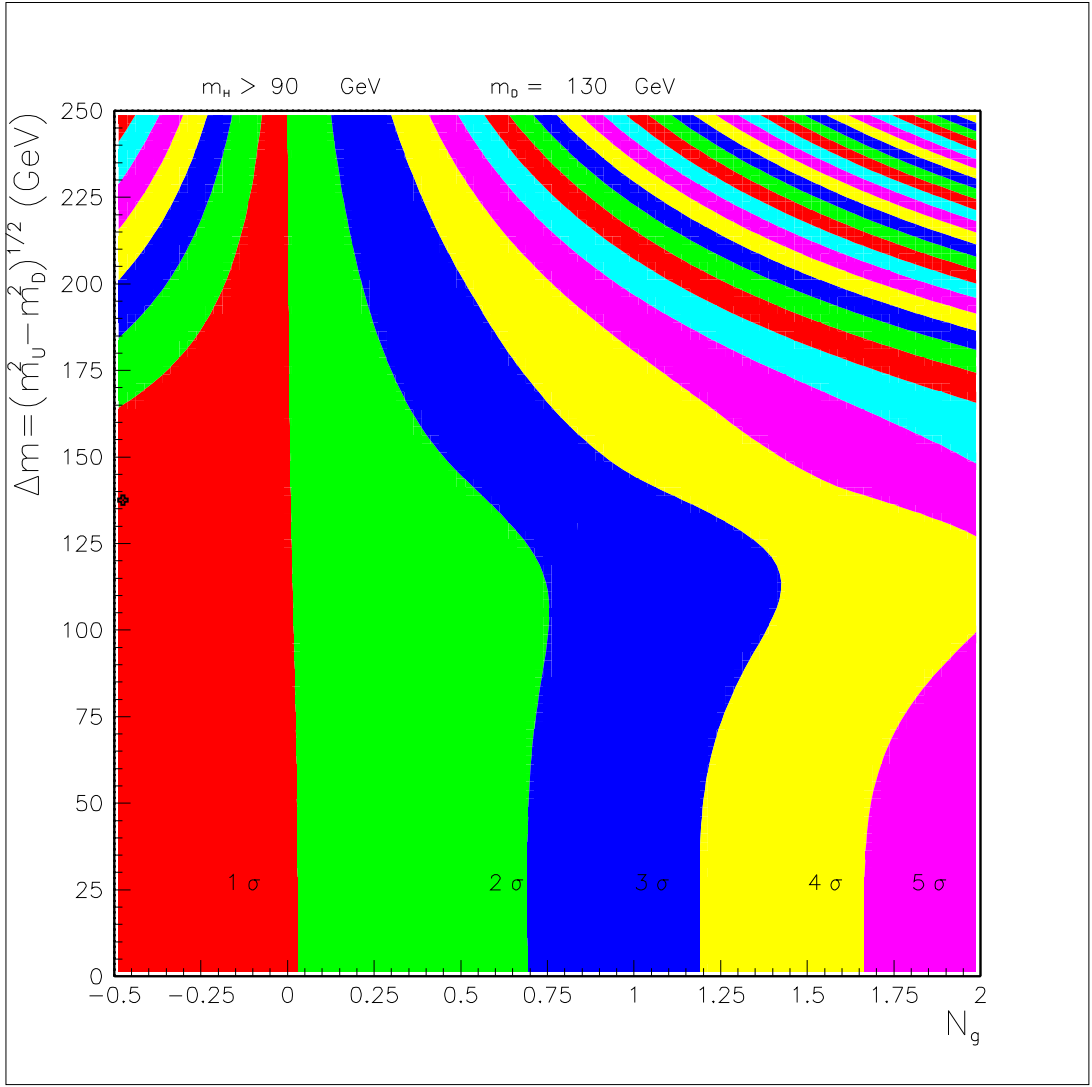}
    \caption[Exclusion plot in the plane $(N_g, \Delta m)$ (horizontal
    degeneracy)]{
      Constraints on the number of extra generations $N_g$ and the mass
      difference $\Delta m$ in the horizontally degenerate case: $(m_N =
      m_U) > (m_E = m_D = 130\GeV)$. The lower bound $m_{U,D} > 130\GeV$
      comes from Tevatron search~\cite{PDG98}. All electroweak precision
      data and $m_H > 90\GeV$ at $95\pCL$~\cite{LEPII99} were used in the
      fit. The cross corresponds to $\chi^2$ minimum; regions show $<
      1\sigma$, $< 2\sigma$, etc.\ allowed domains.}
    \label{fig:GEN.200}
\end{figure}

\begin{figure}[!t] \centering
    \includegraphics[width=0.9\textwidth]{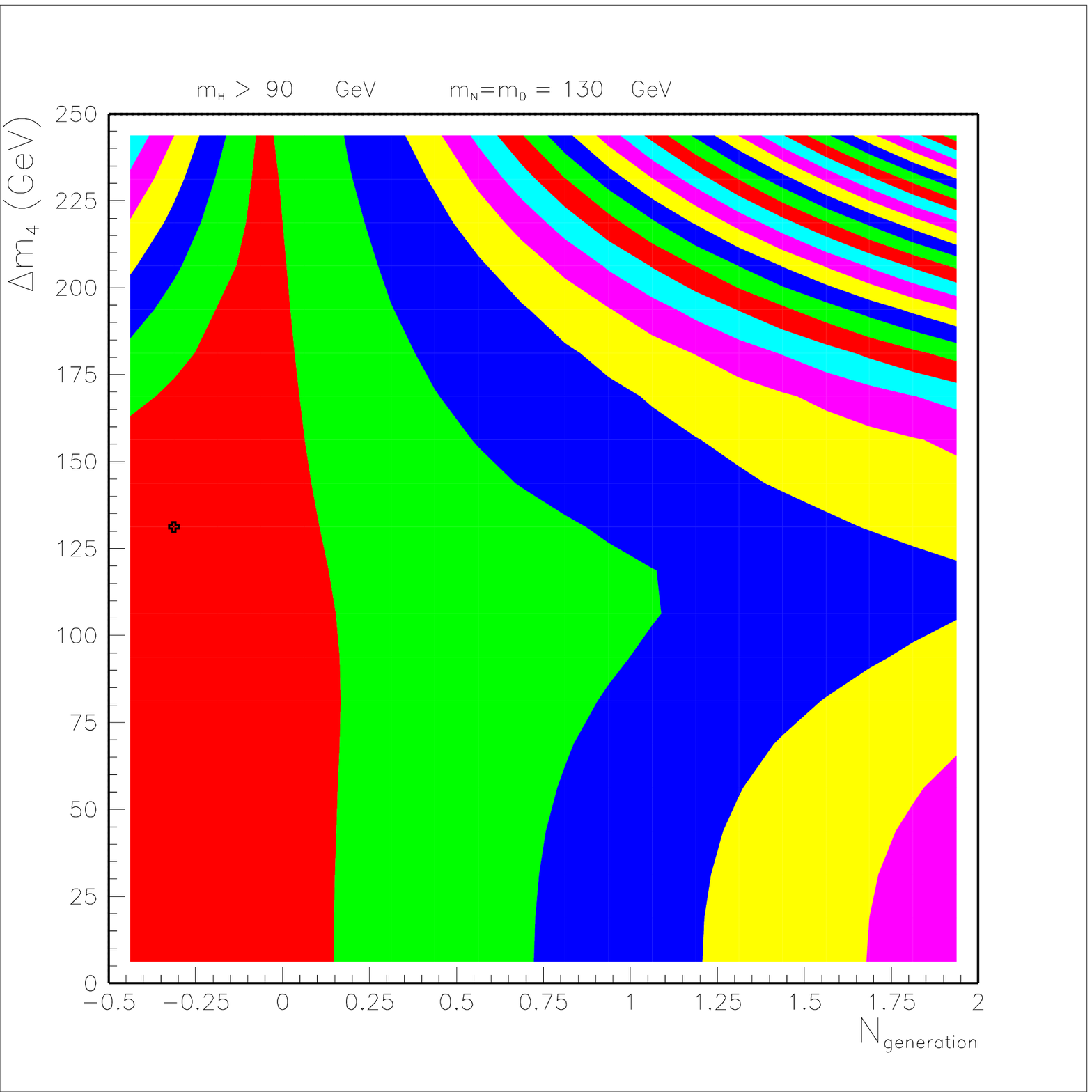}
    \caption[Exclusion plot in the plane $(N_g, \Delta m)$ (cross
    degeneracy)]{
      Constraints on the number of extra generations $N_g$ and the mass
      difference $\Delta m$ in the cross degenerate case: $(m_E = m_U) >
      (m_N = m_D = 130\GeV)$. The lower bound $m_{U,D} > 130\GeV$ comes from
      Tevatron search~\cite{PDG98}. All electroweak precision data and $m_H
      > 90\GeV$ at $95\pCL$~\cite{LEPII99} were used in the fit. The cross
      corresponds to $\chi^2$ minimum; regions show $< 1\sigma$, $<
      2\sigma$, etc.\ allowed domains.}
    \label{fig:GEN.210}
\end{figure}

\begin{figure}[!t] \centering
    \includegraphics[width=0.9\textwidth]{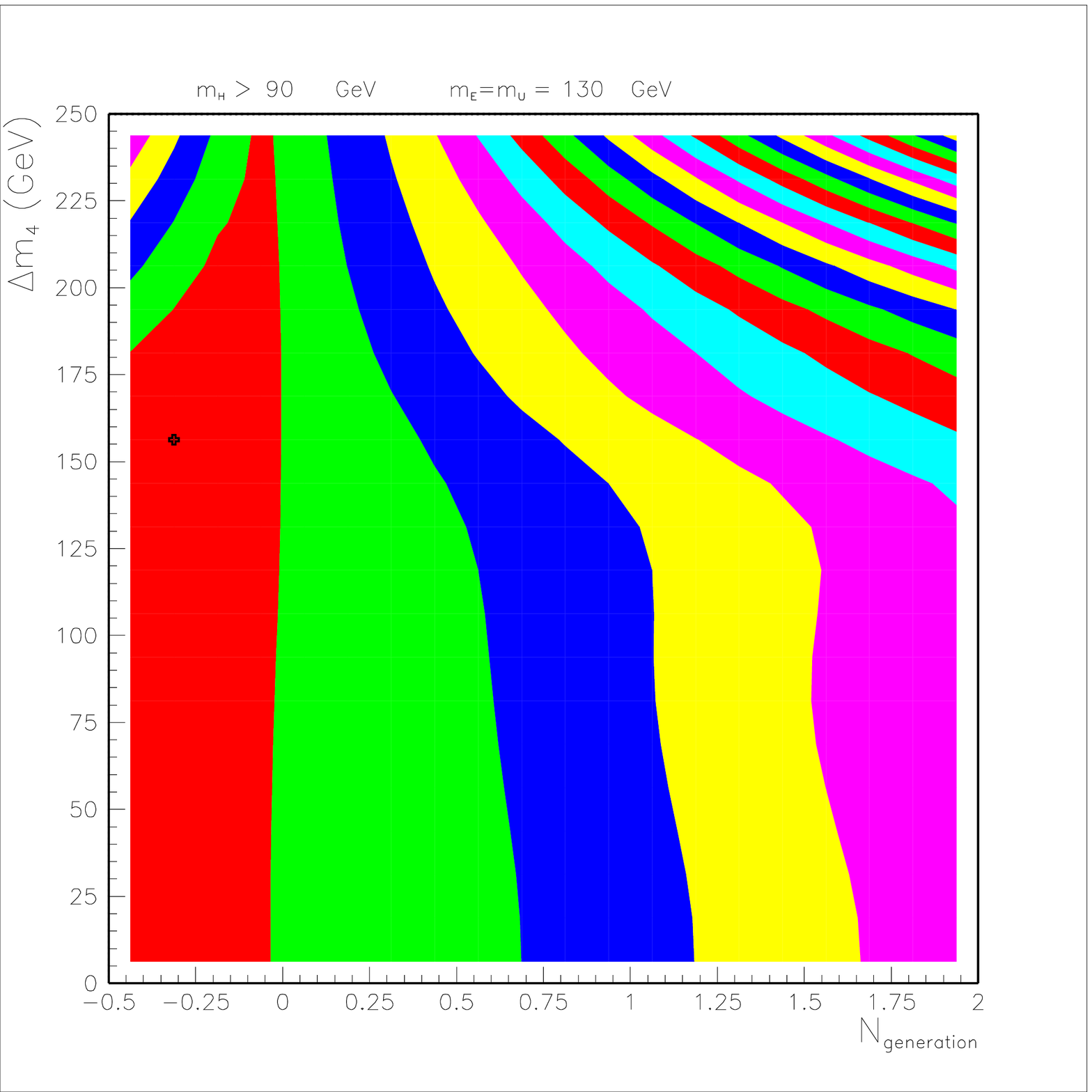}
    \caption[Exclusion plot in the plane $(N_g, \Delta m)$ (anti-cross
    degeneracy)]{
      Constraints on the number of extra generations $N_g$ and the mass
      difference $\Delta m$ in the anti-cross degenerate case: $(m_N = m_D)
      > (m_E = m_U = 130\GeV)$. The lower bound $m_{U,D} > 130\GeV$ comes
      from Tevatron search~\cite{PDG98}. All electroweak precision data and
      $m_H > 90\GeV$ at $95\pCL$~\cite{LEPII99} were used in the fit. The
      cross corresponds to $\chi^2$ minimum; regions show $< 1\sigma$, $<
      2\sigma$, etc.\ allowed domains.}
    \label{fig:GEN.220}
\end{figure}

\begin{figure}[!t] \centering
    \includegraphics[width=0.9\textwidth]{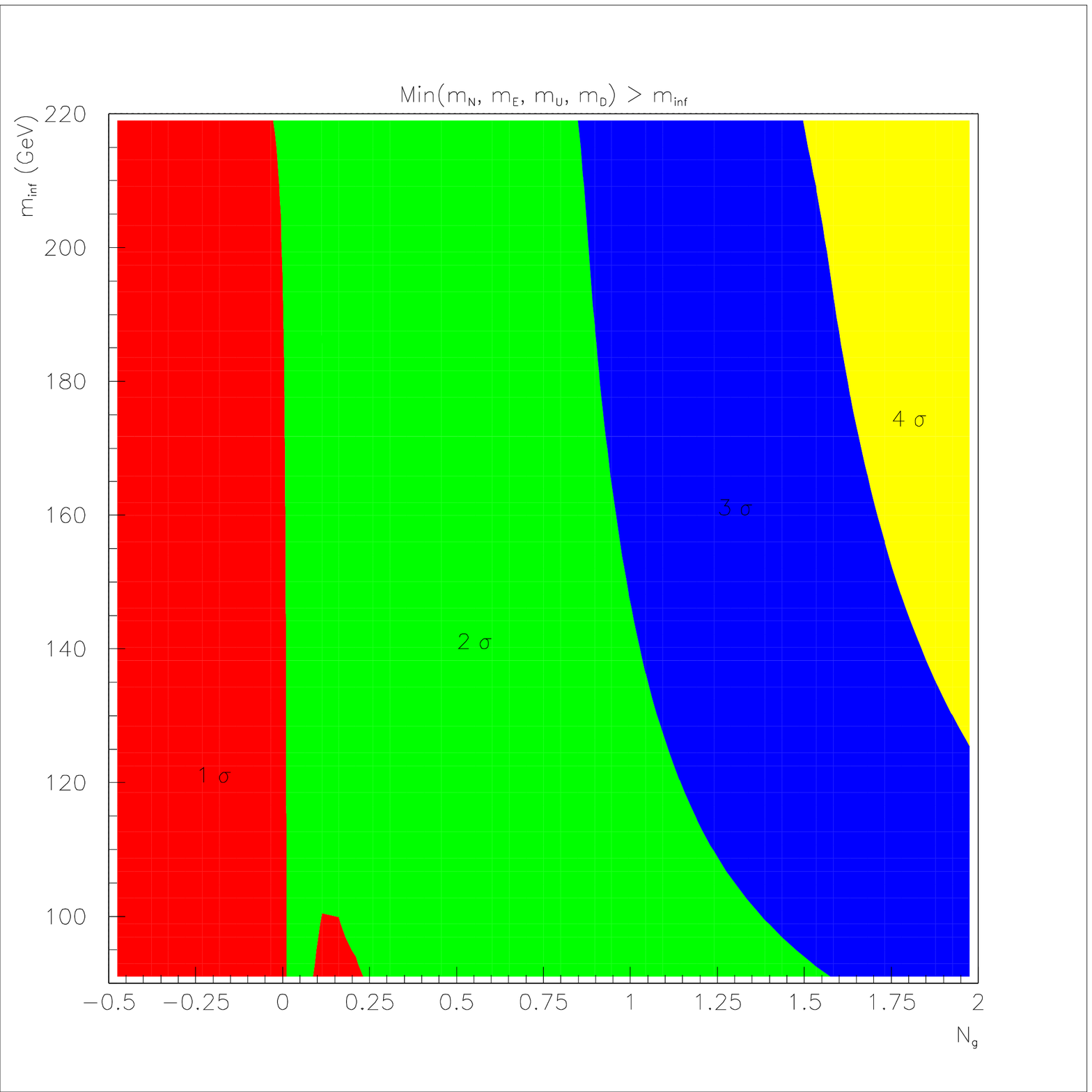}
    \caption[Exclusion plot in the plane $(N_g, m_\mathrm{inf})$]{
      Constraints on the number of extra generations $N_g$ ($x$-axis) as a
      function the lower bound on new fermion mass $m_\mathrm{inf}$
      ($y$-axis). The $\chi^2$ is evaluated using the approximate
      expression~\eqref{eq:GAU.310}; the masses of the four extra fermions
      $m_N$, $m_E$, $m_U$, $m_D$ and the top mass $m_t$ are integrated out.
      For each point $(N_g, m_\mathrm{inf})$ this plot shows the best value
      of the $\chi^2$ which can be obtained having $N_g$ extra generations
      and bounding all the new fermion masses to be heavier than
      $m_\mathrm{inf}$.}
    \label{fig:GEN.230}
\end{figure}

We compare theoretical predictions for the case of extra generations with
experimental data using the computer program \texttt{LEPTOP} (see
Refs.~\cite{NOV95b,NORV99}). We take $m_D = 130\GeV$~- the lowest value
allowed for the new quark mass from Tevatron search~\cite{PDG98}~- and
assume $m_U \gtrsim m_D$. As for the extra leptons, their masses are
independent parameters. To simplify the analysis, we start with the
horizontally degenerate case, setting $m_N = m_U$ and $m_E = m_D$. Any value
of higgs mass above $90\GeV$ is allowed in our fits, however $\chi^2$
appears to be minimal for $m_H = 90\GeV$. In Fig.~\ref{fig:GEN.200} the
excluded domains in coordinates ($N_g$, $\Delta m$) are shown (here $\Delta
m = (m_U^2 - m_D^2)^{1/2}$). Minimum of $\chi^2$ corresponds to $N_g = -0.5$
and the case $N_g=0$ is in the $1$ standard deviation domain. We see that
one extra generation corresponds to $2.5\sigma$ approximately. The behavior
of division lines in Fig.~\ref{fig:GEN.200} can be understood qualitatively
looking at Eq.~\eqref{eq:GEN.60}. For degenerate extra generations the
corrections $\Delta V_i$ are negative. They become positive and large when
$\Delta m$ increases: this is why at large $\Delta m$ division lines
approach $N_g=0$ value. In the intermediate region ($\Delta m \approx
125\GeV$) $\Delta V_i$ cross zero and this explains the turn to the right of
the division lines. However, for different $i$ zero is reached for different
$\Delta m$ values, that is why extra generations are excluded even for
$\Delta m \approx 125\GeV$ (see Ref.~\cite{Novikov95}).

At this point, we have to check whether similar bounds are valid for the
general choice of heavy fermion masses or they are peculiar to the special
case shown in Fig.~\ref{fig:GEN.200}. To see this, one should perform a
general scan of the whole 4-dimensional parameter space ($m_N$, $m_E$,
$m_U$, $m_D$), integrating out fermion masses and looking at the overall
$\chi^2$. However, to do this accurately with \texttt{LEPTOP}, taking
properly into account effects of changing $m_t$, $m_H$ and other SM
parameters, one would require a huge amount of computer time, so this way is
precluded.

To overcome this problem, as a first step we analyzed two more cases which
are somewhat complementary to Fig.~\ref{fig:GEN.200}: the ``cross
degenerate'' case $(m_E = m_U) > (m_N = m_D = 130\GeV)$ and the ``anti-cross
degenerate'' case $(m_N = m_D) > (m_E = m_U = 130\GeV)$. Results are shown
in Figs.~\ref{fig:GEN.210} and~\ref{fig:GEN.220}, respectively. It is
immediate to see that, although in the cross degenerate case the $\chi^2$
is slightly better that in the horizontally degenerate one and bounds on a
fourth generation are consequently weaker, nevertheless the case of an extra
heavy family is excluded almost at $2\sigma$. On the other hand, the
anti-cross degenerate case is strongly disfavored, and this shows that the
mass hierarchy preferred by precision measurements for a new fermion
generation is the same as in the third SM generation. The reason for this
will be explained later.

As already stated, to go beyond these results one need to perform a full
scan of the whole 4-dimensional parameter space. Although this is currently
impossible with \texttt{LEPTOP}, it can be done easily if the $\chi^2$ is
evaluated by means of the approximate expression~\eqref{eq:GAU.310}
introduced in the previous chapter (see Sec.~\ref{sec:GAU.30}). The results
of this search are summarized in Fig.~\ref{fig:GEN.230}, which also includes
effects of changing the top mass (but not the Higgs mass or other SM
parameters). On the $x$-axis we have the number of new generations, as
usual, while on the $y$-axis we report the lower bound $m_\mathrm{inf}$
imposed on the four extra fermion masses (for example: the point $(1; 150)$
corresponds to the minimum value of $\chi^2$ which can be obtained for the
case of $1$ extra generation when all fermion masses are heavier than
$150\GeV$). To compare this figure with the previous ones, we should look at
the horizontal line $m_\mathrm{inf}=130\GeV$ in Fig.~\ref{fig:GEN.230}
(since this is the constraint imposed on all $m_f$ in
Figs.~\ref{fig:GEN.200}-\ref{fig:GEN.220}) and extract a bound for the
number of allowed extra generations. If we do this, we see that $N_g = 1$ is
excluded almost at $2\sigma$, just as it was in the cross degenerate case:
this proves that leaving all the four fermion masses free to vary
independently from one another does not help to improve the quality of the
fit beyond the case of Fig.~\ref{fig:GEN.210}, and the general bound $m_f >
130\GeV$ is enough to exclude the possibility of a fourth generation at
$95\%$~C.L.

Further considerations can be derived qualitatively by means of
Eqs.~(\ref{eq:GEN.40m}-\ref{eq:GEN.40r}). In the rest of this section we
will use these equations to investigate the ``best fit point'' $(\bar{m}_N,
\bar{m}_E, \bar{m}_U, \bar{m}_D)$, showing that many of its properties can
be understood analytically. This is useful to get an idea of how much the
results found in the special cases shown in
Figs.~\ref{fig:GEN.200}-\ref{fig:GEN.220} are representative of the general
choice of extra fermion masses.

Using values from~\eqref{eq:GEN.5}, Eqs.~(\ref{eq:GEN.40m}-\ref{eq:GEN.40r})
can be rewritten in the following way:
\begin{gather}
    \begin{align}
	\label{eq:GEN.200l} \delta m_l & = m_E - m_N, \\
	\label{eq:GEN.200q} \delta m_q & = m_U - m_D,
    \end{align} \\
    \begin{align}
	\label{eq:GEN.200x} x & = \frac{1}{2} \LT[ 
	  \ln\LT( 1 + \frac{\delta m_l}{m_N} \RT) + 
	  \ln\LT( 1 + \frac{\delta m_q}{m_D} \RT) \RT], \\
	\label{eq:GEN.200y} y & = \frac{4}{9} \LT[ 
	  \LT( \frac{\delta m_l}{m_Z} \RT)^2 + 3 
	  \LT( \frac{\delta m_q}{m_Z} \RT)^2 \RT],
    \end{align} \\
    \label{eq:GEN.200v} \LT.
    \begin{aligned}
	\Delta'' V_m & = -\frac{16}{9} s^2 \\
	\Delta'' V_A & = \hspace{4mm} 0 \\
	\Delta'' V_R & = -\frac{8}{9}
    \end{aligned} \RT\} \LT( 1 - x \RT) + y.
\end{gather}
We can now insert Eq.~\eqref{eq:GEN.200v} into~\eqref{eq:GAU.310} and use
Eq.~\eqref{eq:GAU.300} to find the ``best values'' $\bar{x}$ and $\bar{y}$:
\begin{equation} \label{eq:GEN.210}
    \bar{x} \approx 1 + \frac{0.344}{N_g}, \qquad
    \bar{y} \approx -\frac{0.289}{N_g}.
\end{equation}
Of course, we are interested only in the region $N_g > 0$. The first thing
to note is that $\bar{x}$ is always positive. This means (see
Eq.~\eqref{eq:GEN.200x}) that \emph{positive} values for $\delta m_l$ and
$\delta m_q$ are preferred, i.e.\ the overall $\chi^2$ is better when the
mass hierarchy within each extra doublet is $m_E > m_N$ and $m_U > m_D$.
This is just what we have in the second and third SM generation, and is in
perfect agreement with Figs.~\ref{fig:GEN.200}-\ref{fig:GEN.220}, where we
already noted that the cross degenerate case gives the best fit while the
anti-cross degenerate gives the worst.

Another relevant fact is that $\bar{y}$ is always negative. Looking at
Eq.~\eqref{eq:GEN.200y}, it is clear that this condition can never be
realized: as a consequence, the best fit point~\eqref{eq:GEN.210} is
physically unreachable, and we can expect in general a rather poor $\chi^2$.
Also, since the quality of the fit is worsened by a large positive value of
$y$, from Eq.~\eqref{eq:GEN.200x} we learn that $\chi^2$ is better when $x$
is ``less suppressed'', because in this way it is simpler to have $x$ large
keeping $y$ small. So we conclude that the best fit point always has $m_N$
and $m_D$ as small as possible. This has a straightforward consequence: as
soon as direct search from accelerator experiments will raise the lower
bound on extra fermion masses, the case of a fourth generation will be even
strongerly excluded by precision measurements. This is clearly visible in
Fig.~\ref{fig:GEN.230}, where the contour lines strictly approach $N_g = 0$
when $m_\mathrm{inf}$ increases.

\begin{figure}[!t] \centering
    \includegraphics[width=0.9\textwidth]{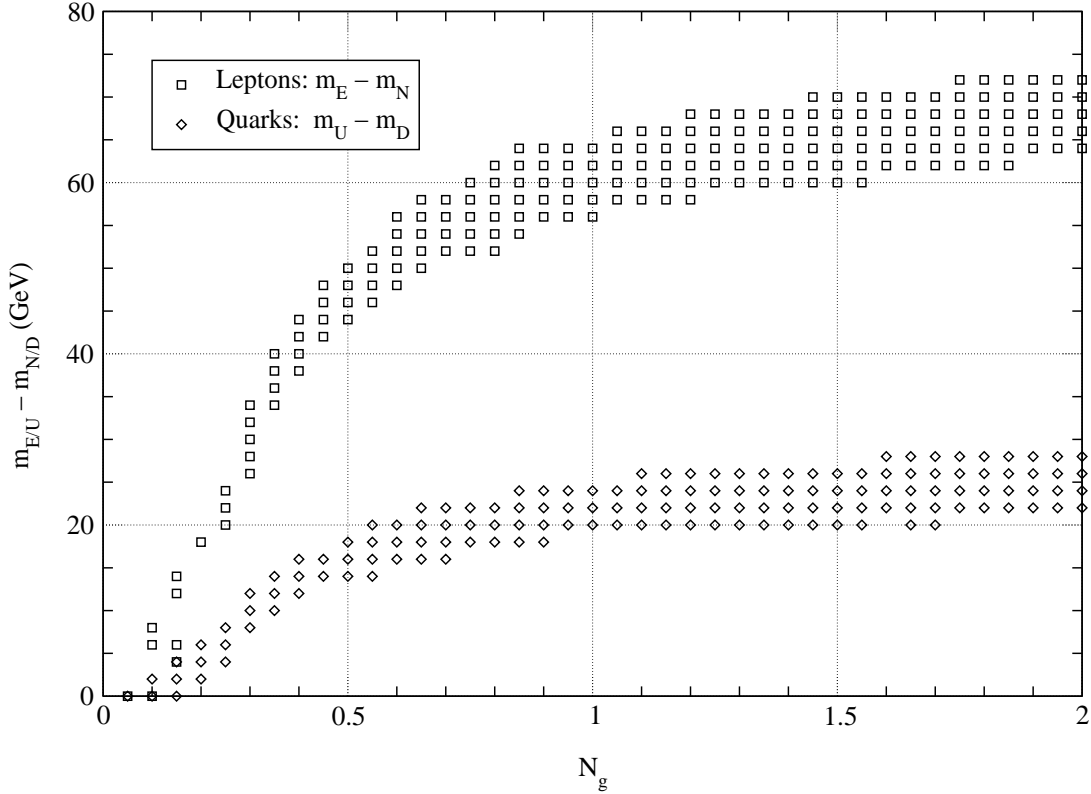}
    \caption[Plot of $m_{E/U} - m_{N/D}$ as a function of $N_g$]{
      Lepton and quark mass splitting $\delta\bar{m}_l$ and
      $\delta\bar{m}_q$ for the best fit point, as a function of the number
      of extra generations $N_g$, for different values of the common mass
      $m_N = m_D = m_\mathrm{inf}$. This figure is derived from the same
      data used to generate Fig.~\ref{fig:GEN.230}; in particular, the
      $\chi^2$ is evaluated using the approximate
      expression~\eqref{eq:GAU.310}.}
    \label{fig:GEN.240}
\end{figure}

The last interesting property of the best fit point $\LT( \delta\bar{m}_l,
\delta\bar{m}_q \RT)$ that we can extract from Eq.~\eqref{eq:GEN.200v} is
more quantitative. Clearly, for any given $m_N$ and $m_D$ the first
derivative of $\chi^2$ with respect to $\delta\bar{m}_l$ and
$\delta\bar{m}_q$ must be zero~- just by definition of ``best fit''. On the
other hand, $\delta m_l$ and $\delta m_q$ enter $\Delta'' V_i$ (and thus
$\chi^2$) \emph{only} through $x$ and $y$, and we know that the only
point~\eqref{eq:GEN.210} where $\chi^2$ is stable under \emph{independent}
variation of $x$ and $y$ is physically unreachable. Therefore, the condition
$\delta \chi^2 = 0$ can only be realized \emph{if} the Jacobian determinant
of the transformation $\LT( \delta m_l, \delta m_q \RT)
\to \LT( x, y \RT)$ is zero:
\begin{equation}
    \begin{vmatrix}
	\dfrac{\partial\LT(x,y\RT)}{\partial\LT(\delta m_l,\delta m_q\RT)}
    \end{vmatrix} =
    \begin{vmatrix} 
	\frac{1}{2 m_E} & \frac{1}{2 m_U} \\[2mm]
	\frac{8}{9} \frac{\delta m_l}{m_Z^2} 
	& \frac{8}{3} \frac{\delta m_q}{m_Z^2}
    \end{vmatrix} 
    = 0.
\end{equation}
After some simple calculation, this gives:
\begin{equation} \label{eq:GEN.220} 
    \delta\bar{m}_l = 3 \frac{\bar{m}_U}{\bar{m}_E} \delta\bar{m}_q.
\end{equation}
If the masses of all the extra fermions are comparable in size, then the
factor $\bar{m}_U / \bar{m}_E$ can be neglected and from
Eq.~\ref{eq:GEN.220} we get that the lepton mass splitting is approximately
three times larger than the quark one. This fact can be immediately verified
by looking at Fig.~\ref{fig:GEN.240}, where we plot the mass differences
$\delta m_l$ and $\delta m_q$ as a function of the number of extra
generations $N_g$: it is straightforward to see that the relation
$\delta\bar{m}_l \approx 3 \delta\bar{m}_q$ is almost always satisfied,
regardless of the number of new generations or of the different value of the
common mass $m_N = m_D$.

Let us conclude this section summarizing the properties of the ``best fit''
$(\bar{m}_N, \bar{m}_E, \bar{m}_U, \bar{m}_D)$ that we found to hold in the
heavy fermion limit:
\begin{itemize}
  \item \emph{electron} is always heavier than \emph{neutrino}, and
    \emph{up-quark} is always heavier than than \emph{down-quark}:
    $\bar{m}_E > \bar{m}_N$, $\bar{m}_U > \bar{m}_D$;

  \item \emph{neutrino} and \emph{down-quark} are as light as possible:
    $\bar{m}_N = m_N^\mathrm{min}$, $\bar{m}_D = m_D^\mathrm{min}$;
    
  \item the lepton mass splitting is approximately three times larger than
    the quark mass splitting: $(\bar{m}_E - \bar{m}_N) \approx 3 (\bar{m}_U
    - \bar{m}_D)$.
\end{itemize}

\section{Comparison with experimental data: $m_N < m_Z$} \label{sec:GEN.30}

\begin{figure}[!t] \centering
    \includegraphics[width=0.9\textwidth]{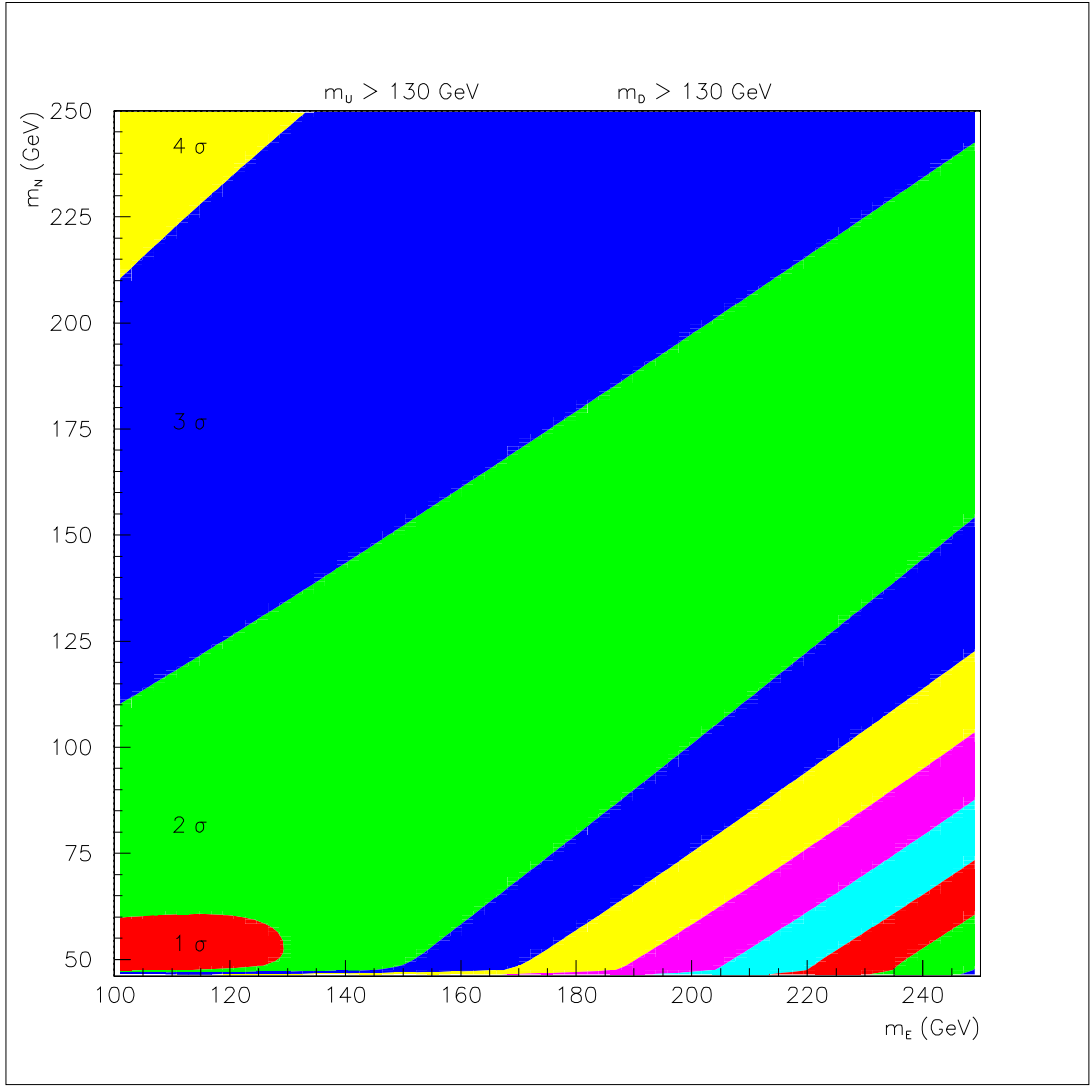}
    \caption[Exclusion plot in the plane $(m_E, m_N)$]{
      Constraints on the masses of the neutral heavy lepton $m_N$ and the
      charged heavy lepton $m_E$. The $\chi^2$ is evaluated using the
      approximate expression~\eqref{eq:GAU.310}. The extra quark masses
      $m_U$ and $m_D$ and the top mass $m_t$ are integrated out, and the
      Tevatron bound $m_q > 130\GeV$ is assumed. Regions show $< 1\sigma$,
      $< 2\sigma$, etc.\ allowed domains.}
    \label{fig:GEN.300}
\end{figure}

\begin{figure}[!t] \centering
    \includegraphics[width=0.9\textwidth]{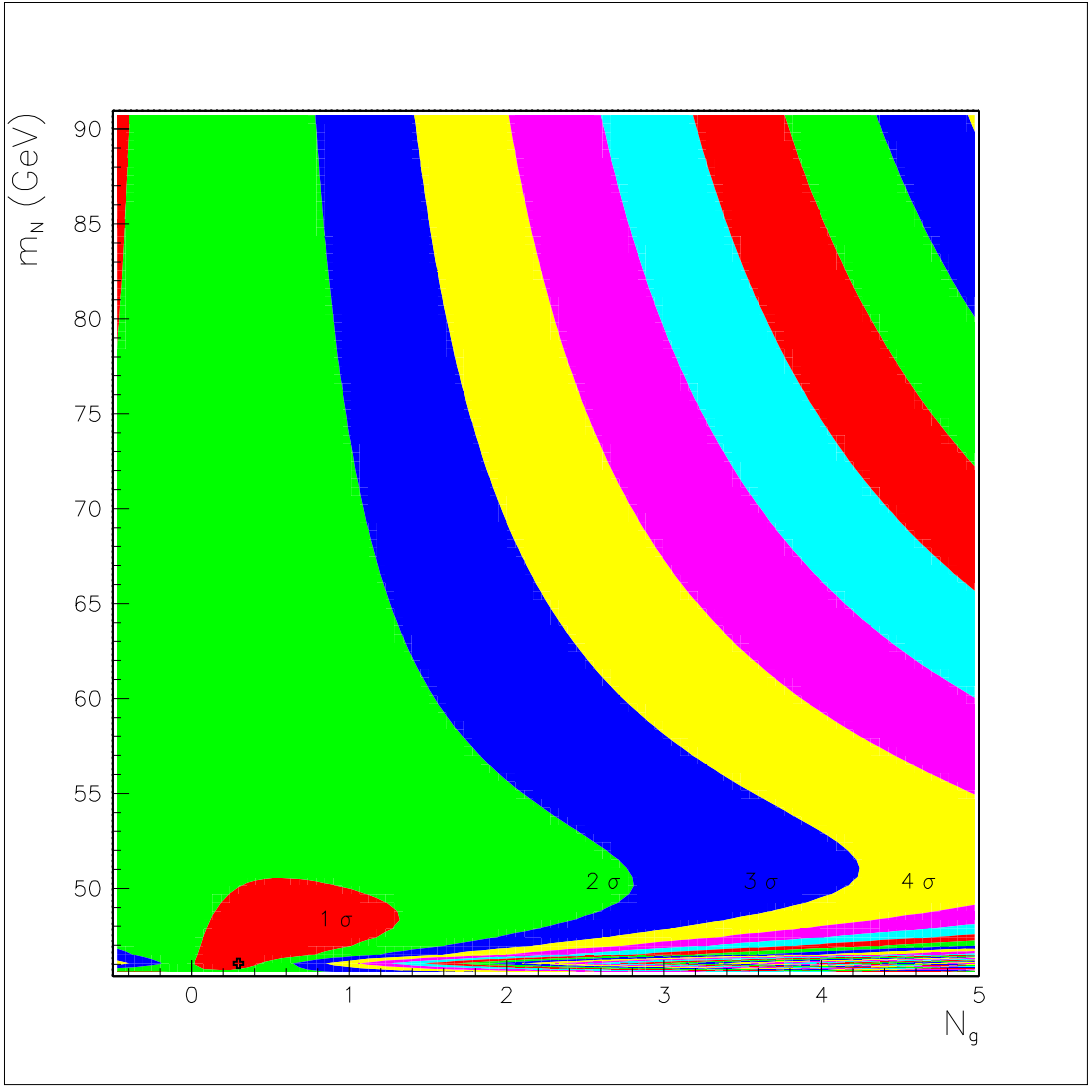}
    \caption[Exclusion plot in the plane $(N_g, m_N)$]{
      Constraints on the number of extra generations $N_g$ and the mass of
      the neutral heavy lepton $m_N$. The values $m_E=100\GeV$,
      $m_U=220\GeV$, $m_D=200\GeV$ were used. All electroweak precision data
      and $m_H > 90\GeV$ at $95\pCL$ from LEP~II~\cite{LEPII99} were used in
      the fit. The cross corresponds to $\chi^2$ minimum; regions show $<
      1\sigma$, $< 2\sigma$, etc.\ allowed domains.}
    \label{fig:GEN.310}
\end{figure}

According to Ref.~\cite{PDG98}, the lower bound on $m_E$ from LEP~II is
approximately $80\GeV$. However, quasi-stable neutral lepton $N$ can be
considerably lighter. From LEP~II searches of the decays $N\to lW^{\ast}$,
where $W^{\ast}$ is virtual while $l$ is $e$, $\mu$ or $\tau$, it follows
that $m_N > 70 \div 80\GeV$ if the mixing angle between $N$ and the three
known neutrinos is larger than $10^{-6}$~\cite{Ackerstaff98}. Thus let us
take in this section this mixing to be less than $10^{-6}$: in this case
only DELPHI bound $m_N > 45\GeV$~\cite{Abreu92} from the measurement of the
$Z$-boson width is applicable. If $m_N$ is larger than $m_Z/2$, searches at
LEP~II of the reaction $e^+ e^- \to N\bar{N}\gamma$ should bound $m_N$. The
observation of a ``lonely photon'' was suggested long time ago as a method
to study cross section of the $e^+e^-$ annihilation into
neutrinos~\cite{Dolgov72}. DELPHI collaboration performed such a search at
$E\leq 183\GeV$ and found that the total number of neutrinos is $N_{\nu} =
2.92 \pm 0.25\mathrm{(stat)} \pm 0.14\mathrm{(syst)}$~\cite{Ferrari98};
however, most of the events correspond to the production of a \emph{real}
$Z$-boson in reaction $e^+e^- \rightarrow \gamma Z \rightarrow\gamma \nu
\bar{\nu}$, so bounds of Ref.~\cite{Ferrari98} are inapplicable to
reaction $e^+ e^- \rightarrow \gamma Z^* \rightarrow \gamma N \bar{N}$ for
$m_N > m_Z/2$. Also, although the cross section for single photon production
in $e^+ e^-$ annihilation at LEP~II is big, it is saturated by $t$-channel
diagrams, where $\nu_e$ are produced, so there is no chance to observe $N$
production with reasonable statistics using the present data.

For particles with masses of the order of $m_Z/2$ oblique corrections
drastically differ from what we have in the heavy fermion limit. In
particular, renormalization of $Z$-boson wave function produces large
negative contribution to $V_A$. From the analysis of the initial set of
precision data in Refs.~\cite{Evans94,Bamert95} (published in years
1994-1995) it was found that the existence of additional light fermions with
masses around $50\GeV$ is allowed. Now analyzing all precision data and
using bounds from direct searches we conclude that the \emph{only} presently
allowed light fermion is neutral lepton $N$. This is clearly visible in
Fig.~\ref{fig:GEN.300}, where we show the $1\sigma$, $2\sigma$, etc.\
allowed domain as a function of $m_N$ and $m_E$. This figure was generated
using the approximate $\chi^2$ given in Eq.~\eqref{eq:GAU.310}, just as we
did in the previous section, and $m_U$, $m_D$, $m_t$ are correctly
integrated out. It is immediate to see that the $1\sigma$ region include
only a small area around $m_N = 45 \div 60\GeV$, $m_E = 100 \div 130\GeV$,
and no other solutions are allowed by precision measurements.

As a further example, we take $m_U = 220\GeV$, $m_D = 200\GeV$, $m_E =
100\GeV$ and draw the exclusion plot in coordinates $(N_g, m_N)$ using the
program \texttt{LEPTOP}. Results are shown in Fig.~\ref{fig:GEN.310} and
looking at it we can see that if the neutrino is light enough the
description of the data is not worse than in the Standard Model: for
$m_N\approx 50\GeV$ even two new generations are allowed within $1.5\sigma$.

\section{The case of SUSY} \label{sec:GEN.40}

\begin{figure}[!t] \centering
    \includegraphics[width=0.9\textwidth]{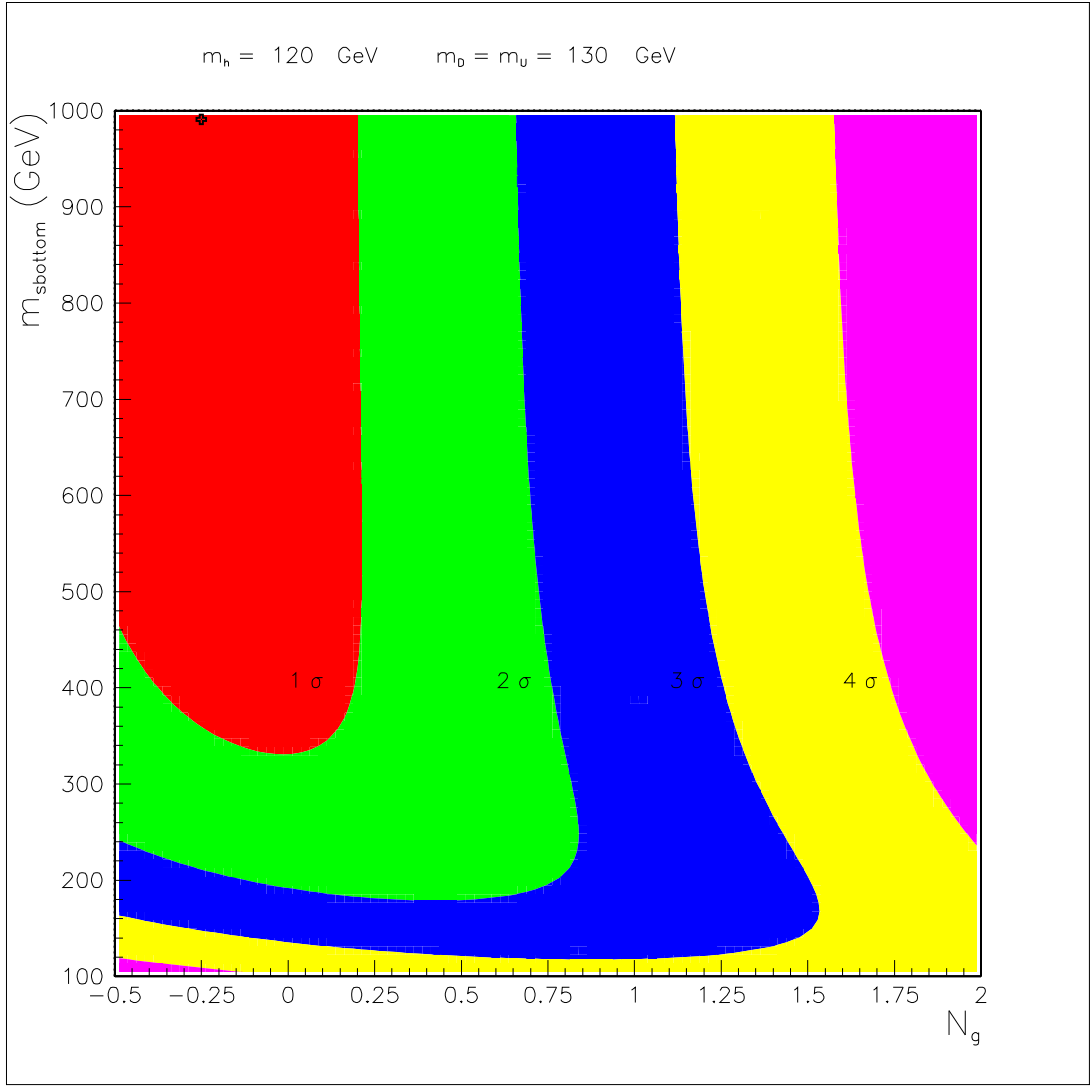}
    \caption[Exclusion plot in the plane $(N_g, m_\mathrm{sbottom})$]{
      The 2-dimensional exclusion plot for the $N_g$ degenerate extra
      generations and the mass of sbottom $m_{\tilde{b}}$ in SUSY models and
      for the choice $m_D = m_U = m_E = m_N = 130\GeV$, using $m_H =
      120\GeV$, $m_{\tilde{g}} = 200\GeV$ and assuming the absence of
      $\tilde {t}_L - \tilde {t}_R$ mixing. Little cross corresponds to
      $\chi^2$ minimum; regions show $< 1\sigma$, $< 2\sigma$, etc.\ allowed
      domains.}
    \label{fig:GEN.400}
\end{figure}

\begin{figure}[!t] \centering
    \includegraphics[width=0.9\textwidth]{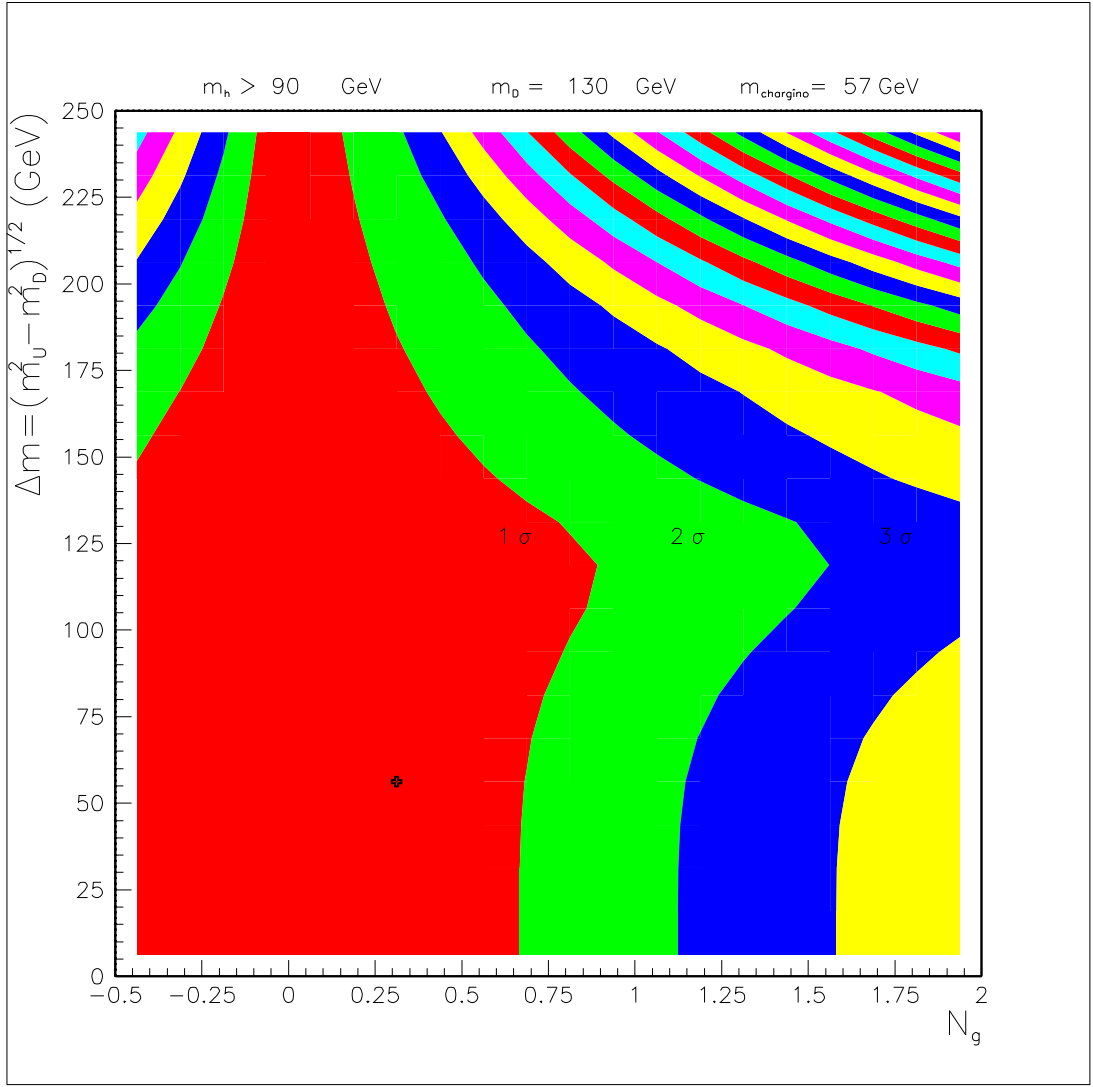}
    \caption[Exclusion plot in the plane $(N_g, \Delta m)$
    (higgsino-dominated case)]{
      Constraints on the number of extra generations $N_g$ and the mass
      difference in the extra generations $\Delta m$ in case of $57\GeV$
      higgsino-dominated quasi degenerate chargino and neutralino. The
      lowest allowed value $m_D=130\GeV$ from Tevatron search~\cite{PDG98}
      was used and $m_E=m_D$, $m_N=m_U$ was assumed. All electroweak
      precision data and $m_H > 90\GeV$ at $95\pCL$~\cite{LEPII99} were used
      in the fit. The cross corresponds to $\chi^2$ minimum; regions show $<
      1\sigma$, $< 2\sigma$, etc.\ allowed domains.}
    \label{fig:GEN.410}
\end{figure}

\begin{figure}[!t] \centering
    \includegraphics[width=0.9\textwidth]{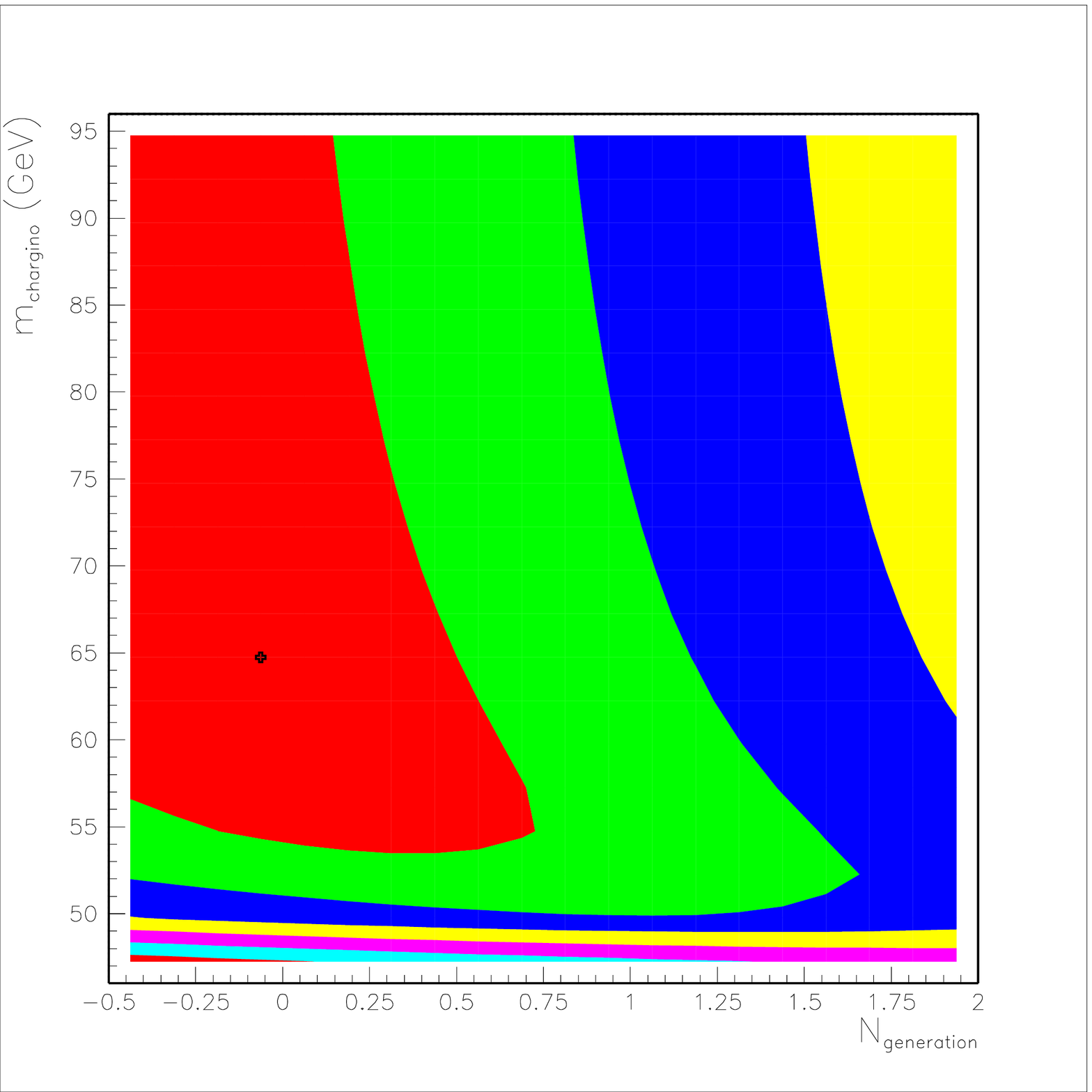}
    \caption[Exclusion plot in the plane $(N_g, m_{\tilde{\chi}})$
    (higgsino-dominated case)]{
      Constraints on the number of extra generations $N_g$ and the gaugino
      mass $m_{\tilde{\chi}}$ in case of higgsino-dominated quasi degenerate
      chargino and neutralino. The lowest allowed value $m_D = 130\GeV$ from
      Tevatron search~\cite{PDG98} was used and $m_E = m_D$, $m_N = m_U
      \approx 180\GeV$ ($\Delta m = 125\GeV$) was assumed. All electroweak
      precision data and $m_H > 90\GeV$ at $95\pCL$~\cite{LEPII99} were used
      in the fit. The cross corresponds to $\chi^2$ minimum; regions show $<
      1\sigma$, $< 2\sigma$, etc.\ allowed domains.}
    \label{fig:GEN.420}
\end{figure}

\begin{figure}[!t] \centering
    \includegraphics[width=0.9\textwidth]{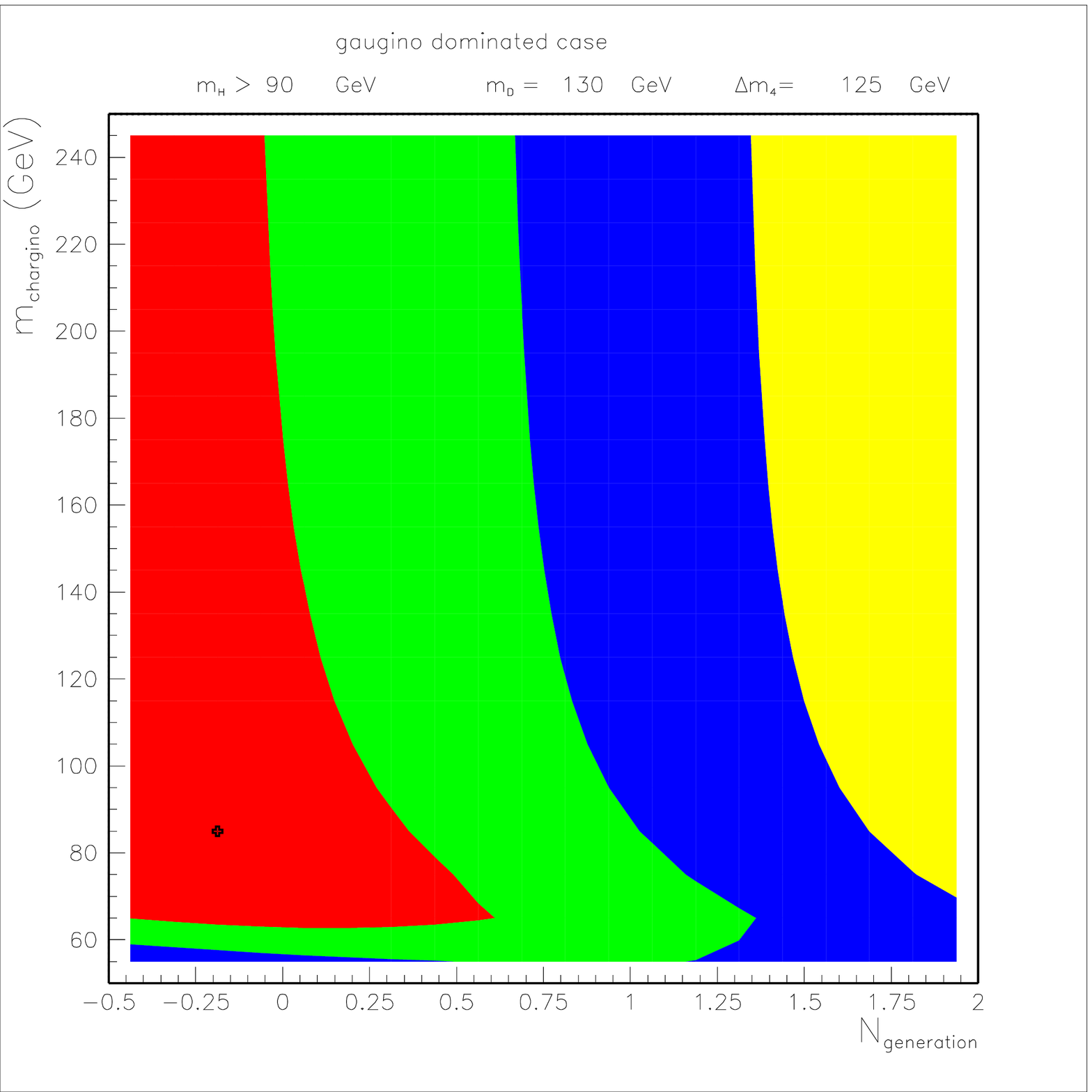}
    \caption[Exclusion plot in the plane $(N_g, m_{\tilde{\chi}})$
    (wino-dominated case)]{
      Constraints on the number of extra generations $N_g$ and the gaugino
      mass $m_{\tilde{\chi}}$ in case of wino-dominated quasi degenerate
      chargino and neutralino. The lowest allowed value $m_D = 130\GeV$ from
      Tevatron search~\cite{PDG98} was used and $m_E = m_D$, $m_N = m_U
      \approx 180\GeV$ ($\Delta m = 125\GeV$) was assumed. All electroweak
      precision data and $m_H > 90\GeV$ at $95\pCL$~\cite{LEPII99} were used
      in the fit. The cross corresponds to $\chi^2$ minimum; regions show $<
      1\sigma$, $< 2\sigma$, etc.\ allowed domains.}
    \label{fig:GEN.430}
\end{figure}

In this section we investigate bounds on extra generations which occur in
SUSY extensions. When SUSY particles are heavy they decouple (i.e.\ their
contributions to electroweak observables become power suppressed) and the
same Standard Model exclusion plots shown in
Figs.~\ref{fig:GEN.200}-\ref{fig:GEN.310} are valid. The present lower
bounds on the sparticle masses from direct searches close most of the
regions which may be interesting for radiative corrections, leaving
available mainly only this decoupled domain. One possible exception is a
contribution of the third generation squark doublet, enhanced by large
stop-sbottom splitting: in this way we get noticeable positive contributions
to functions $V_i$~\cite{Gaidaenko98,Gaidaenko99}, which may help to
compensate negative contributions of degenerate extra generations.

We analyze the simplest case of the absence of $\tilde{t}_L - \tilde{t}_R $
mixing in Fig.~\ref{fig:GEN.400}. In this figure, extra fermions are assumed
to be degenerate with common mass $130\GeV$, and according to
Ref.~\cite{Gaidaenko98} the masses of the stop and sbottom are related by
the MSSM formula:
\begin{equation} \begin{split}
    m_{\tilde{t}_L}^2 - m_{\tilde{b}_L}^2 & = 
      m_t^2 - m_b^2 + m_W^2 \cos 2\beta \\
    & \approx m_t^2 \text{~(for $\tan\beta \approx 2$)}.
\end{split} \end{equation}
The exclusion plot presented in Fig.~\ref{fig:GEN.400} is in coordinates
$(N_{g}, m_{\tilde{b}})$: it is straightforward to see that also with
inclusion of SUSY corrections new heavy generations are disfavored.

To understand why adding sfermions does not help to improve the quality of
the fit, let us consider the case of horizontally degenerate heavy fermions.
In this limit, the expressions for the $\Delta V_i$ corrections coming from
an extra generation are well approximate by Eq.~\eqref{eq:GEN.60}.
Concerning sfermions contributions, according to Ref.~\cite{Gaidaenko98}
they are almost universal, and after some simple calculations we get the
approximate relation:
\begin{equation}
    \Delta'' V_i = \frac{4}{3} \LT(
    \frac{m_{\tilde{t}} - m_{\tilde{b}}}{m_Z} \RT)^2.
\end{equation}
Comparing this formula with Eq.~\eqref{eq:GEN.60}, it is clear why the
quality of the fit is unaffected: contributions of sfermions simply mimic
those of non-degenerate extra fermions, and since the latters are excluded
by precision measurements (as we discussed in detail in
Sec.~\ref{sec:GEN.20}) also the formers are.

The case of SUSY sfermions clearly proves that to change the fate of extra
generations we must search for New Physics giving \emph{non-universal}
contributions. From Chapter~\ref{sec:GAU}, we know that a chargino almost
degenerate with the lightest neutralino and close in mass to $m_Z/2$ induces
a large negative correction into $V_A$, just as a $50\GeV$ neutrino does, so
in this case we can expect a general improvement of the quality of the fit.
To check whether this really happens, let us consider once more the case of
horizontal degeneracy and make some numerical estimates. Subtracting the
experimental values $\overline{\delta V_i}$ given in~\eqref{eq:GAU.300} from
Eq.~\eqref{eq:GEN.60} we get:
\begin{equation} \label{eq:GEN.400}
    \delta^\mathrm{th-exp}[V_m,\; V_A,\; V_R] =
    [-0.340,\; +0.327,\; -0.894] + [y,\; y,\; y],
\end{equation}
where $y$ is defined in Eq.~\eqref{eq:GEN.200y}. Clearly, the overall
$\chi^2$ will be better when all the $\delta^\mathrm{th-exp} V_i$ vanish.
The main obstacle to this is the large splitting between negative
$\delta^\mathrm{th-exp} V_{m,R}$ and positive $\delta^\mathrm{th-exp} V_A$:
universal $SU(2)_V$ breaking contributions described by $y$ cannot
compensate it. Now let us include effects of almost degenerate
chargino-neutralino and give some numerical example. In the
higgsino-dominated case, for $m_\tdC = 57\GeV$ we have:
\begin{equation} \label{eq:GEN.410h}
    \delta^\tdH [V_m,\; V_A,\; V_R] = [+0.185,\; -0.678,\; +0.237],
\end{equation}
while in the wino-dominated case with $m_\tdC = 63\GeV$ we find:
\begin{equation} \label{eq:GEN.410w}
    \delta^\tdW [V_m,\; V_A,\; V_R] = [+0.209,\; -0.816,\; +0.177].
\end{equation}
Comparing these expressions with Eq.~\eqref{eq:GEN.400}, we see that
adding~\eqref{eq:GEN.410h} or~\eqref{eq:GEN.410w} to it partially
compensates the large differences $\delta^\mathrm{th-exp} \LT( V_m - V_A
\RT)$ and $\delta^\mathrm{th-exp} \LT( V_R - V_A \RT)$, so in both these
cases the overall $\chi^2$ will be smaller. Also, since we have
$\delta^\mathrm{th-exp} \LT( V_m - V_R \RT) > 0$, $\delta^\tdH \LT( V_m -
V_R \RT) < 0$ and $\delta^\tdW \LT( V_m - V_R \RT) > 0$ (see also
Figs.~\ref{fig:GAU.20} and~\ref{fig:GAU.30}), we conclude that the quality
of the fit will be slightly better in the higgsino-dominated case than in
the wino-dominated case. This result is in agreement with
Figs.~\ref{fig:GEN.420} and~\ref{fig:GEN.430}.

Fig.~\ref{fig:GEN.410} demonstrates how presence of chargino-neutralino pair
(dominated by higgsino) with mass $57\GeV$ relaxes the bounds shown on
Fig.~\ref{fig:GEN.200}: we see that one extra generation of heavy fermions
is now allowed within $1.5 \sigma$ domain. Also, in Figs.~\ref{fig:GEN.420}
and~\ref{fig:GEN.430} we draw the exclusion plots in coordinates $(N_g,
m_\tdC)$ for both higgsino and wino domination, respectively: it is evident
how precision measurements favor a light value of chargino-neutralino mass
as soon as the number of extra generations increases.

\section{Conclusions} \label{sec:GEN.50}

As we saw in Sec.~\ref{sec:GEN.30}, inclusion of new generations in Standard
Model is not excluded by precision data if new neutral leptons are rather
light having mass of the order of $50\GeV$ (see Fig.~\ref{fig:GEN.310}).
Mixing of new leptons with leptons from three known generations should be
small, $\theta \lesssim 10^{-6}$, to avoid bounds from direct search at
LEP~II. We can not exclude stability of one of these new neutrinos; in this
case it becomes interesting for cosmology. If the early universe was charge
symmetric, annihilation of $N\bar{N}$ in primordial plasma bounds the
present abundance $\Omega_N$ of these particles to be less than $10^{-3}$
(formula for heavy neutrino abundance in case $m_N \ll m_Z$ was obtained
in~\cite{Vysotsky77}). If the early universe was charge asymmetric, the
abundance of relic $N$'s is larger. However their contribution to mass
density of the halo of our galaxy can not be larger than $0.01 \div 0.1$~-
otherwise they would be already detected in laboratory searches for dark
matter~\cite{Caldwell94}. Even this small admixture of $50\GeV$ neutrino in
the halo of our galaxy can help to explain gamma background through
$N\bar{N}$ annihilation into $e^+ e^-$ with subsequent scattering of
electrons and positrons on optical photons~\cite{Fargion99}.

Concerning SUSY extensions: if masses of sparticles are of the order of
several hundred GeV or larger their contribution to electroweak radiative
corrections is negligible, hence the above statements remain valid. However
in the case of quasi degenerate chargino and neutralino with masses about
$60\GeV$ extra generations of heavy fermions appear to be less forbidden
than without SUSY.

In order to experimentally investigate the case of $m_N < 50\GeV$ a special
post-LEP~II run of LEP~I' measuring the Z-line shape slightly above the
Z-peak is needed. In this way the bound~\cite{Abreu92} will be improved. For
$m_N > 50\GeV$ search for the reaction $e^+ e^- \rightarrow \gamma Z^*
\rightarrow \gamma N \bar{N}$ with larger statistics than that
of~\cite{Ferrari98} and improved systematics is needed. Finally, further
experimental search for light chargino and neutralino~\cite{Abreu99} is of
interest. These searches could close the existing windows of ``light'' extra
particles, or open a door into a realm of New Physics.

\clearemptydoublepage

\nochapter{Conclusions}

In this thesis, we considered effects of radiative corrections on
electroweak precision measurements, using them to put bounds on New Physics
parameters. Here we summarize the main results obtained:

\begin{itemize}
  \item In Chapter~\ref{sec:ANG} we investigated the origin of the numerical
    closeness among three different definitions of the electroweak mixing
    angle, finding that the degeneracy between $\hat s^2$ and $s^2$, as well
    as that between $s_l^2$ and $s^2$, occurs only for $m_t \sim 170\GeV$
    and therefore is merely accidental. Conversely, for the case of $\hat
    s^2$ and $s_l^2$ the dependence of their difference on the $m_t$ in only
    logarithmic, so their closeness would have occurred even if the top
    quark mass would have been quite different from its actual value. Also,
    we provided an explicit and very simple relation between the
    phenomenological quantity $s^2$ and the $\MS$ parameter $\hat s^2$.

  \item In Chapter~\ref{sec:RUN} we discussed in detail how both the
    decoupling and the non-decoupling approach to the running of coupling
    constants in the $\MS$ scheme produce the same numerical value of
    $m_\GUT$, despite of the different initial conditions. The concept of
    threshold was introduced and used, and the dependence of $\hat\alpha_s$
    on the SM parameters $\hat\alpha$, $\hat s^2$ and $\hat m_t$ and the
    MSSM quantities $m_\SUSY$ and $\tan\beta$ was studied. Combining the
    results obtained in this chapter with analysis of $\hat s^2$ carried out
    in the previous chapter it is straightforward to evaluate the impact of
    the numerical value of the physical observables $\albar$, $s^2$ and
    $m_t$ on the prediction of $\hat\alpha_s$ from the demand of SUSY Grand
    Unification.

  \item In Chapter~\ref{sec:GAU} we analyzed in detail the effects of
    radiative corrections for the case of a chargino almost degenerate with
    the lightest neutralino on electroweak precision measurements. Both the
    higgsino-dominated and the wino-dominated scenario were studied, and for
    these limits simple analytical formulae were derived. For the case of
    wino domination, our bound $m_\tdC \gtrsim 56\GeV$ is presently the
    strongest constraint which can be imposed without making any assumption
    on the mass spectrum of other superpartners.

  \item In Chapter~\ref{sec:GEN} we investigated the effects of new fermion
    generations on precision measurements, showing that even 1 extra
    generation with all particles heavier that $Z$ boson is strongly
    disfavored by present experimental data. However, for the specific case
    of the extra neutrino around $50\GeV$ in mass, the situation change and
    the quality of the fit is not worse than the SM. Concerning SUSY
    extensions, contributions of sfermions do not affect bounds on extra
    generations, while for the case of almost degenerate chargino/neutralino
    these bounds are relaxed.
\end{itemize}

The results discussed in this thesis are presented in the published
papers~\cite{Maltoni98b,Maltoni99a,Maltoni99b} and in
Ref.~\cite{Maltoni99c}, currently submitted for publication on PLB. Also,
results quoted in Chapters~\ref{sec:GAU} and~\ref{sec:GEN} were presented at
the conference PASCOS99, and will appear in the
proceedings~\cite{Maltoni99d,Maltoni99e}.

\clearemptydoublepage

\nochapter{Acknowledgments}

It is a pleasure for me to express my gratitude to Mikhail Vysotsky for
introducing me in the field of Particle Physics and Phenomenology, for his
continuous support during my Ph.D.\ activity and for his friendship.

I am also grateful to Victor Novikov, Lev Okun and Alexander Rozanov,
together with whom the work presented in Chapter~\ref{sec:GEN} was
developed, for their kindness and constant availability, and to Giovanni
Fiorentini for always supporting me during the last three years.

A special thanks to all the researchers of the theoretical group of Ferrara
University and of the `Istituto TeSRE' of Bologna for their friendship and
kindness, and to my family and friends for their constant patience and
encouragement.

Finally, I am grateful to Lidia for leading me to look at things from a
different point of view.

\clearemptydoublepage

\appendix
\chapter{The $V_i$ functions} \label{sec:APP}

In this appendix we give formulas for the $V_i$ functions in terms of gauge
boson self-energies, vertex and box diagrams. We only summarize the final
results, addressing the reader to Ref.~\cite{NOV93} for a more complete
discussion.

The expression for $V_m$ can be obtained comparing Eq.~\ref{eq:DEF.600m}
with~\ref{eq:ANG.90}:
\begin{equation} \begin{split} \label{eq:APP.10m}
    V_m = \frac{16\pi s^4}{3\albar} \bigg[
    & \frac{c^2}{s^2} \Pi_Z(m_Z^2) 
    + \LT( 1 - \frac{c^2}{s^2} \RT) \Pi_W(m_W^2) \\
    & - \Pi_W(0) - \Pi_\gamma(m_Z^2) - 2 \frac{s}{c} \Pi_\GZ(0) - \altw - D
    \bigg].
\end{split} \end{equation}

The function $V_A$ is given in Eq.~(77) of Ref.~\cite{NOV93}:
\begin{equation} \label{eq:APP.10a}
    V_A = \frac{16\pi c^2 s^2}{3\albar} \LT[ \Pi_Z(m_Z^2) - \Pi_W(0) 
    - \Sigma'_Z(m_Z^2) - D - 8cs F_A^{Ze} \RT],
\end{equation}
where $\Sigma'_Z(m_Z^2)$ comes from $Z$-boson wave-function renormalization
and $F_A^{Ze}$ is the axial part of the $Zll$ vertex (see
Sec.~\ref{sec:ANG.10.40}).

The function $V_R$ can be derived comparing Eqs.~\ref{eq:DEF.600r},
\ref{eq:ANG.30l} and~\ref{eq:ANG.170}:
\begin{equation} \begin{split} \label{eq:APP.10r}
    V_R = \frac{16\pi c^2 s^2}{3\albar} \bigg[
    & - \frac{c^2 - s^2}{cs} \LT[ F_V^{Ze} - \LT( 1 - 4s^2 \RT) F_A^{Ze} +
    \Pi_\GZ(m_Z^2) \RT] \\
    & - \Pi_\gamma(m_Z^2) + \Pi_Z(m_Z^2) - \Pi_W(0) 
    - 2 \frac{s}{c} \Pi_\GZ(0) - \altw - D \bigg],
\end{split} \end{equation}
where $F_V^{Ze}$ is the vector part of the $Zll$ vertex (see
Sec.~\ref{sec:ANG.10.40}).

These functions were introduced in Ref.~\cite{NOV93} to study the dependence
of precision measurements on the top and higgs masses, and the overall
numerical coefficients were chosen in such a way that the leading top
contribution into $V_i$ is simply $(m_t/m_Z)^2$. However, since extra
particles which occur in many extensions of the SM (like for example the
MSSM) participate to $V_i$ through radiative corrections, these functions
can also be used as a convenient parameterization of New Physics. In this
case, it is more practical to write $V_i = V_i^\SM + \delta^\NP V_i$ (see
also Sec.~\ref{sec:GAU.30}), since in general the expressions for the
$\delta^\NP V_i$ are simpler than Eqs.~(\ref{eq:APP.10m}-\ref{eq:APP.10r}).
In particular:
\begin{itemize}
  \item vertex and box contributions can usually be neglected;
    
  \item the combination $\Pi_\gamma(m_z^2) + \altw$ simply reduces to
    $\Sigma'_\gamma(0)$;

  \item $\Pi_\GZ(0)$ gets contributions only from the gauge sector of the
    SM, so it vanishes when considering effects of New Physics (at least for
    what concerns the MSSM and extra generations, the two cases we are
    interested in).
\end{itemize}
Therefore, we have:
\begin{align}
    \label{eq:APP.20m} \delta^\NP V_m & = \frac{16\pi c^2 s^2}{3\albar} 
      \LT[ \Pi_Z(m_Z^2) - \LT( 1 - \frac{s^2}{c^2} \RT) \Pi_W(m_W^2)
      - \frac{s^2}{c^2}\Pi_W(0) - \frac{s^2}{c^2}\Sigma'_\gamma(0)
      \RT]_\NP, \\
    \label{eq:APP.20a} \delta^\NP V_A & = \frac{16\pi c^2 s^2}{3\albar}
      \LT[ \Pi_Z(m_Z^2) - \Pi_W(0) - \Sigma'_Z(m_Z^2) \RT]_\NP, \\
    \label{eq:APP.20r} \delta^\NP V_R & = \frac{16\pi c^2 s^2}{3\albar}
      \LT[ \Pi_Z(m_Z^2) - \Pi_W(0) - \Sigma'_\gamma(0)
      - \frac{c^2 - s^2}{cs} \Pi_\GZ(m_Z^2) \RT]_\NP.
\end{align}

It is also useful to give here explicit formulas relating the $\delta^\NP
V_i$ functions to the $S$, $T$, $U$ parameters introduced by Peskin and
Takeuchi~\cite{Peskin90}:
\begin{align}
    S & = \frac{3}{4\pi} \LT( \delta^\NP V_A - \delta^\NP V_R \RT), \\
    T & = \frac{3}{16\pi c^2 s^2} \delta^\NP V_A, \\
    U & = -\frac{3}{4\pi (c_2 - s^2)}
      \LT[ \LT( \delta^\NP V_A - \delta^\NP V_m \RT)
      - 2 s^2 \LT( \delta^\NP V_A - \delta^\NP V_R \RT) \RT].
\end{align}

Let us conclude defining some functions which are widely used in
Chapters~\ref{sec:GAU} and~\ref{sec:GEN}:
\begin{align}
    \begin{split} \label{eq:APP.40w}
	F(m_W^2, m_U^2, m_D^2) &= -1 + \frac{m_U^2 + m_D^2}{m_U^2 - m_D^2}
	  \ln\LT(\frac{m_U}{m_D}\RT) \\
	& - \int\limits_0^1 \ln \frac{t^2 m_W^2 
	  - t \LT( m_W^2 + m_U^2 - m_D^2 \RT) + m_U^2}{m_U m_D} \, dt,
    \end{split} \\
     \label{eq:APP.40f} F(x) \equiv F(m_Z^2, m_Z^2 x, m_Z^2 x) & = 
    \begin{cases}
	2 \LT[ 1 - \sqrt{4x-1} \arcsin \LT( \frac{1}{\sqrt{4x}} \RT) \RT]
	  & x > \frac{1}{4},\\[8pt]
	2 \LT[ 1 - \sqrt{1-4x} \ln \LT( \frac{1+\sqrt{1-4x}}{\sqrt{4x}} \RT) \RT]
	  & x < \frac{1}{4},
    \end{cases} \\
    \label{eq:APP.40p} F'(x) \equiv -x \frac{d}{dx} F(x) 
    & = \frac{1- 2x F(x)}{4x - 1}.
\end{align}

The following relation is useful in deriving
Eqs.~(\ref{eq:GEN.20m}-\ref{eq:GEN.20r}):
\begin{equation}
    \int\limits^1_0 (t^2 - t) \ln\LT( t^2 - t + x \RT) \, dt =
    \frac{1+2x}{6} F(x) - \frac{1}{18} - \frac{1}{6} \ln x.
\end{equation}

\clearemptydoublepage

\end{document}